\documentclass[final,3p,nopreprintline]{elsarticle}
\ifdefined\pdfmapfile\pdfmapfile{+lm.map}\pdfmapfile{+cm.map}\pdfmapfile{+symbols.map}\pdfmapfile{+euler.map}\fi

\usepackage[T1]{fontenc}
\usepackage[utf8]{inputenc}
\usepackage{lmodern,microtype}
\usepackage{graphicx,tikz}
\usetikzlibrary{arrows.meta,calc,patterns}
\usepackage{amsmath,amssymb,amsthm,bm,mathtools}
\usepackage{booktabs,tabularx,array,enumitem,longtable}
\usepackage{needspace,float}
\usepackage[section]{placeins}
\usepackage{xcolor}
\usepackage{listings}
\usepackage[colorlinks=true,linkcolor=blue!45!black,citecolor=blue!45!black,urlcolor=blue!45!black]{hyperref}
\biboptions{numbers,sort&compress}
\setlist[itemize]{leftmargin=*,itemsep=5pt,topsep=5pt}
\setlist[enumerate]{leftmargin=*,itemsep=4pt}
\numberwithin{equation}{section}
\newtheorem{theorem}{Theorem}[section]
\newtheorem{proposition}{Proposition}[section]
\newtheorem{lemma}{Lemma}[section]
\newtheorem{corollary}{Corollary}[section]
\theoremstyle{definition}
\newtheorem{remark}{Remark}[section]
\newtheorem{example}{Example}[section]
\newenvironment{recalled}[2][]{\par\medskip\noindent\textbf{#2}\if\relax\detokenize{#1}\relax\else\ \textbf{(#1)}\fi\textbf{.}\ \itshape}{\par\medskip}
\newcounter{algorithm}
\newenvironment{algorithmbox}[1]{\par\medskip\noindent\rule{\linewidth}{0.8pt}\par\nobreak\vspace{1pt}\refstepcounter{algorithm}\noindent\textbf{Algorithm~\thealgorithm.}\ #1\par\nobreak\vspace{-5pt}\noindent\rule{\linewidth}{0.4pt}\par\nobreak\smallskip}{\par\vspace{-4pt}\noindent\rule{\linewidth}{0.8pt}\par\medskip}
\newcommand{\pv}{\operatorname{PV}}
\newcommand{\HH}{\mathcal H_I}
\newcommand{\dd}{\mathrm d}
\newcommand{\norm}[1]{\left\lVert#1\right\rVert}
\newcommand{\doi}[1]{\href{https://doi.org/#1}{\nolinkurl{doi:#1}}}
\journal{Applied Mathematical Modelling}
\begin{document}
\begin{frontmatter}
\title{The Airfoil Equation Beyond Zero Thickness: Exact Solutions for Aeroelastic Analysis}
\author[am,mme]{Mohamed A. Serry}
\author[mme]{Sean D. Peterson}
\affiliation[am]{organization={Department of Applied Mathematics, University of Waterloo},
 city={Waterloo}, state={Ontario}, country={Canada}}
\affiliation[mme]{organization={Department of Mechanical and Mechatronics Engineering, University of Waterloo},
 city={Waterloo}, state={Ontario}, country={Canada}}
\begin{abstract}
The classical airfoil integral equation, which relates the pressure jump across a zero-thickness airfoil to its downwash, underlies much of the mathematical theory of subsonic aeroelasticity, but it discards the thickness of real sections. We study its counterpart for airfoils with prescribed camber and small, nonzero thickness in steady subsonic flow: a thickness-modified airfoil equation, a coupled finite Hilbert transform system for the jumps of the disturbance velocity, in which thickness and camber multiply the unknowns. A similar system was derived earlier and solved by collocation, without guarantees of existence, uniqueness, or accuracy. We revisit its formulation, extend it to compressible flow, and treat it analytically. Adapting techniques for systems of singular integral equations, we give the solution in weighted $L^2$ spaces by a closed-form integral formula, and bound the solution error by the surface-slope approximation error. For polynomial camber and two weighted-polynomial thickness families that uniformly approximate continuous sections with closed edges, the solution is a finite weighted-rational expression, with poles from polynomial root finding and amplitudes from a well-posed linear system; a MATLAB implementation is provided in an appendix. Comparisons with published wind-tunnel pressures for symmetric and cambered NACA sections show accuracy comparable to inviscid XFOIL. The loads are explicit in the incidence, enabling direct coupling with structural models. Applied to an elastic cantilever wing and to an elastic airfoil with a camber degree of freedom, each for twenty undeformed NACA sections, the approach predicts divergence speeds and camber divergence pressures close to those obtained with XFOIL-based loads.
\end{abstract}
\end{frontmatter}

\section{Introduction}
When designing aerial systems, analyzing the coupling between lifting-surface deformations and aerodynamic loads is essential to ensuring the desired aerodynamic performance and structural safety across different flight regimes. This analysis requires repeated evaluations of the aerodynamic model in parameter sweeps over geometry and flight conditions, iterative calculations of static equilibrium, and sensitivity analyses for optimization. The model should therefore provide accurate loads efficiently and admit a unique solution for every admissible configuration.

Existence and uniqueness are practical requirements. When the model equations admit no solution, a numerical method may nevertheless return an approximate load, potentially producing a spurious instability or optimum. When multiple solutions exist, the computed load may depend on the discretization, initial guess, or collocation points unless an appropriate solution-selection criterion is imposed. A unique solution ensures that the model assigns a single, well-defined set of aerodynamic loads to each configuration. Moreover, because airfoil geometries are often represented by fitted profiles and may deviate from their intended shapes during manufacturing, bounds on the propagation of geometry errors to aerodynamic loads are needed to quantify load uncertainty and establish geometric accuracy requirements. Explicit dependence of the loads on geometric parameters further clarifies how design changes affect aerodynamic loading and aeroelastic response.

For a zero-thickness section in subsonic flow, the Possio integral equation provides an established framework relating the pressure jump across the chord to the prescribed downwash~\cite{Kussner1941,Balakrishnan2003}. Its derivation from the linearized field equations, solvability, and application to aeroelastic stability have been studied extensively~\cite{BalakrishnanShubov,HodgesPierce}. In the steady limit, it reduces to the classical airfoil equation, for which finite Hilbert transform theory provides inversion formulas, endpoint conditions, and solvability classes~\cite{Sohngen1954,OkadaProssdorf1997,GohbergKrupnik,CoifmanSteinerberger,CurberaOkadaRicker2019}. Related steady pressure--downwash formulations have been used to determine the divergence of wings near the ground~\cite{serry2017subsonic} and plates in axial flow~\cite{SerryTuffaha2018}.

A zero-thickness model does not distinguish between the upper and lower surfaces of an airfoil and therefore omits thickness effects on aerodynamic loads, including the influence of the thickness distribution on pitching moment and aeroelastic divergence. Several analytical approaches account for thickness. Classical thin-airfoil theory represents an airfoil by a vortex sheet on its chord, yielding explicit relations between camber, incidence, and loading~\cite{Acheson1990,MITThinAirfoil}; thickness enters through a decoupled source distribution that contributes no lift. Conformal-mapping formulations account for the full section geometry and provide two-dimensional incompressible solutions for surface velocities, lift, and pitching moments~\cite{Theodorsen411,TheodorsenGarrick452}. However, they require a geometry-specific map between the exterior of the airfoil and the exterior of a circle, which complicates their extension to compressible flow and their application to deforming sections. Systematic second-order subsonic theory incorporates thickness, camber, incidence, and edge corrections through successive singular integrals, transferring the surface boundary condition to the chord by an asymptotic expansion~\cite{VanDyke1274}; explicit results are available for selected families, whereas general profiles require geometry-dependent integrations. Thickness-dependent singular integral models have supported aeroelastic stability analyses of wedge-shaped sections, whose special geometry permits explicit solutions~\cite{Tuffaha2009,BalakrishnanTuffaha}. Analytical solutions that retain thickness, camber and incidence have also been obtained for single sections~\cite{Ayton2017Geometry} and, by Riemann--Hilbert methods, for cascades of airfoils~\cite{BaddooAyton2018}, and analytical treatments exist for porous airfoils~\cite{HajianJaworski}. In these solutions the boundary condition is linearized so that the surface slopes multiply only the freestream velocity, and the thickness and lifting problems decouple.

The closest predecessor of the present work is the formulation of Mateescu and Abdo~\cite{MateescuAbdo2005}. They retain the no-penetration condition on both faces of the section, including the products of the surface slopes with the disturbance velocity. These products couple the conditions on the two faces. The result is a pair of coupled singular integral equations on the chord for the jumps of the disturbance velocity across it, in which the thickness and camber slopes multiply the unknowns; for zero thickness the pair reduces to the classical airfoil equation. Their steady formulation builds on a velocity-singularity approach also developed for unsteady flows past oscillating rigid and flexible thin airfoils~\cite{MateescuAbdo2003,Mateescu2011MESA}. They solve the system approximately: the solution is assumed to be a polynomial multiplied by a prescribed endpoint factor, and the polynomial coefficients are determined by collocation at selected points on the chord. The polynomial degrees and the collocation points are choices of the user.

The system of Mateescu and Abdo is a system of singular integral equations with Cauchy kernels. Existence and uniqueness for such systems, and for the equivalent Riemann--Hilbert and linear conjugation problems, have been established under various conditions on the coefficients and on the class of solutions~\cite{Muskhelishvili1953,Vekua1967,Peters1972,Porter1986,KhuskivadzeKokilashviliPaatashvili1998,KokilashviliPaatashviliSamko2005,KokilashviliPaatashvili2007}. These results are in general not constructive, but they are useful: they identify the conditions under which a solution exists and is unique, and they supply techniques for reducing a system to simpler problems. In the present study we adapt some of these techniques, namely the Cauchy integral of the unknowns~\cite{Peters1972} and its factorization by an analytic function~\cite{Porter1986}, to the conditions imposed on the airfoil profiles, whose slopes may be discontinuous or unbounded. The reduction leads to the system of a flat plate in weighted $L^2$ spaces that contain the Kutta condition, and gives the solution as an integral formula.

This leaves two questions open. First, the available solutions are approximate: the explicit second-order results~\cite{VanDyke1274} are truncated expansions for selected families of sections, and the collocation solutions carry no estimate of their error; to our knowledge, no exact solutions of the coupled system with nonzero thickness have been reported. Second, sections are in practice represented by fitted profiles, and constructing camber and thickness distributions from measured or tabulated ordinates requires particular care near a rounded leading edge~\cite{TorresMarques2024}; no bound is available on how errors of the fitted geometry propagate to the aerodynamic solution. To our knowledge, no other derivation of this coupled system has been reported.

The objective of this paper is to place the thickness-modified airfoil equation on the same mathematical footing as its zero-thickness counterpart, in a form suitable for aeroelastic computation. Specifically, we aim to
\begin{enumerate}
 \item[(i)] rederive the coupled finite Hilbert transform system for steady subsonic flow so that compressibility is retained in the field equation, rather than introduced through a compressibility correction of the incompressible solution, and so that a prescribed deformation of the section is included;
 \item[(ii)] give the solution in closed form, in weighted spaces that exclude the trailing-edge singularity and so encode the Kutta condition: as an integral formula for general section geometry with piecewise H\"older-continuous slopes, including corners and a rounded leading edge, and as a finite rational expression for families of sections that approximate general profiles, including sections with a rounded leading edge, so that the loads are explicit functions of the incidence and the Mach number;
 \item[(iii)] bound the error of the solution in terms of the error of the fitted geometry;
 \item[(iv)] assess the resulting pressures and loads against inviscid XFOIL computations and published wind-tunnel measurements, and demonstrate their use in aeroelastic divergence analysis.
\end{enumerate}

The paper is organized as follows. Section~\ref{sec:geometry} defines the geometry, deformation variables and flow assumptions in the secant frame. Section~\ref{sec:airfoil-derivation} states the modified airfoil system and its normalized form. Section~\ref{sec:existence} gives the solution in weighted $L^2$ spaces as an integral formula, together with the conditions on the geometry under which it is unique. Section~\ref{sec:families} introduces the flat-endpoint and rounded-leading-edge profile families and their approximation properties. Section~\ref{sec:exact-solutions} constructs the finite rational solutions for these families. Section~\ref{sec:loads} gives the pressures and loads and their explicit dependence on the incidence and the Mach number. Section~\ref{sec:geometry-stability} bounds the error of the solution by the errors of the surface slopes. Section~\ref{sec:comparison-pressure-loads} validates the model against inviscid XFOIL and against measurements for symmetric and cambered NACA sections. Section~\ref{sec:aeroelastic-analysis} couples the loads to an elastic wing and to a section with a camber degree of freedom and determines their divergence. The final sections discuss limitations and conclusions. Derivations, proofs, and other supplementary materials are given in the appendices; Appendix~\ref{app:matlab} lists a MATLAB implementation of the construction for symmetric NACA sections, from the section number and the flow conditions to the surface pressures.

\section{Problem setup}\label{sec:geometry}\label{main:flow}
\begin{figure}[!htbp]\centering
\input{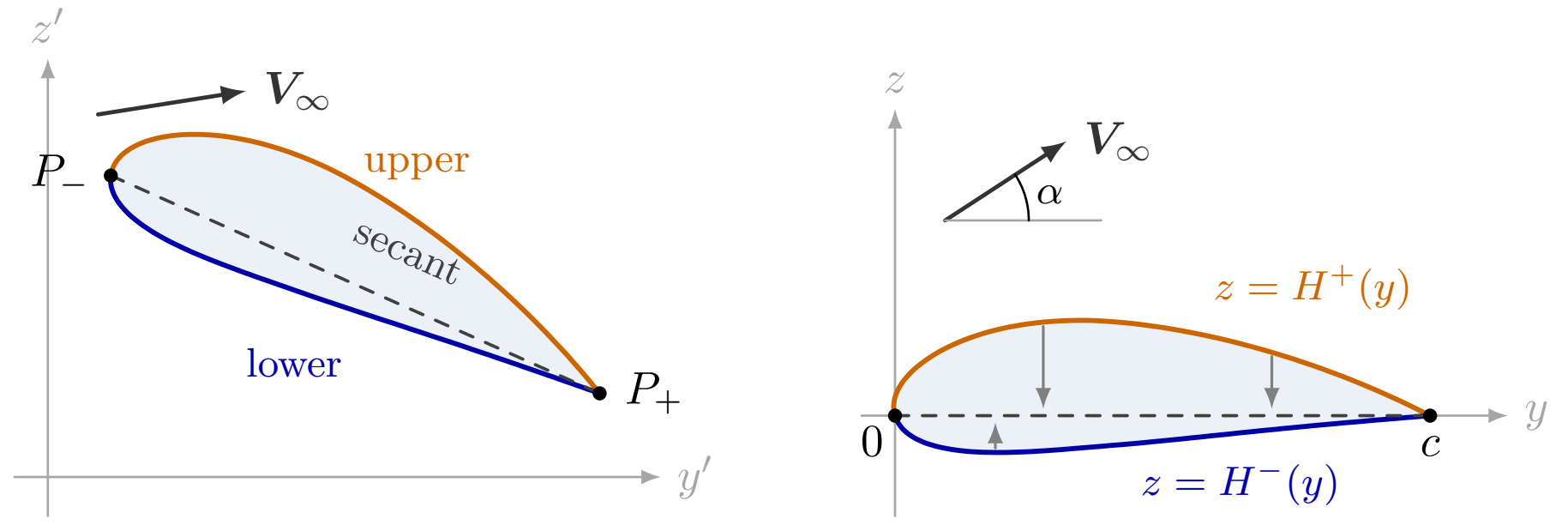}
\caption{Secant frame. (a) A section in arbitrary sectional coordinates, with the secant through its endpoints (dashed). (b) After translation and rotation: endpoints at $(0,0)$ and $(c,0)$, surfaces $z=H^\pm(y)$, and freestream at the angle $\alpha$ to the $y$ axis. Schematic.}
\label{fig:wing-schematic}
\end{figure}

We consider the flow past an airfoil section in sectional coordinates $(y,z)$. By a translation and an orthogonal rotation of the coordinates, the secant line through the leading and trailing edges is aligned with the $y$ axis, with the leading edge at the origin and the trailing edge at $(c,0)$; see Figure~\ref{fig:wing-schematic}. The problem is formulated in this secant frame from the outset. The freestream speed $U$, Mach number $M$ and density $\rho_\infty$ are unaffected by the transformation, and $\alpha$ denotes the angle between the freestream direction and the $y$ axis.

\subsection{Geometry}
On the chord interval $I=(y_-,y_+)=(0,c)$, the upper and lower surfaces are prescribed by single-valued functions $z=H^\pm(y)$, with $H^+\ge H^-$ and both functions zero at the endpoints. The reference $z=0$ is therefore the endpoint secant. We write the profiles as
\begin{equation}
 H^\pm=f^\pm+h_e,\qquad
 f^\pm(0)=f^\pm(c)=h_e(0)=h_e(c)=0.
 \label{main:imposed-perturbation}
\end{equation}
Here $f^\pm$ are the undeformed faces and $h_e$ is a deformation common to both faces. For a directly prescribed profile one may take $h_e=0$ and $f^\pm=H^\pm$. Define
\begin{equation}
 f_c=\tfrac12(f^++f^-),\quad f_t=\tfrac12(f^+-f^-)\ge0,
 \qquad g_c=f_c',\quad g_t=f_t',\quad g^\pm=(f^\pm)'.
 \label{geo:profile-slopes}
\end{equation}
The camber $f_c$ need not vanish in the interior. Upper--lower symmetry means $f_c=0$. The face and mean slopes are
\begin{equation}
 k^\pm=(H^\pm)',\label{main:straight-incidence}
\end{equation}
\begin{equation}
 k_c=\tfrac12(k^++k^-)=g_c+h_e',\qquad
 \tfrac12(k^+-k^-)=g_t.
 \label{air:effective-slopes}
\end{equation}
Endpoint compatibility gives $\int_0^c k_c(y)\,\mathrm dy=0$. For a rigid section pitched nose-up by $\theta$ from an attitude at which the freestream angle is $\alpha_\infty$, the secant frame rotates with the section, so that
\begin{equation}
 \alpha=\alpha_\infty+\theta,
 \qquad h_e=0,\qquad k_c=g_c.
 \label{air:rigid-perturbation}
\end{equation}

\subsection{Two-dimensional potential flow and disturbance variables}
Assume steady inviscid, irrotational, isentropic subsonic flow, with
\begin{equation}
 \bm V_\infty=U(\cos\alpha,\sin\alpha),\qquad
 M=U/a_\infty<1,\qquad \beta=\sqrt{1-M^2}.
\end{equation}
Here $a_\infty$ is the freestream sound speed and $\alpha$ is measured from the secant. Define $\nabla=(\partial_y,\partial_z)^T$ and $\nabla^2=\partial_{yy}+\partial_{zz}$. Write the total velocity potential as
\begin{equation}
 \Phi=\Phi_0+\varphi,\qquad
 \Phi_0=U(y\cos\alpha+z\sin\alpha),\qquad
 \bm V=\nabla\Phi=\bm V_\infty+\nabla\varphi.
 \label{main:potential-decomposition}
\end{equation}
The potential $\Phi_0$ is that of the uniform freestream. The disturbance $\varphi$ carries all effects of the prescribed profile, of the incidence relative to the secant and of the deformation.

Linearizing the equations of steady isentropic potential flow, conservation of mass and Bernoulli's equation, about the uniform freestream gives, for the density disturbance $\tilde\rho=\rho-\rho_\infty$ and the disturbance potential~\cite{AshleyLandahl1965,BisplinghoffAshleyHalfman1955},
\begin{equation}
 \begin{aligned}
 \rho_\infty\nabla^2\varphi
 +\bm V_\infty\cdot\nabla\tilde{\rho}&=0,\\
 \frac{a_\infty^2}{\rho_\infty}\tilde{\rho}
 +\bm V_\infty\cdot\nabla\varphi&=0,
 \end{aligned}
 \label{main:uniform-linearization}
\end{equation}
and hence the linearized compressible potential equation
\begin{equation}\label{eq:PotentialEquation}
 \nabla^2\varphi
 -\frac{1}{a_\infty^2}
  (\bm V_\infty\cdot\nabla)^2\varphi=0.
\end{equation}

\subsection{Surface tangency and far field boundary conditions}
The velocity at each face is tangent to it. With $\bm V=\bm V_\infty+\nabla\varphi$ and the unit normal proportional to $(-k^\pm,1)$, this gives the exact graph-surface condition
\begin{equation}
 [\varphi_z-k^\pm\varphi_y]_{z=H^\pm(y)}
 =k^\pm U\cos\alpha-U\sin\alpha.
 \label{main:total-exact-slip}
\end{equation}

We now collapse the upper and lower surfaces onto the reference plane $z=0$, while retaining their distinct slopes $k^+$ and $k^-$ in the boundary data. The flow is represented by separate upper- and lower-half-plane fields, with one-sided limiting velocities
\begin{equation}
 u^\pm(y)=\lim_{z\to0^\pm}\varphi_y(y,z),\qquad
 v^\pm(y)=\lim_{z\to0^\pm}\varphi_z(y,z).
 \label{main:one-sided-traces}
\end{equation}

Evaluated at $z=0^\pm$ instead of on the surfaces, the tangency condition~\eqref{main:total-exact-slip} becomes
\begin{equation}
 v^\pm-k^\pm u^\pm
 =k^\pm U\cos\alpha-U\sin\alpha,
 \qquad y\in I.
 \label{air:tangency}
\end{equation}

Far from the airfoil, the total velocity approaches the uniform
freestream. The disturbance velocities therefore satisfy
\begin{equation}
 \lim_{\|(y,z)\|\to\infty}\nabla\varphi(y,z)=\bm 0.
 \label{air:farfield}
\end{equation}

The disturbance potential satisfies the linearized equation~\eqref{eq:PotentialEquation} separately in the upper and lower half-planes. With
\begin{equation}
 \begin{aligned}
 A_\alpha&=1-M^2\cos^2\alpha,\qquad
 C_\alpha=M^2\sin\alpha\cos\alpha,\\
 D_\alpha&=1-M^2\sin^2\alpha,\qquad
 A_\alpha D_\alpha-C_\alpha^2=\beta^2>0,\\
a&=\frac{C_\alpha}{D_\alpha},\qquad\qquad
 b=\frac{\beta}{D_\alpha}>0,
 \end{aligned}
 \label{air:coefficients}
\end{equation}
it reads
\begin{equation}
 \label{main:flow-strip}
 A_\alpha\varphi_{yy}
 -2C_\alpha\varphi_{yz}
 +D_\alpha\varphi_{zz}=0,
 \qquad z\ne0.
\end{equation}
Since $M<1$, the coefficient matrix is positive definite, and the
equation is elliptic in each half-plane.

\subsection{Total pressure jump on the collapsed surfaces}
The derivation uses Bernoulli's relation linearized about the uniform freestream, so that $p^\pm$ and $p_{\rm tot}$ denote the linearized face pressures and their jump; Section~\ref{sec:validation-pressure-variants} expresses them in terms of the computed velocities. They satisfy
\begin{equation}
 \frac{p^\pm-p_\infty}{\rho_\infty}
 +\bm V_\infty\cdot( u^\pm,v^\pm)=0.
 \label{main:pressure-bernoulli-step}
\end{equation}
Thus
\begin{equation}
 p^\pm-p_\infty=-\rho_\infty U
 (\cos\alpha\,u^\pm+\sin\alpha\,v^\pm).
 \label{main:bernoulli}
\end{equation}
The total pressure jump is lower-surface pressure minus upper-surface pressure:
\begin{equation}
 p_{\rm tot}=p^- -p^+
 =\rho_\infty U[\cos\alpha(u^+-u^-)+\sin\alpha(v^+-v^-)].
 \label{main:pressure-jump-expanded}
\end{equation}
Positive $p_{\rm tot}$ denotes an upward pressure difference.

\section{The modified airfoil system}\label{sec:airfoil-derivation}
The unknowns are the jumps of the disturbance velocity across the reference plane. With the one-sided velocities~\eqref{main:one-sided-traces}, define the jumps and averages
\begin{equation}
 d=u^+-u^-,\quad n=v^+-v^-,\qquad
 \bar u=\tfrac12(u^++u^-),\quad \bar v=\tfrac12(v^++v^-),
 \label{air:jumps}
\end{equation}
and the finite Hilbert transform
\begin{equation}
 (\HH \psi )(y)=\frac1\pi\pv\int_{y_-}^{y_+}
 \frac{\psi (\eta)}{y-\eta}\,\dd\eta,\qquad y\in I.
 \label{air:zero-extension-H}
\end{equation}
The one modeling assumption beyond those of Section~\ref{sec:geometry} is that the jumps vanish outside $I$: on the reference plane, normal-velocity continuity gives $n=0$ off $I$, and zero pressure jump gives $\cos\alpha\,d+\sin\alpha\,n=0$, so that $d=0$ there when $\cos\alpha\ne0$. Eliminating the field by a Fourier transform in $y$, with the far-field condition~\eqref{air:farfield}, and imposing the tangency conditions~\eqref{air:tangency} on the chord gives the coupled system (Appendix~\ref{app:derivation})
\begin{equation}
 \begin{aligned}
 b\HH d+g_t d+\frac{k_c-a}{b}\HH(n-ad)&=F,\\
 n-k_c d-\frac{g_t}{b}\HH(n-ad)&=G,
 \qquad y\in I,
 \end{aligned}
 \label{main:total-integral}
\end{equation}
with the forcings
\begin{equation}
 F=2\bigl[U\sin\alpha-k_cU\cos\alpha\bigr],\qquad
 G=2g_tU\cos\alpha,
 \label{air:total-forcing}
\end{equation}
the flow coefficients $a$, $b$ of~\eqref{air:coefficients} and the slopes $k_c$, $g_t$ of~\eqref{air:effective-slopes} and~\eqref{geo:profile-slopes}. The pressure jump then follows from~\eqref{main:pressure-jump-expanded}.

For a zero-thickness plate without deformation, $k_c=g_t=0$, the system reduces to $n=0$ and
\begin{equation}
 \frac{a^2+b^2}{b}\HH d=2U\sin\alpha,
\end{equation}
the classical vortex-sheet airfoil equation for an inclined flat plate~\cite[Eq.~(6.19)]{Liburdy2021}. We refer to~\eqref{main:total-integral} as the thickness-modified airfoil system.

\subsection{Normalized system}\label{sec:normalized-system}
The system is analyzed and solved in a normalized chord coordinate $\eta$; $y$ remains the dimensional coordinate. Set
\begin{equation}
 \begin{gathered}
 \eta=\frac{2(y-y_-)}c-1,\qquad t=\frac{1+\eta}{2},\\
 w(\eta)=\sqrt{1-\eta^2},\qquad
 r(\eta)=\sqrt{\frac{1-\eta}{1+\eta}}.
 \end{gathered}
 \label{total:coordinate-weights}
\end{equation}
The finite Hilbert transform is invariant under this affine change. All chordwise quantities retain their existing symbols after reparameterization by $\eta$; in particular, $F,G$ remain the forcings in~\eqref{air:total-forcing}. For a cambered and deformed profile, the common slope from~\eqref{air:effective-slopes} is
\begin{equation}
 k_c(\eta)=(f_c'+h_e')\!\left(y_-+\frac c2(1+\eta)\right),\qquad
 \lambda(\eta)=a-k_c(\eta),\qquad s=a-ib.
 \label{total:parameters}
\end{equation}
Primes on $f_c,h_e$ denote differentiation in dimensional $y$; equivalently, $k_c=(2/c)\partial_\eta(f_c+h_e)$ for the reparameterized functions.
In these variables the system~\eqref{main:total-integral} reads
\begin{equation}
 \begin{aligned}
 b\HH d+g_t d-\frac\lambda b\HH(n-ad)&=F,\\
 n-k_c d-\frac{g_t}{b}\HH(n-ad)&=G.
 \end{aligned}
 \label{total:coupled-system}
\end{equation}

\section{Existence and uniqueness}\label{sec:existence}\label{sec:total-solutions}\label{sec:existence-theorem}
This section gives conditions for existence and uniqueness of solutions of~\eqref{total:coupled-system}, for general bounded slopes and for rounded leading edges, and the solution as an integral formula, without assuming polynomial geometry. Both coefficients $\lambda$ and $g_t$ of the system may vary along the chord, and camber enters both the operator, through $\lambda$, and the forcing.

Put $\ell=(n-ad)/b$, so that~\eqref{total:coupled-system} reads
\begin{equation}
 b\HH d-\lambda\HH\ell+g_td=F,\qquad
 b\ell+\lambda d-g_t\HH\ell=G,
 \label{stab:system}
\end{equation}
with $F,G$ from~\eqref{air:total-forcing} and $\lambda$ from~\eqref{total:parameters}. In matrix operator form,
\begin{equation}
 \mathsf A\begin{pmatrix}d\\ \ell\end{pmatrix}=\begin{pmatrix}F\\ G\end{pmatrix},\qquad
 \mathsf A=\begin{pmatrix} b\HH+g_t&-\lambda\HH\\ \lambda&b-g_t\HH\end{pmatrix},
 \label{stab:matrix-form}
\end{equation}
where $\lambda$ and $g_t$ act by multiplication. Let
\begin{equation}
 X=L^2\bigl((-1,1);r^{-1}\dd\eta\bigr),\qquad
 \norm{f}_X^2=\int_{-1}^1\frac{f(\eta)^2}{r(\eta)}\,\dd\eta,
 \label{stab:space}
\end{equation}
where $r$ is the weight in~\eqref{total:coordinate-weights}. Membership in $X$ permits the inverse-square-root leading-edge behavior $O(r)$ and excludes an inverse-square-root trailing-edge singularity, so that the Kutta condition is part of the space. The closed-form jumps~\eqref{total:flat-ansatz} and~\eqref{total:rounded-ansatz} of Section~\ref{sec:exact-solutions} belong to $X$. For rounded leading edges we also use
\begin{equation}
 Y=L^2\bigl((-1,1);w^{-1}\dd\eta\bigr),\qquad
 \norm{(d,\ell)}_Y=\bigl(\norm{d}_Y^2+\norm{\ell}_Y^2\bigr)^{1/2},
 \label{stab:space-Y}
\end{equation}
which in addition excludes the inverse-square-root singularity at the leading edge.

The operator $\mathsf A$ separates the flow coefficients from the geometry:
\begin{equation}
 \begin{gathered}
 \mathsf A=\mathsf E\,(b\,\mathsf J+\mathsf K)\,\mathsf D,\\[2pt]
 \mathsf E=\begin{pmatrix}0&-1\\1&0\end{pmatrix},\quad
 \mathsf J=\begin{pmatrix}0&\HH^{-1}\\-\HH&0\end{pmatrix},\quad
 \mathsf K=\begin{pmatrix}\lambda&-g_t\\-g_t&\lambda\end{pmatrix},\quad
 \mathsf D=\begin{pmatrix}1&0\\0&\HH\end{pmatrix}.
 \end{gathered}
 \label{stab:factorization}
\end{equation}
The operators $\mathsf J$ and $\mathsf D$ are considered on $X\times X$. On the unweighted space $L^2(-1,1)$ the finite Hilbert transform is injective, but its range is a proper dense subspace, so that its inverse is unbounded~\cite[Lemma~3.1]{OkadaElliott1991}. On $X$, by contrast, $\HH$ is a unitary operator, with the inverse $\HH^{-1}g=-r\,\HH[g/r]$ defined on all of $X$ (Lemma~\ref{lem:H-unitary} in Appendix~\ref{app:proof-geometry-stability}, which rests on~\cite{AstalaPaivarintaSaksman1996}). Hence $\mathsf D$ and $\mathsf J$ are unitary on $X\times X$, and $\mathsf J^2=-\mathsf I$. The thin-airfoil operator is $\mathsf E\,b\mathsf J\,\mathsf D$, which is inverted exactly since $(b\mathsf J)^{-1}=-\mathsf J/b$, and the geometry enters only through the multiplication operator $\mathsf K$. The eigenvalues of the matrix $\mathsf K$ are $\lambda\mp g_t$. The geometry enters the solvability conditions through the angles
\begin{equation}
 \phi^\pm(\eta)=\arctan\frac{\pm g_t(\eta)-\lambda(\eta)}{b}\in\bigl(-\tfrac\pi2,\tfrac\pi2\bigr).
 \label{stab:angles}
\end{equation}
The angles $\phi^+$ and $\phi^-$ belong to the upper and lower faces: $\pm g_t-\lambda$ is the slope of the face, shifted by $a$. For bounded slopes define also
\begin{equation}
 \Lambda=\norm{\lambda}_\infty,\qquad\Gamma=\norm{g_t}_\infty,\qquad
 \mu=b-\Lambda-\Gamma,
 \label{stab:margin}
\end{equation}
with suprema over the open chord.

\begin{theorem}[Existence and uniqueness]\label{thm:existence}
Fix $U>0$, $0\le M<1$ and $\alpha$, and hence $a,b$.

\emph{(i) Small bounded slopes.} Let $\lambda,g_t\in L^\infty(-1,1)$ with $\mu>0$. Then $b\mathsf J+\mathsf K$ has a bounded inverse on $X\times X$, given by the Neumann series
\begin{equation}
 (b\mathsf J+\mathsf K)^{-1}=-\frac1b\sum_{n\ge0}\Bigl(\frac{\mathsf J\mathsf K}{b}\Bigr)^{n}\mathsf J,
 \label{stab:neumann}
\end{equation}
which converges in the operator norm, and $\mathsf A^{-1}=\mathsf D^{-1}(b\mathsf J+\mathsf K)^{-1}\mathsf E^{-1}$. In particular,~\eqref{stab:system} has exactly one solution $(d,\ell)\in X\times X$, and $\norm{d}_X+\norm{\ell}_X\le\mu^{-1}\bigl(\norm{F}_X+\norm{G}_X\bigr)$.

In parts~(ii) and~(iii) a function on $[-1,1]$ is called piecewise H\"older continuous if there are finitely many points $\eta_k$ of the open chord such that the function is H\"older continuous on each of the closed intervals into which these points divide $[-1,1]$, with one-sided limits at the points $\eta_k$. At these points the angles~\eqref{stab:angles} are required to satisfy
\begin{equation}
 \bigl|\phi^\pm(\eta_k+0)-\phi^\pm(\eta_k-0)\bigr|<\frac\pi2\qquad\text{for every }k\text{ and for both faces,}
 \label{stab:corner-angles}
\end{equation}
that is, each face turns by less than a right angle at each corner.

\emph{(ii) Bounded slopes.} Let $\lambda$ and $g_t$ be piecewise H\"older continuous with~\eqref{stab:corner-angles}, and let the angles between the two faces at the edges satisfy
\begin{equation}
 \bigl|\phi^+(-1)-\phi^-(-1)\bigr|<\frac\pi2,\qquad\bigl|\phi^+(1)-\phi^-(1)\bigr|<\frac\pi2 .
 \label{stab:edge-angles}
\end{equation}
Then $\mathsf A$ has a bounded inverse on $X\times X$. In particular,~\eqref{stab:system} has exactly one solution $(d,\ell)\in X\times X$.

\emph{(iii) Rounded leading edge.} Let
\begin{equation}
 g_t=\frac{S}{\sqrt{1+\eta}},\qquad S(-1)>0,
 \label{stab:wedge-class}
\end{equation}
with piecewise H\"older-continuous $\lambda$ and $S$ and with~\eqref{stab:corner-angles}, so that the thickness slope at the trailing edge, $g_t(1)=S(1)/\sqrt2$, is finite, and let the slope angles of the two faces at the trailing edge satisfy
\begin{equation}
 \varDelta_{\rm te}:=\bigl|\phi^+(1)-\phi^-(1)\bigr|<\frac\pi2 .
 \label{stab:te-angle}
\end{equation}
Then~\eqref{stab:system} has exactly one solution $(d,\ell)\in Y\times Y$. This part contains the sections whose faces have a common tangent at the trailing edge,
\begin{equation}
 g_t=\frac{R}{w},\qquad R(\eta)=(1-\eta)R_1(\eta),\qquad R_1(-1)>0,
 \label{stab:rounded-class}
\end{equation}
with piecewise H\"older-continuous $\lambda$ and $R_1$: for them $S=\sqrt{1-\eta}\,R_1$ and $\varDelta_{\rm te}=0$.

In parts~(ii) and~(iii) the solution is given by the integral formula~\eqref{wp:solution}.
\end{theorem}
The Kutta condition enters through the spaces and is not stated separately. Both $X$ and $Y$ exclude the inverse-square-root trailing-edge singularity, so that uniqueness in these spaces is the selection made by the Kutta condition, in the sense of integrability. For the families of Section~\ref{sec:exact-solutions} the condition holds pointwise: the jumps vanish at the trailing edge (part~(iii) of Theorems~\ref{thm:flat-family} and~\ref{thm:rounded-family}).

\begin{figure}[!htbp]\centering
\includegraphics[width=\linewidth]{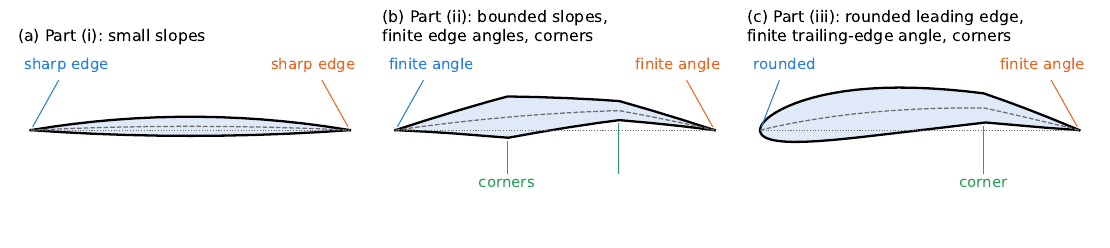}
\caption{Illustrative sections covered by Theorem~\ref{thm:existence}, drawn with camber (dashed: mean line). (a) Part~(i): small slopes. (b) Part~(ii): bounded slopes, with finite edge angles and corners. (c) Part~(iii): rounded leading edge, with a finite trailing-edge angle and a corner.}\label{fig:section-classes}
\end{figure}

Figure~\ref{fig:section-classes} shows an example of the sections covered by each part.

The proof is given in Appendix~\ref{app:proof-geometry-stability}. The three parts are proved by two arguments.

\emph{Part~(i)} treats the geometry as a perturbation of the thin-airfoil operator. In the factorization~\eqref{stab:factorization}, $b\mathsf J$ is inverted exactly and $\mathsf K$ has norm at most $\Lambda+\Gamma<b$, so that the Neumann series~\eqref{stab:neumann} converges: the solution is the thin-airfoil solution followed by successive corrections for thickness and camber. A contraction argument of the same kind was used for the steady Possio equation in ground effect~\cite{SerryTuffahaGround2017}. 

\emph{Parts~(ii) and~(iii)} remove the smallness condition by a different argument. On each face, the tangency condition prescribes the real part of the complex disturbance velocity multiplied by a coefficient whose argument is the slope angle $\phi^\pm$ of that face. A function that is analytic off the chord, and whose argument on each side of the chord is $-\phi^\pm$, is constructed explicitly from the two angles. Dividing the complex velocity by this function removes the angles from the two conditions, and what remains is the system of a flat plate, which is solved by the classical inversion of the finite Hilbert transform. This reduction adapts to the present setting the approach of Peters~\cite{Peters1972} and Porter~\cite{Porter1986} for systems of singular integral equations, in which the system is posed for the boundary values of the Cauchy integral of the unknowns and this Cauchy integral is factored by an analytic function chosen so that the system decouples. Alternatively, the slit plane can be mapped conformally onto the unit disk, where the two face conditions become a Riemann--Hilbert problem with a piecewise continuous coefficient on the unit circle, and existence and uniqueness then follow from the theory of that problem in weighted classes of Cauchy-type integrals~\cite{Muskhelishvili1953,KhuskivadzeKokilashviliPaatashvili1998,KokilashviliPaatashviliSamko2005,KokilashviliPaatashvili2007}. We use the reduction on the chord because it gives the solution formula directly and the canonical function used for the explicit bounds of Section~\ref{sec:geometry-stability}. 

The flat-endpoint family of Section~\ref{sec:families} falls under part~(ii) with zero edge angles, the rounded-leading-edge family under part~(iii) with $\varDelta_{\rm te}=0$, and the NACA four- and five-digit sections with a closed trailing edge under part~(iii) with a finite trailing-edge angle. The closed-form solutions of Section~\ref{sec:exact-solutions} do not include the last group, and Theorem~\ref{thm:te-angle} bounds the error made when such sections are approximated by the rounded-leading-edge family. The proof also gives an integral formula for the solution,~\eqref{wp:solution}, which holds for every section covered by parts~(ii) and~(iii).

Part~(iv) of Theorems~\ref{thm:flat-family} and~\ref{thm:rounded-family} identifies the closed-form solutions of Section~\ref{sec:exact-solutions} with the unique solutions given here.

\section{Two families of airfoil profiles}\label{sec:families}
The closed-form solutions of Section~\ref{sec:exact-solutions} are constructed for polynomial camber and two weighted-polynomial thickness families: one with vanishing endpoint thickness slopes and one with a rounded leading edge and a sharp trailing edge. This section first defines the two families (Sections~\ref{sec:camber-family} to~\ref{sec:rounded-family}) and then shows that they approximate general closed profiles and, under additional regularity, their slopes (Section~\ref{sec:approximation-properties}). The families are defined in the secant frame, since a rotation of the coordinates does not preserve the weighted polynomial form. The flat-endpoint family has continuous slopes with zero thickness slope at both edges and falls under part~(ii) of Theorem~\ref{thm:existence}; the rounded-leading-edge family falls under part~(iii). Neither requires a smallness condition (part~(iv) of Theorems~\ref{thm:flat-family} and~\ref{thm:rounded-family}).

\begin{figure}[!htbp]\centering
\includegraphics[width=\linewidth]{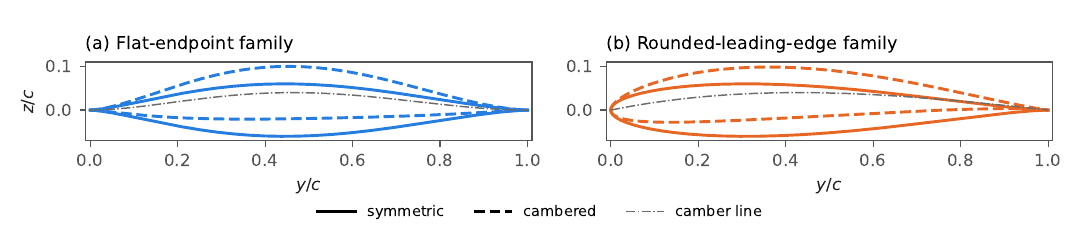}
\caption{Illustrative members of the two profile families, symmetric and cambered. (a) Flat-endpoint family~\eqref{total:flat-profile-slope}: both faces are tangent to the chord at both edges. (b) Rounded-leading-edge family~\eqref{total:rounded-profile-slope}: rounded nose and zero thickness slope at the trailing edge.}\label{fig:profile-families}
\end{figure}

Figure~\ref{fig:profile-families} shows representative members of both families.

\subsection{Coordinates and polynomial camber}\label{sec:camber-family}
The surfaces are functions of the dimensional chord coordinate $y$, and the slopes $k_c$ and $g_t$ of the system are their derivatives in $y$. The families are written in the normalized coordinate $\eta=\eta(y)$ of~\eqref{total:coordinate-weights}, for which
\begin{equation}
 y=y_-+\frac c2(1+\eta),\qquad \frac{\dd}{\dd y}=\frac2c\,\frac{\dd}{\dd\eta}.
 \label{total:chain-rule}
\end{equation}
In this section a prime denotes differentiation in $\eta$. Both families use a polynomial camber that vanishes at the two edges,
\begin{equation}
 f_c(y)=\frac c2\,\mathcal C(\eta),\qquad
 \mathcal C(\eta)=(1-\eta^2)\mathcal A(\eta),\quad \mathcal A\in\mathbb R[\eta];
 \label{total:camber-family}
\end{equation}
every polynomial camber vanishing at $\eta=\pm1$ has this form. By~\eqref{total:chain-rule}, $\dd f_c/\dd y=\mathcal C'$, so that
\begin{equation}
 k_c=\mathcal C'+\frac{\dd h_e}{\dd y},\qquad \lambda=a-\mathcal C'-\frac{\dd h_e}{\dd y}.
 \label{total:camber-mean-slope}
\end{equation}
Camber may change sign, and no condition $|f_c|\le f_t$ is required to keep the faces ordered.

In both families the half-thickness $f_t$ is $c/2$ times a weight multiplied by a polynomial $P$ that is nonnegative on $[-1,1]$, so that its slope $g_t=\dd f_t/\dd y$ is the derivative of that product in $\eta$. The polynomials $\mathcal A$ and $P$ are dimensionless. The undeformed faces and their slopes are
\begin{equation}
 f^\pm(y)=f_c(y)\pm f_t(y),\qquad
 \frac{\dd f^\pm}{\dd y}=\mathcal C'(\eta)\pm g_t(\eta),
 \label{total:family-faces}
\end{equation}
and a deformation adds $\dd h_e/\dd y$ to both slopes. The families differ in the weight, a power of $w$ from~\eqref{total:coordinate-weights}, and hence in the behavior of the faces at the two edges.

\subsection{Flat-endpoint family}\label{sec:flat-family}
The half-thickness of the first family and its slope are
\begin{equation}
 f_t(y)=\frac c2\,w^3P(\eta),\qquad
 g_t=\bigl(w^3P\bigr)'=wQ,\qquad Q=(1-\eta^2)P'-3\eta P.
 \label{total:flat-profile-slope}
\end{equation}
The thickness slope vanishes at both edges, so that by~\eqref{total:family-faces} the two faces have a common tangent there. They are tangent to the chord when the camber is taken in the smaller class
\begin{equation}
 \mathcal C(\eta)=(1-\eta^2)^2 \mathcal A_{\rm f}(\eta),\qquad
 \mathcal A_{\rm f}\in\mathbb R[\eta],
 \label{total:flat-camber}
\end{equation}
for which $\mathcal C(\pm1)=\mathcal C'(\pm1)=0$. Higher-order contact at the edges is obtained by taking $P$ with a factor that is a power of $1-\eta^2$.

\subsection{Rounded-leading-edge family}\label{sec:rounded-family}
The half-thickness of the second family and its slope are
\begin{equation}
 f_t(y)=\frac c2\,wP(\eta),\qquad
 g_t=\bigl(wP\bigr)'=\frac{R}{w},\qquad R=(1-\eta^2)P'-\eta P,
 \label{total:rounded-profile-slope}
\end{equation}
with $P(-1)>0$ and $P(1)=0$. The first condition gives a rounded leading edge, at which the slopes of the two faces are unbounded and opposite. The second gives $R(1)=0$: the thickness slope vanishes at the trailing edge, and the two faces have a common tangent there. The camber may be any member of~\eqref{total:camber-family}.

\subsection{Approximation properties}\label{sec:approximation-properties}
Both families approximate general closed sections. Write $T(\eta)=2f_t/c$ for the dimensionless thickness of a given section, so that $g_t=T'$ by~\eqref{total:chain-rule}; it is approximated by $w^3P_N$ in the first family and by $wP_N$ in the second, uniformly and, under additional regularity, together with its slope. Primes again denote differentiation in $\eta$.

\begin{theorem}[Approximation by flat-endpoint profiles]\label{thm:flat-approximation}
Let $w$ be the weight in~\eqref{total:coordinate-weights}, and let primes denote derivatives in $\eta$.

\emph{(i) Profiles.} For every real $T\in C[-1,1]$ with $T(-1)=T(1)=0$, there exist real polynomials $P_N$ such that
\begin{equation}
 T_N=w^3P_N,\qquad \|T_N-T\|_\infty\longrightarrow0.
 \label{approx:flat-uniform}
\end{equation}
If $T\ge0$, the polynomials can be chosen nonnegative on $[-1,1]$.

\emph{(ii) Slopes.} If, in addition, $T\in C^1[-1,1]$ and $T'(-1)=T'(1)=0$, the polynomials can be chosen so that
\begin{equation}
 \|T_N-T\|_{C^1}:=\|T_N-T\|_\infty+\|T_N'-T'\|_\infty
 \longrightarrow0,
 \label{approx:flat-slopes}
\end{equation}
with the same positivity property.
\end{theorem}
The proof is given in Appendix~\ref{app:proof-approximation}; it combines a cutoff near the endpoints with the convergence of Bernstein polynomials and of their derivatives. Part~(ii) gives uniform approximation of the half-thickness slope $g_t=T'$. The conditions $T'(\pm1)=0$ are necessary for convergence in the $C^1$ norm, since every $w^3P_N$ has zero endpoint derivatives. Uniform approximation of the profile alone, as in part~(i), does not imply approximation of its slope or convergence of the aerodynamic solution.

\begin{theorem}[Approximation by rounded-leading-edge profiles]\label{thm:rounded-approximation}
Let $w$ be the weight in~\eqref{total:coordinate-weights}, and let primes denote derivatives in $\eta$.

\emph{(i) Profiles.} For every real $T\in C[-1,1]$ with $T(-1)=T(1)=0$, there exist real polynomials $P_N$ with $P_N(1)=0$ such that
\begin{equation}
 T_N=wP_N,\qquad \|T_N-T\|_\infty\longrightarrow0.
 \label{approx:rounded-uniform}
\end{equation}
If $T\ge0$, the polynomials can be chosen nonnegative on $[-1,1]$.

\emph{(ii) Slopes.} If, in addition, $P_T=T/w$ extends from $(-1,1)$ to a function in $C^1[-1,1]$ with $P_T(1)=0$, the polynomials can be chosen so that $P_N(1)=0$, $P_N(-1)=P_T(-1)$ and
\begin{equation}
 \|T_N-T\|_\infty+
 \|w(T_N'-T')\|_\infty\longrightarrow0,
 \label{approx:rounded-slopes}
\end{equation}
with the same positivity property. The weighted derivatives are understood by their continuous extensions to the endpoints.
\end{theorem}
The proof is given in Appendix~\ref{app:proof-approximation}. By part~(i), even the subclass with $P_N(1)=0$ is uniformly dense in the continuous profiles that vanish at both endpoints. As for the first family, $g_t=T'$, so that part~(ii) gives convergence of $wg_t$, and ordinary uniform slope convergence on every compact subinterval of $(-1,1)$. The weighted norm permits the square-root slope singularity at the leading edge, and $P_N(-1)=P_T(-1)$ keeps the nose rounded when the target is. The regular-factor assumption on $P_T$ is sufficient for part~(ii) and is not required for part~(i). The condition $P_N(1)=0$ enforces a vanishing trailing-edge slope, so that uniform approximation of the slope up to the trailing edge is not claimed for sections whose slope there is nonzero. Sections with a finite trailing-edge angle are approximated in an integral norm instead (Theorem~\ref{thm:te-angle}).

\section{Exact solutions for the two families}\label{sec:exact-solutions}
We now restrict to polynomial camber $f_c$ and polynomial deformation $h_e$, with the thickness in one of the families of Section~\ref{sec:families}. For these data the solutions of~\eqref{total:coupled-system} are weighted-rational functions whose poles and multiplicities are determined by the operator. Camber enters both the operator and the forcing; the constructions therefore cover cambered profiles, and no Hilbert transform is evaluated numerically.

The constructions use explicit finite Hilbert transforms of the weighted pole functions, the matching of polynomial coefficients, and a confluent representation for repeated poles. These ingredients, including the branch functions $W$ and $K$, the background term $N_0$, the degree ${\mathsf p}$ and the operators $\mathcal D_\zeta$ and $\mathcal L_{\zeta,\nu}$ used below, are collected in Appendix~\ref{sec:transform-identities}. The camber $f_c$ and the deformation $h_e$ enter only through their combined slope $k_c$, so that the theorems are stated for an arbitrary polynomial $k_c$.

\subsection{Flat-endpoint solutions}\label{sec:flat-solutions}
Throughout this subsection, let $Q\not\equiv0$ be the real profile polynomial defined by~\eqref{total:flat-profile-slope}, with ${\mathsf m}=\deg Q$. Prescribe polynomial $k_c$ as in~\eqref{total:parameters}, and let ${\mathsf p}$ be defined by~\eqref{total:deformation-degree}. Use $a,b$ from~\eqref{air:coefficients}, $\lambda$ and $s$ from~\eqref{total:parameters}, $F,G$ from~\eqref{air:total-forcing}, the weights from~\eqref{total:coordinate-weights}, and $W,K$ from~\eqref{total:branch-functions}. Set ${\mathsf q}=\max\{{\mathsf m},{\mathsf p}-1\}$. Let $N_0$ be defined by~\eqref{total:velocity-background}.

Algorithm~\ref{alg:flat} computes the solution, and Theorem~\ref{thm:flat-family} shows that each of its steps is well defined and that its output solves the coupled system.

\begin{algorithmbox}{Closed-form solution for a flat-endpoint profile}\label{alg:flat}
\noindent\textbf{Input:} the flow data $U$, $M$, $\alpha$, and hence $a$, $b$ and $s$; the polynomials $Q$ and $k_c$.\\
\textbf{Output:} the velocity jumps $d$ and $n$ on the chord.
\begin{enumerate}[label=\textbf{Step \arabic*.},leftmargin=*,itemsep=4pt]
\item \emph{Poles and multiplicities.}
Compute the roots of the candidate polynomial
\begin{equation}
 (\zeta ^2-1)Q(\zeta )^2-[b+i\lambda(\zeta )]^2,
 \label{total:flat-candidate-system}
\end{equation}
and retain the distinct roots $\zeta _j$, $1\le j\le J$, in either half-plane, that are zeros of the pole function
\begin{equation}
 \Psi_{\rm f}(\zeta )=Q(\zeta )W(\zeta )-b-i\lambda(\zeta ).
 \label{total:flat-poles}
\end{equation}
Let $\mathsf r_j$ be the order of $\zeta _j$ as a zero of $\Psi_{\rm f}$, obtained as in Section~\ref{sec:multiplicities}, and set
\begin{equation}
 \mathcal I_{\rm f}=\{(j,\nu):1\le j\le J,\ 1\le\nu\le \mathsf r_j\}.
\end{equation}

\item \emph{Amplitudes.}
With $\mathcal D_\zeta$ from~\eqref{total:divided-difference} and $\mathcal L_{\zeta ,\nu}$ from~\eqref{total:confluent-operator}, form the polynomials in $\eta$, of degree at most ${\mathsf q}$,
\begin{equation}
 \begin{aligned}
 E_1(\zeta ,\eta)&=-\mathcal D_\zeta [(1-\eta)Q]
       -\frac{(s-a)K(\zeta )}b\mathcal D_\zeta k_c,\\
 E_2(\zeta ,\eta)&=b\mathcal D_\zeta k_c
       +(s-a)K(\zeta )\mathcal D_\zeta [(1+\eta)Q],\\
 E_{k,j\nu}(\eta)&=
 \left.\mathcal L_{\zeta ,\nu}E_k(\zeta ,\eta)\right|_{\zeta =\zeta _j},
 \qquad k=1,2.
 \end{aligned}
\end{equation}
and, in a fixed ordering of $\mathcal I_{\rm f}$, solve the real linear system
\begin{equation}
 \begin{aligned}
 \bm e_{j\nu}&=\begin{pmatrix}[E_{1,j\nu}]_{\mathsf q}\\{}[E_{2,j\nu}]_{\mathsf q}\end{pmatrix},\qquad
 \bm x_{\rm f}=(\Re A_{j\nu},\Im A_{j\nu})_{(j,\nu)\in\mathcal I_{\rm f}}^T,\\
 \bm M_{\rm f}&=\bigl[\,2\Re\bm e_{j\nu}\ \ -2\Im\bm e_{j\nu}\,\bigr]_{(j,\nu)\in\mathcal I_{\rm f}}
 \in\mathbb R^{(2{\mathsf q}+2)\times(2{\mathsf q}+2)},\\
 \bm h_{\rm f}&=\begin{pmatrix}
 {}[2(U\sin\alpha-aU\cos\alpha)]_{\mathsf q}\\
 {}[-bN_0]_{\mathsf q}
 \end{pmatrix},\qquad
 \bm M_{\rm f}\bm x_{\rm f}=\bm h_{\rm f}.
 \end{aligned}
 \label{total:flat-real-system}
\end{equation}

\item \emph{Jumps.}
Evaluate
\begin{equation}
 \begin{aligned}
 d&=2r\Re\sum_{(j,\nu)\in\mathcal I_{\rm f}}
 \frac{A_{j\nu}}{(\zeta _j-\eta)^\nu},\\
 n&=r\left[N_0+2\Re\sum_{(j,\nu)\in\mathcal I_{\rm f}}
 \frac{sA_{j\nu}}{(\zeta _j-\eta)^\nu}\right].
 \end{aligned}
 \label{total:flat-ansatz}
\end{equation}
\end{enumerate}
\end{algorithmbox}

\begin{theorem}[Computable confluent rational solution for flat-endpoint profiles]\label{thm:flat-family}
Under the assumptions stated at the beginning of Section~\ref{sec:flat-solutions}, the following hold.

\emph{(i) Pole count.} The function $\Psi_{\rm f}$ in~\eqref{total:flat-poles} has exactly ${\mathsf q}+1$ zeros counted with multiplicity, all off $[-1,1]$. Hence $\sum_{j=1}^J\mathsf r_j={\mathsf q}+1$, and $\bm M_{\rm f}$ in~\eqref{total:flat-real-system} is square.

\emph{(ii) Amplitudes.} The matrix $\bm M_{\rm f}$ is invertible. Thus the amplitudes exist uniquely and satisfy
\begin{equation}
 2\Re\sum_{(j,\nu)\in\mathcal I_{\rm f}}A_{j\nu}E_{1,j\nu}
 =2(U\sin\alpha-aU\cos\alpha),\qquad
 2\Re\sum_{(j,\nu)\in\mathcal I_{\rm f}}A_{j\nu}E_{2,j\nu}
 =-bN_0.
 \label{total:flat-amplitudes}
\end{equation}

\emph{(iii) Solution.} The jumps~\eqref{total:flat-ansatz} solve~\eqref{total:coupled-system}. They are real and integrable, satisfy $d,n=O(r)$ at the leading edge, and vanish at the trailing edge; their transforms are bounded there. The amplitudes and jumps are unique within this fixed-pole confluent representation.

\emph{(iv) Existence and uniqueness in $X\times X$.} With $\ell$ as in~\eqref{stab:system}, the pair $(d,\ell)$ belongs to $X\times X$, defined in~\eqref{stab:space}, and solves~\eqref{stab:system}; it is the unique solution in $X\times X$ given by part~(ii) of Theorem~\ref{thm:existence}, so the jumps~\eqref{total:flat-ansatz} are unique in all of $X\times X$, not only within the fixed-pole representation.
\end{theorem}
The proof and transform formulas are given in Appendix~\ref{app:proof-flat-family}. Since $N_0=-2bU\cos\alpha$, the two right-hand sides in~\eqref{total:flat-amplitudes} are the constants $2U(\sin\alpha-a\cos\alpha)$ and $2b^2U\cos\alpha$. The coefficient system enforces polynomial identities, rather than boundary conditions at selected chord points. For $\mathsf r_j=1$ at every pole, $E_{k,j1}=E_k(\zeta _j,\eta)$ and $\Phi_1=K(\zeta _j)/(\zeta _j-\eta)$ recover the simple-pole formulas. Invertibility does not by itself bound the condition number of the system. Appendix~\ref{app:examples} works through the construction for a symmetric section with two simple poles (Example~\ref{ex:flat-poles-amplitudes}).

\subsection{Rounded-leading-edge solutions}\label{sec:rounded-solutions}
Throughout this subsection, let the prescribed dimensionless profile polynomial $P$ satisfy $P(-1)>0$ and $P(1)=0$, and determine $R$ by~\eqref{total:rounded-profile-slope}, with ${\mathsf m}=\deg R\ge2$. Prescribe polynomial $k_c$ as in~\eqref{total:parameters}, and use ${\mathsf p}$ from~\eqref{total:deformation-degree}. Use $a,b$ from~\eqref{air:coefficients}, $\lambda$ and $s$ from~\eqref{total:parameters}, $F,G$ from~\eqref{air:total-forcing}, the weights from~\eqref{total:coordinate-weights}, and $W,K$ from~\eqref{total:branch-functions}. Set ${\mathsf q}_1=\max\{{\mathsf m}-1,{\mathsf p}\}$, and ${\mathsf q}_2=\max\{{\mathsf m},{\mathsf p}+1\}$. Define the polynomial forcings
\begin{equation}
 F_{\rm r}=2U\bigl(\sin\alpha-k_c\cos\alpha\bigr),\qquad
 G_{\rm r}=2U\cos\alpha\,R.
 \label{total:rounded-forcings}
\end{equation}
Thus $F_{\rm r}=F$ and $G_{\rm r}=wG$. Here $s$ fixes the ratio of the pole amplitudes in $n$ and $d$ on the selected branch.

Algorithm~\ref{alg:rounded}, which has the same three steps as Algorithm~\ref{alg:flat}, computes the solution, and Theorem~\ref{thm:rounded-family} establishes that each step is well defined and that the output solves the coupled system.

\begin{algorithmbox}{Closed-form solution for a rounded-leading-edge profile}\label{alg:rounded}
\noindent\textbf{Input:} the flow data $U$, $M$, $\alpha$, and hence $a$, $b$ and $s$; the polynomials $R$ and $k_c$.\\
\textbf{Output:} the velocity jumps $d$ and $n$ on the chord.
\begin{enumerate}[label=\textbf{Step \arabic*.},leftmargin=*,itemsep=4pt]
\item \emph{Poles and multiplicities.}
Compute the roots of the candidate polynomial
\begin{equation}
 R(\zeta )^2-[b+i\lambda(\zeta )]^2(\zeta ^2-1),
 \label{total:rounded-candidate-system}
\end{equation}
whose only root on $[-1,1]$ is the simple root $\zeta =1$, which is not a pole. Retain the distinct off-chord roots $\zeta _j$, $1\le j\le J$, in either half-plane, that are zeros of the pole function
\begin{equation}
 \Psi_{\rm r}(\zeta )=R(\zeta )+[b+i\lambda(\zeta )]W(\zeta ).
 \label{total:rounded-poles}
\end{equation}
Let $\mathsf r_j$ be the order of $\zeta _j$ as a zero of $\Psi_{\rm r}$, obtained as in Section~\ref{sec:multiplicities}, and set
\begin{equation}
 \mathcal I_{\rm r}=\{(j,\nu):1\le j\le J,\ 1\le\nu\le \mathsf r_j\}.
\end{equation}

\item \emph{Amplitudes.}
Form the polynomials in $\eta$, of degrees at most ${\mathsf q}_1$ and ${\mathsf q}_2$ respectively,
\begin{equation}
 \begin{aligned}
 E(\zeta ,\eta)&=-\mathcal D_\zeta R(\eta)-b-i\lambda(\eta)-iW(\zeta )\mathcal D_\zeta \lambda(\eta),\\
 \mathcal T(\zeta ,\eta)&=(\zeta +\eta)[\lambda(\zeta )-ib]
 -(1-\eta^2)\mathcal D_\zeta \lambda(\eta)
 -iR(\eta)-iW(\zeta )\mathcal D_\zeta R(\eta),\\
 E_{j\nu}&=\left.\mathcal L_{\zeta ,\nu}E(\zeta ,\eta)\right|_{\zeta =\zeta _j},\qquad
 \mathcal T_{j\nu}=\left.\mathcal L_{\zeta ,\nu}\mathcal T(\zeta ,\eta)\right|_{\zeta =\zeta _j}.
 \end{aligned}
\end{equation}
and, in a fixed ordering of $\mathcal I_{\rm r}$, solve the real linear system
\begin{equation}
 \begin{aligned}
 \bm t_{j\nu}&=\begin{pmatrix}[E_{j\nu}]_{{\mathsf q}_1}\\{}[\mathcal T_{j\nu}]_{{\mathsf q}_2}\end{pmatrix},\qquad
 \bm x_{\rm r}=(\Re A_{j\nu},\Im A_{j\nu})_{(j,\nu)\in\mathcal I_{\rm r}}^T,\\
 \bm M_{\rm r}&=\bigl[\,2\Re\bm t_{j\nu}\ \ -2\Im\bm t_{j\nu}\,\bigr]_{(j,\nu)\in\mathcal I_{\rm r}}
 \in\mathbb R^{({\mathsf q}_1+{\mathsf q}_2+2)\times2|\mathcal I_{\rm r}|},\\
 \bm h_{\rm r}&=\begin{pmatrix}[F_{\rm r}]_{{\mathsf q}_1}\\{}[G_{\rm r}]_{{\mathsf q}_2}\end{pmatrix},\qquad
 \bm M_{\rm r}\bm x_{\rm r}=\bm h_{\rm r}.
 \end{aligned}
 \label{total:rounded-real-system}
\end{equation}

\item \emph{Jumps.}
With
\begin{equation}
 \mathcal Z=\sum_{(j,\nu)\in\mathcal I_{\rm r}}
 \frac{A_{j\nu}}{(\zeta _j-\eta)^\nu},
\end{equation}
evaluate
\begin{equation}
 d=2w\Re\mathcal Z,\qquad n=2w\Re(s\mathcal Z).
 \label{total:rounded-ansatz}
\end{equation}
\end{enumerate}
\end{algorithmbox}

\begin{theorem}[Computable confluent solution for rounded-leading-edge profiles]\label{thm:rounded-family}
Under the assumptions stated at the beginning of Section~\ref{sec:rounded-solutions}, the following hold; the construction includes repeated off-chord roots.

\emph{(i) Pole count.} The function $\Psi_{\rm r}$ in~\eqref{total:rounded-poles} has exactly ${\mathsf q}_1+1$ off-chord zeros counted with multiplicity, so that $|\mathcal I_{\rm r}|=\sum_{j=1}^J\mathsf r_j={\mathsf q}_1+1$.

\emph{(ii) Amplitudes.} The system~\eqref{total:rounded-real-system} is consistent and has full column rank. More precisely, with ${\mathsf q}={\mathsf q}_1$ and ${\mathsf q}_2={\mathsf q}+1$, its $2{\mathsf q}+3$ coefficient equations have the single redundant relation $\mathcal T(1)=0$ in the second polynomial block. Representing that block in $(\eta-1)\mathbb R[\eta]_{\le {\mathsf q}}$ gives an invertible $(2{\mathsf q}+2)\times(2{\mathsf q}+2)$ system. The unique amplitudes satisfy
\begin{equation}
 2\Re\sum_{(j,\nu)\in\mathcal I_{\rm r}}A_{j\nu}E_{j\nu}=F_{\rm r},\qquad
 2\Re\sum_{(j,\nu)\in\mathcal I_{\rm r}}A_{j\nu}\mathcal T_{j\nu}=G_{\rm r}.
 \label{total:rounded-amplitudes}
\end{equation}

\emph{(iii) Solution.} The jumps~\eqref{total:rounded-ansatz} solve~\eqref{total:coupled-system}. The jumps are real and satisfy $d,n=O(w)$ at both endpoints. The amplitudes and jumps are unique within this complete confluent representation. No separate consistency, full-column-rank or simplicity assumption is needed under the stated endpoint conditions. No completeness claim in larger endpoint classes is made.

\emph{(iv) Existence and uniqueness in $Y\times Y$.} The profile satisfies~\eqref{stab:rounded-class}. With $\ell$ as in~\eqref{stab:system}, the pair $(d,\ell)$ belongs to $Y\times Y$, defined in~\eqref{stab:space-Y}, and is the unique solution in $Y\times Y$ given by part~(iii) of Theorem~\ref{thm:existence}; no smallness condition on the slopes is required.
\end{theorem}
The endpoint conditions cannot be dropped. Without a sharp trailing edge, the same finite-pole ansatz need not be solvable: for $P=1$, $M=0$, $k_c=0$ and $\alpha=0$, one has $R=-\eta$, $b=1$ and $\Psi_{\rm r}(\zeta )=W(\zeta )-\zeta $, which has no off-chord zero. The pole sum is empty, while the second forcing is $G=-2U\eta/w\not\equiv0$. This example lies outside the stated endpoint conditions and shows why solvability is not asserted for arbitrary $P$.

The proof and transform formulas are given in Appendix~\ref{app:proof-rounded-family}, and a MATLAB implementation for symmetric NACA sections is listed in Appendix~\ref{app:matlab}. Both constructions reduce to their simple-pole forms when every $\mathsf r_j=1$. Example~\ref{ex:rounded-poles-amplitudes} in Appendix~\ref{app:examples} carries out the construction for a symmetric section with a cubic pole equation.

\subsection{Multiplicities of the pole functions}\label{sec:multiplicities}
We call $\Psi_{\rm f}$ and $\Psi_{\rm r}$ the pole functions, because their zeros $\zeta _j$ are the poles of the rational factors of the jumps~\eqref{total:flat-ansatz} and~\eqref{total:rounded-ansatz}: a term $A/(\zeta -\eta)^\nu$ leaves a non-polynomial remainder in the coupled system unless the pole function vanishes at $\zeta $ to order at least $\nu$. They contain the branch function $W$ and are not polynomials, so their zeros are computed from the candidate polynomials~\eqref{total:flat-candidate-system} and~\eqref{total:rounded-candidate-system}. The orders $\mathsf r_j$ are not lost in this step. Since $W^2=\zeta ^2-1$, each candidate polynomial is the product of its pole function and a companion on the other branch of the square root,
\begin{equation}
 (\zeta ^2-1)Q^2-(b+i\lambda)^2=\Psi_{\rm f}\,\widetilde\Psi_{\rm f},\qquad
 R^2-(b+i\lambda)^2(\zeta ^2-1)=\Psi_{\rm r}\,\widetilde\Psi_{\rm r},
 \label{total:candidate-factorization}
\end{equation}
with $\widetilde\Psi_{\rm f}=QW+b+i\lambda$ and $\widetilde\Psi_{\rm r}=R-(b+i\lambda)W$. All four factors are analytic off the chord, so the multiplicity of an off-chord root of a candidate polynomial is the sum of its orders as a zero of the two factors. A retained root satisfies $\Psi_{\rm f}(\zeta _j)=0$, that is, $Q(\zeta _j)W(\zeta _j)=b+i\lambda(\zeta _j)$, respectively $\Psi_{\rm r}(\zeta _j)=0$, that is, $[b+i\lambda(\zeta _j)]W(\zeta _j)=-R(\zeta _j)$. Substituting these relations into the companions gives
\begin{equation}
 \widetilde\Psi_{\rm f}(\zeta _j)=2[b+i\lambda(\zeta _j)]=2Q(\zeta _j)W(\zeta _j),\qquad
 \widetilde\Psi_{\rm r}(\zeta _j)=2R(\zeta _j)=-2[b+i\lambda(\zeta _j)]W(\zeta _j).
\end{equation}
If this value is nonzero, the companion does not vanish, and $\mathsf r_j$ equals the multiplicity of $\zeta _j$ as a root of the candidate polynomial. The value is zero only if $\zeta _j$ is a common root of $b+i\lambda$ and of $Q$, respectively $R$. The multiplicity is then shared between the two factors, and $\mathsf r_j$ is the smallest $k\ge1$ for which the $k$th derivative of the pole function at $\zeta _j$ is nonzero; these derivatives are explicit, since $W'=\zeta /W$.

\section{Pressures and loads}\label{sec:loads}
This section gives the pressures and loads that follow from the velocity jumps, and their explicit dependence on the incidence and the Mach number. The derivations are given in Appendix~\ref{app:loads-derivation}.

\subsection{Pressures and load coefficients}\label{sec:validation-pressure-variants}\label{sec:load-coefficients}
The disturbance velocities on the two collapsed faces are
\begin{equation}
 u^\pm=\frac{\HH(n-ad)}{2b}\pm\frac d2,\qquad
 v^\pm=\frac{a\HH(n-ad)}{2b}-\frac b2\HH d\pm\frac n2.
 \label{comparison:face-pressure-mapping}
\end{equation}
With the dynamic pressure and the pressure coefficients
\begin{equation}
 q_\infty=\tfrac12\rho_\infty U^2,\qquad
 C_p^\pm=\frac{p^\pm-p_\infty}{q_\infty},\qquad
 \Delta C_p=C_p^--C_p^+,
 \label{comparison:pressure-normalization}
\end{equation}
the linearized Bernoulli relation~\eqref{main:bernoulli} gives the face pressures and their jump,
\begin{equation}
 C_p^\pm=-\frac2U(\cos\alpha\,u^\pm+\sin\alpha\,v^\pm),\qquad
 \Delta C_p=\frac2U(\cos\alpha\,d+\sin\alpha\,n).
 \label{comparison:linear-bernoulli}
\end{equation}
The loads are those of the pressure jump acting on the chord. With the chord fraction $t$ of~\eqref{total:coordinate-weights}, the coefficients of the normal force, of the lift and of the nose-up moment about the chord point $e$ are
\begin{equation}
 C_N=\int_0^1\Delta C_p\,\dd t,\qquad
 C_L=C_N\cos\alpha,\qquad
 C_m=\int_0^1(e-t)\,\Delta C_p\,\dd t.
 \label{comparison:physical-pressure-force}
\end{equation}

\subsection{Explicit dependence on the incidence and the Mach number}\label{sec:symbolic-velocity}\label{sec:symbolic-incidence}\label{sec:separation-summary}
An aeroelastic analysis requires the loads and their slopes as functions of the incidence. Fix the geometry and the flow coefficients $a$ and $b$ of~\eqref{air:coefficients}, and let $(d^{(1)},n^{(1)})$ and $(d^{(2)},n^{(2)})$ be the solutions obtained when the freestream components $(U\cos\alpha,U\sin\alpha)$ are replaced by $(1,0)$ and $(0,1)$. The loads are then quadratic forms in the cosine and sine of the incidence,
\begin{equation}
 C_X(M,\alpha)=(\bm Q_X)_{11}\cos^2\alpha+2(\bm Q_X)_{12}\sin\alpha\cos\alpha+(\bm Q_X)_{22}\sin^2\alpha,\qquad X\in\{N,m\},
 \label{sym:quadratic-form}
\end{equation}
with the symmetric $2\times2$ matrices
\begin{equation}
 \bm Q_N=\int_0^1\bm\Delta\,\dd t,\qquad
 \bm Q_m=\int_0^1(e-t)\,\bm\Delta\,\dd t,\qquad
 \bm\Delta=\begin{pmatrix}2d^{(1)}&d^{(2)}+n^{(1)}\\ d^{(2)}+n^{(1)}&2n^{(2)}\end{pmatrix}.
 \label{sym:load-matrices}
\end{equation}
The matrices depend on the geometry, on $e$ and on $a$, $b$. At $M=0$, where $a=0$ and $b=1$ for every incidence, they depend on the geometry and on $e$ only: they are computed once per section, and three numbers per load give the load at every incidence. For zero thickness and camber,
\begin{equation}
 \bm Q_N^{0}=\frac{\pi b}{a^2+b^2}\begin{pmatrix}0&1\\ 1&0\end{pmatrix},\qquad
 \bm Q_m^{0}=\bigl(e-\tfrac14\bigr)\bm Q_N^{0},
 \label{sym:thin-matrices}
\end{equation}
which are the thin-airfoil loads.

The slope of a load with respect to the incidence at zero incidence is
\begin{equation}
 \left.\frac{\partial C_X}{\partial\alpha}\right|_{\alpha=0}=2(\bm Q_X)_{12}+M^2\,\partial_a(\bm Q_X)_{11},\qquad (a,b)=(0,\beta),
 \label{sym:slope-zero}
\end{equation}
and its dependence on the Mach number is given, to second order in the surface slopes, by
\begin{equation}
 \left.\frac{\partial C_X}{\partial\alpha}\right|_{\alpha=0}\simeq\frac{C_{X\alpha}(0)}{\beta}\Bigl(1+\kappa_X\frac{M^2}{\beta^2}\Bigr),\qquad
 \kappa_X=\left.\frac{\partial_a(\bm Q_X)_{11}-(\bm Q_X)_{12}-\partial_b(\bm Q_X)_{12}}{2(\bm Q_X)_{12}}\right|_{(a,b)=(0,1)},
 \label{sym:pg2}
\end{equation}
where $C_{X\alpha}(0)=2(\bm Q_X)_{12}$ is the incompressible slope. The slope at every subsonic Mach number thus follows, approximately, from two numbers of the section, $C_{X\alpha}(0)$ and $\kappa_X$, both computed at $M=0$; $\kappa_X$ vanishes for zero thickness and camber, for which~\eqref{sym:pg2} is the Prandtl--Glauert rule. For five of the sections of Section~\ref{sec:section-dependence}, the rule departs from the slope of the analytical solution by at most $0.03\%$ at $M=0.5$ and $0.2\%$ at $M=0.7$. The derivatives with respect to $a$ and $b$ follow from the derivatives of the poles and of the amplitude systems, without new root finding (Appendix~\ref{sec:separation}).

\section{Approximation errors of the solution}\label{sec:geometry-stability}
Theorems~\ref{thm:flat-approximation} and~\ref{thm:rounded-approximation} concern the geometry only, and Theorem~\ref{thm:existence} the solvability for a fixed geometry. We now quantify how an error in the surface slopes propagates to the velocity jumps. Quantities carrying an asterisk refer to a target geometry $(k_c^*,g_t^*)$, and
\begin{equation}
 \delta k=k_c-k_c^*,\qquad\delta g=g_t-g_t^*.
\end{equation}
For the target, $\mu^*$ denotes the margin~\eqref{stab:margin}.

\begin{corollary}[Geometry-error bounds]\label{thm:geometry-stability}
Fix $U>0$, $0\le M<1$ and $\alpha$, and hence $a,b$. The target and the approximating geometry share these flow data.

\emph{(i) Bounded slopes.} Let the target slopes $k_c^*,g_t^*\in L^\infty(-1,1)$ satisfy $\mu^*>0$, with solution $(d^*,\ell^*)$ from part~(i) of Theorem~\ref{thm:existence}. Let $(d,\ell)\in X\times X$ solve~\eqref{stab:system} for other bounded slopes $(k_c,g_t)$, and set $\epsilon_g=\norm{\delta k}_\infty+\norm{\delta g}_\infty$. Then
\begin{equation}
 \norm{d-d^*}_X+\norm{\ell-\ell^*}_X
 \le\frac{\epsilon_g}{\mu^*}
 \Bigl(2\sqrt\pi\,U|\cos\alpha|+\norm{d}_X+\norm{\ell}_X\Bigr).
 \label{stab:aposteriori}
\end{equation}
Moreover $\norm{n-n^*}_X\le|a|\norm{d-d^*}_X+b\norm{\ell-\ell^*}_X$. In particular, the flat-endpoint solutions of Theorem~\ref{thm:flat-family} converge in $X$, at the rate of the slope error, whenever their slopes $(k_c,g_t)$ converge uniformly to a target with $\mu^*>0$.

\emph{(ii) Rounded leading edges.} Let the target be of the form~\eqref{stab:rounded-class} with H\"older-continuous $k_c^*$ and $R_1^*$ on $[-1,1]$, and let the solution be $(d^*,\ell^*)\in Y\times Y$. Then there exist constants $\mathcal N^*>0$ and $\underline\varpi^*>0$, depending only on the target geometry and the flow data, with the following property. Let $(d,\ell)\in Y\times Y$ solve~\eqref{stab:system} for another geometry with bounded $k_c$ and bounded $\sqrt{1+\eta}\,g_t$, and set
\begin{equation}
 \epsilon_r=\norm{\delta k}_\infty+\norm{\sqrt{\tfrac{1+\eta}2}\,\delta g}_\infty.
 \label{stab:rounded-error-measure}
\end{equation}
Then
\begin{equation}
 \norm{(d-d^*,\ell-\ell^*)}_Y\le \frac{\mathcal N^*}{\underline\varpi^*}\,\epsilon_r
 \Bigl(2\sqrt\pi\,U|\cos\alpha|+\norm{d}_Y+\norm{\ell}_Y\Bigr).
 \label{stab:rounded-aposteriori}
\end{equation}
In particular, the rounded-family solutions of Theorem~\ref{thm:rounded-family} converge in $Y$ to the target solution, at the rate of $\epsilon_r$, whenever $\norm{\delta k}_\infty+\norm{\sqrt{1+\eta}\,\delta g}_\infty\to0$.
\end{corollary}
The proof is given in Appendix~\ref{app:proof-corollary}. A small error in the surface slopes thus produces a proportionally small error in the solution. The slopes, not the ordinates, enter the equations, so that a fit must reproduce the slopes of the section; for a rounded leading edge the slope error is measured with the weight $\sqrt{1+\eta}$, which removes the singularity at the nose. Both parts are proved by subtracting the two systems: the difference of the solutions solves the system of the target, with a right-hand side equal to the slope error times the other solution. In part~(i) both are measured in $X$; in part~(ii) the right-hand side is measured in $X$ and the solution in $Y$, which absorbs the singular slope at the nose. The approximants of Theorems~\ref{thm:flat-approximation} and~\ref{thm:rounded-approximation} satisfy these conditions, and the corresponding solutions converge to that of the target section. The constants of part~(ii) are computable but conservative, and both parts compare two solutions of the same model equations: they do not measure how close the model is to the real flow.

\subsection{Sections with a finite trailing-edge angle}\label{sec:te-angle}
Sections of practical interest usually have a finite trailing-edge angle, whereas every member of the rounded-leading-edge family has zero thickness slope at the trailing edge, so that the supremum norm of Corollary~\ref{thm:geometry-stability} cannot measure the error of approximating them. The slope error is confined to a neighbourhood of the trailing edge, and an integral norm of it both tends to zero and bounds the distance between the two solutions.

\begin{theorem}[Approximation of sections with a finite trailing-edge angle]\label{thm:te-angle}
Fix $U>0$, $0\le M<1$ and $\alpha$. Let the target geometry $(k_c^*,g_t^*)$ be of the form~\eqref{stab:wedge-class} with H\"older-continuous $k_c^*$ and $S^*$ on $[-1,1]$ and with~\eqref{stab:te-angle}, with slope angles $\phi^{*\pm}$ from~\eqref{stab:angles}, trailing-edge jump $\varDelta_{\rm te}$, and solution $(d^*,\ell^*)\in Y\times Y$, whose chordwise face velocities are $u^{*\pm}$ as in~\eqref{comparison:face-pressure-mapping}. For a second geometry $(k_c,g_t)$ with slope angles $\phi^\pm$, and with $\delta k$ and $\delta g$ as above, define
\begin{equation}
 \norm{\delta}_*=\Biggl(\sum_\pm\int_{-1}^1\frac{|\delta k\pm\delta g|^2}{(a-k_c\mp g_t)^2+b^2}\,\bigl(U\cos\alpha+u^{*\pm}\bigr)^2\,\frac{\dd\eta}{w}\Biggr)^{1/2},
 \label{stab:target-weighted-norm}
\end{equation}
\begin{equation}
 \varDelta_{\rm ang}=\max_\pm\,\sup_{-1<\eta<1}\Bigl|2\bigl(\phi^\pm-\phi^{*\pm}\bigr)\pm\bigl(\phi^{*+}(1)-\phi^{*-}(1)\bigr)\Bigl(1-\frac{\arccos\eta}{\pi}\Bigr)\Bigr| .
 \label{stab:angle-distance}
\end{equation}

\emph{(i) Distance between the solutions.} Let the second geometry have a bounded mean slope $k_c$ and a bounded $\sqrt{1+\eta}\,g_t$, let $(d,\ell)\in Y\times Y$ be a solution of~\eqref{stab:system} for it, and let $\varDelta_{\rm ang}<\pi/2$. Then
\begin{equation}
 \norm{(d-d^*,\ell-\ell^*)}_Y\le\frac{\mathcal N_{\rm c}^*}{\cos\varDelta_{\rm ang}}\,\norm{\delta}_*,
 \label{stab:integral-bound}
\end{equation}
where the constant $\mathcal N_{\rm c}^*$, given by~\eqref{app:target-constant}, depends on the target geometry and the flow data only.

\emph{(ii) Approximation by the rounded-leading-edge family.} Write the target thickness as $T^*=(f^+-f^-)/c=wP^*$, and suppose that $P^*\in C^1(-1,1)$ is continuous on $[-1,1]$ with $P^*(1)=0$ and $P^*(-1)>0$, and that $wP^{*\prime}$ extends continuously to $[-1,1]$. Then there are polynomials $P_N$, with $P_N(1)=0$ and, for all sufficiently large $N$, $P_N(-1)>0$, and polynomials $k_N$ with $\int_{-1}^1k_N\,\dd\eta=0$, such that the sections of the rounded-leading-edge family with thickness $T_N=wP_N$ and mean slope $k_N$ satisfy
\begin{equation}
 \norm{T_N-T^*}_\infty\longrightarrow0,\qquad
 \norm{\delta_N}_*\longrightarrow0,\qquad
 \limsup_{N\to\infty}\varDelta_{{\rm ang},N}\le\varDelta_{\rm te}<\frac\pi2 .
 \label{stab:integral-approximation}
\end{equation}

\emph{(iii) Convergence.} The solutions $(d_N,\ell_N)$ of Theorem~\ref{thm:rounded-family} for these sections converge to the solution for the target, $\norm{(d_N-d^*,\ell_N-\ell^*)}_Y\to0$, and the distance is bounded by~\eqref{stab:integral-bound}.
\end{theorem}
The proof is given in Appendix~\ref{app:proof-te-angle}. The weight of the norm~\eqref{stab:target-weighted-norm} is the square of the total chordwise velocity on the faces of the target section. Part~(ii) is an existence statement for a particular sequence of approximants; it gives no rate of convergence and does not assert that the least-squares fits of Section~\ref{sec:comparison-pressure-loads} are such a sequence.

\section{Aerodynamic analysis}\label{sec:comparison-pressure-loads}\label{sec:rerun-comparisons}\label{sec:sqrt-results}
We compare the solutions of the modified airfoil system with two references. Inviscid XFOIL on the contour of each section tests the collapsed-boundary, small-disturbance model together with the polynomial fit of the geometry, and published pressure measurements test both models against the physical flow; the two questions are not equivalent. All measurements retained here have chord Reynolds numbers of order $10^6$ or larger; the Reynolds number labels the experiment and is not a parameter of either inviscid prediction. The comparisons use the rounded-leading-edge family; the flat-endpoint family remains an analytical result. Section~\ref{sec:experimental-data} lists the data sets, and Sections~\ref{sec:symmetric-results} and~\ref{sec:cambered-results} present the results section by section.

\subsection{Approach}\label{sec:xfoil-protocol}
The numerical reference is XFOIL~\cite{DrelaXfoil,KatzPlotkin2001}, run in inviscid mode: a panel method on the actual contour, which involves neither the collapse of the boundary onto the chord nor the linearization of the flow equation. Its pressures follow from the full Bernoulli relation on the contour, with the K\'arm\'an--Tsien correction at nonzero Mach number. Its contour is that of the original formula of each section (Appendix~\ref{app:section-formulas}), with the trailing edge closed. The present solution uses a member of the rounded-leading-edge family fitted to the same section at order 30, with the pressures~\eqref{comparison:linear-bernoulli}; no aerodynamic data enter the fit. The nominal sections have a finite trailing-edge angle and fall under part~(iii) of Theorem~\ref{thm:existence}; the fitted sections approximate them in the sense of Theorem~\ref{thm:te-angle}. In the figures, blue solid curves denote the present solution and red dotted curves inviscid XFOIL. The settings of XFOIL, the fitting procedure and the numerical checks are given in Appendix~\ref{app:section-formulas}.

\subsection{Choice of the fit order}\label{sec:fit-order}
\begin{figure}[!htbp]\centering
\includegraphics[width=\linewidth]{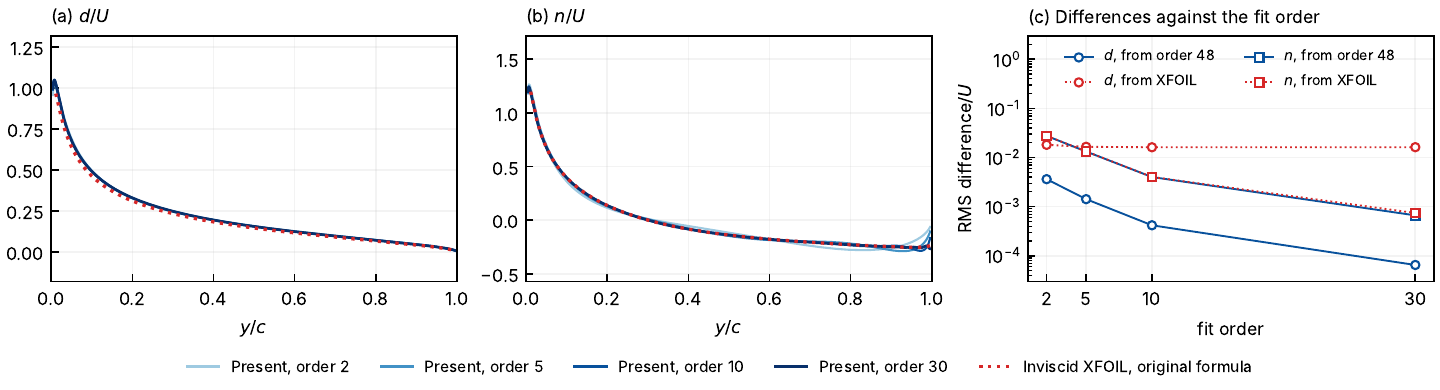}
\caption{NACA 0012 at $M=0$ and $\alpha=4^\circ$: velocity jumps of the present solutions for fits of order 2, 5, 10 and 30 of the rounded-leading-edge family, and jumps reconstructed from inviscid XFOIL on the contour of the original formula. (a) $d/U$; (b) $n/U$; (c) root-mean-square differences on $0.02\le y/c\le0.95$ from the order-48 solution and from XFOIL, against the fit order.}\label{fig:val-jumps}
\end{figure}

Before the comparisons with measurements, we examine how the solution depends on the order of the fit, and how it compares with XFOIL on the original contour, for NACA 0012 at $M=0$. Figure~\ref{fig:val-jumps} shows the velocity jumps $d$ and $n$ of the solutions of Theorem~\ref{thm:rounded-family} for fits of order 2, 5, 10 and 30, at an incidence of $4^\circ$. The XFOIL jumps are reconstructed from its surface pressures on the contour of the original formula: at $M=0$ the speed on the surface is $U\sqrt{1-C_p}$, directed along the local tangent and reversed ahead of the stagnation point, and the jumps are the differences of the disturbance-velocity components on the two faces at the same chord position.

The solutions converge rapidly with the order, and already at very low order they are close to the converged solution. On $0.02\le y/c\le0.95$, the root-mean-square difference of $n/U$ from the order-48 solution is $2.8\times10^{-2}$, $1.3\times10^{-2}$, $4.0\times10^{-3}$ and $6.6\times10^{-4}$ at orders 2, 5, 10 and 30, and that of $d/U$ is $3.6\times10^{-3}$, $1.4\times10^{-3}$, $4.2\times10^{-4}$ and $6\times10^{-5}$. The normal-velocity jump agrees closely with XFOIL: the difference falls with the order, to $7.4\times10^{-4}$ at order 30 and $4.8\times10^{-4}$ at order 48, so that it is governed by the fit and not by the model. The chordwise jump differs from that of XFOIL by $1.6\times10^{-2}$ at every order from 5 upwards, and by $1.8\times10^{-2}$ at order 2: this difference is essentially independent of the fit and is the error of the collapsed-boundary model, largest near the leading edge. Even the order-2 fit, which has three coefficients, reproduces both jumps over most of the chord; the visible departures are in $n$, over the rear half of the chord and near the trailing edge, where the low-order fits oscillate about the finite slope of the original section. All comparisons below use the order-30 fit; at this order the jumps differ from those of the order-48 fit by less than $10^{-3}U$, well below the differences between the models.

\subsection{Experimental data sets}\label{sec:experimental-data}\label{sec:established-benchmarks}\label{sec:extended-naca-validation}\label{sec:cambered-validation}\label{sec:additional-asymmetric}
Table~\ref{tab:validation-sources} lists every experimental data set used in the comparisons. Angles are used as tabulated. For NACA 4412 and NACA 23012 these are the source's effective section angles $\alpha_0$; no further downwash correction or fitted incidence shift is applied. The model Mach number is the reported value where the source gives one, a nominal reconstruction from the chord, Reynolds number and a standard air state otherwise (Appendix~\ref{app:pressure-source-details}), and $M=0$ where the selected data do not establish a unique test Mach number. NASA TM-100526, NASA TP-3151 and NREL/TP-442-6472 present data uncorrected for wall interference. Symbols read from published curves carry reading allowances, which are shown as bars and are not experimental confidence intervals. Pressure errors are RMSE values at the observation points on $0.02\le y/c\le0.95$, evaluated identically for the present solution and for XFOIL. At $M>0$, a panel is labeled supercritical when the measured minimum pressure coefficient lies below the isentropic sonic value
\begin{equation}
 C_p^*=\frac{2}{\gamma M^2}\left[\left(\frac{2+(\gamma-1)M^2}{\gamma+1}\right)^{\gamma/(\gamma-1)}-1\right].
 \label{comparison:cp-star}
\end{equation}
Such panels contain a local supersonic region that neither potential-flow prediction is designed to represent; they are shown but excluded from the aggregate errors. The results are grouped by section: each figure shows the section geometry above the pressure comparisons for all data sets of that section.

\begin{table}[!htbp]\centering\footnotesize
\caption{Experimental pressure data sets. $M$ is the Mach number used by both predictions: reported (r), reconstructed from the nominal test conditions (n), or set to zero for low-speed data. ASPIRE denotes the transcription of~\cite{ASPIREPressureData}; uncorrected data carry no wall corrections.}\label{tab:validation-sources}\label{tab:cambered-pressure-sources}
\setlength{\tabcolsep}{4pt}
\begin{tabularx}{\textwidth}{@{}l>{\raggedright\arraybackslash\hsize=1.05\hsize}Xlll>{\raggedright\arraybackslash\hsize=0.95\hsize}X@{}}
\toprule
Section & Source & $Re_c/10^6$ & $M$ & $\alpha$ (deg) & Notes\\
\midrule
NACA 0006 & NACA TN 3797~\cite{Gambucci1956} & 4.5 & 0.15 (r) & 4.18 & both faces\\
NACA 0008 & NACA TN 3524~\cite{McCullough1955} & 6.0 & 0 & 8 & transcribed data\\
NACA 0009 & NACA Report 637~\cite{SilversteinBecker637} & 1.73--5.02 & 0 & 0 & upper face; run $Re_c$ unknown\\
NACA 0012 & Gregory and O'Reilly~\cite{Gregory1970} & 2.88 & 0 & 0, 10 & upper face\\
 & NASA TM-100526~\cite{LadsonHillJohnson1987}, ASPIRE & 3.0 & 0.3--0.7 (r) & 0, 2, 4 & uncorrected\\
NACA 0015 & NASA TP-3151~\cite{McAlisterTakahashi1991} & 1.5, 2.5 & 0.13, 0.22 (n) & 0, 6, 12 & $y/s=0.69$; uncorrected\\
NACA 0018 & NACA Report 637~\cite{SilversteinBecker637} & 1.73--5.02 & 0 & 0 & upper face; run $Re_c$ unknown\\
\addlinespace
NACA 4412 & NACA Report 563~\cite{Pinkerton1936} & 3.1 & 0 & $-0.5$, 2.9, 6.4 & tabulated; effective $\alpha_0$\\
NACA 4415 & NREL/TP-442-6472~\cite{Reuss1995} & 2.0 & 0.192 (n) & 4.12 & clean surface; uncorrected\\
NACA 23012 & NACA Report 614~\cite{Wenzinger1938} & 1.22 & 0.105 (n) & 2.70 & effective $\alpha_0$; no flap\\
NACA 64A410 & NACA TN 3162~\cite{Stivers1954}, ASPIRE & 1.0--1.7 & 0.32--0.67 (r) & 0, 2, 4 & tabulated ordinates fitted\\
\bottomrule
\end{tabularx}
\end{table}

\subsection{Symmetric sections}\label{sec:symmetric-results}
\begin{figure}[!htbp]\centering
\includegraphics[width=\linewidth]{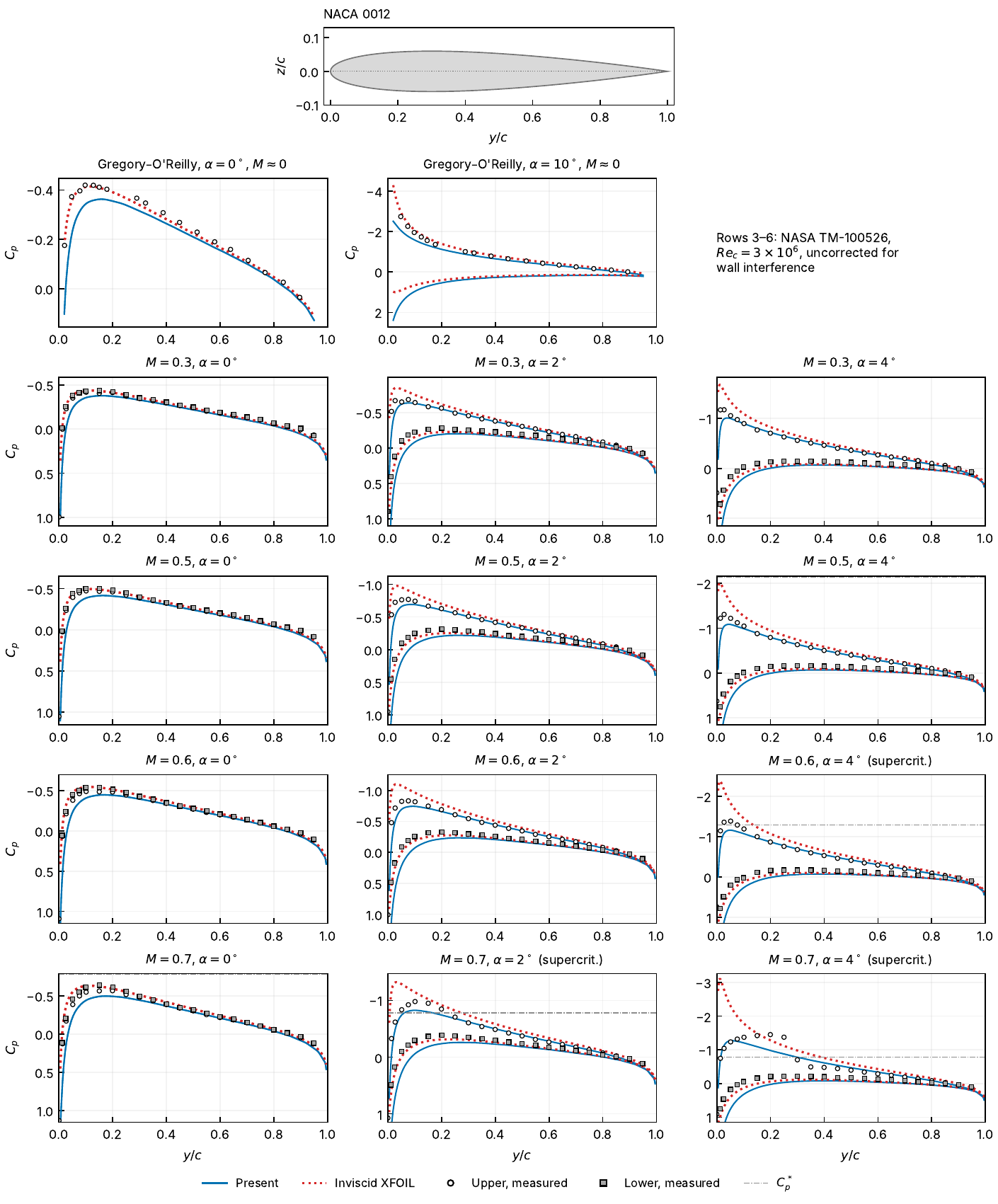}
\caption{NACA 0012. Top: section geometry from the original formula. Row 2: Gregory--O'Reilly upper-surface pressures, $Re_c\simeq2.88\times10^6$, $M=0$. Rows 3--6: NASA TM-100526, $Re_c=3\times10^6$, at $M=0.3$, $0.5$, $0.6$, $0.7$ and $\alpha=0^\circ$, $2^\circ$, $4^\circ$. The present solution (blue solid), computed with the order-30 fit, and inviscid XFOIL (red dotted), computed on the contour of the original formula, use the same Mach number. The dash-dotted line is $C_p^*$ from~\eqref{comparison:cp-star}, drawn when it lies in the plotted range.}\label{fig:val-naca0012}\label{fig:pressure-validation-matrix}\label{fig:mach-naca0012}
\end{figure}

\emph{NACA 0012} (Figure~\ref{fig:val-naca0012}). For the low-speed data the RMSE of the present solution is $0.081$ at $0^\circ$ and $0.252$ at $10^\circ$, against $0.012$ and $0.103$ for XFOIL. Over the nine subcritical panels of the Mach-number sweep the aggregate RMSE is $0.108$ for the present solution and $0.079$ for XFOIL.

\begin{figure}[!htbp]\centering
\includegraphics[width=\linewidth]{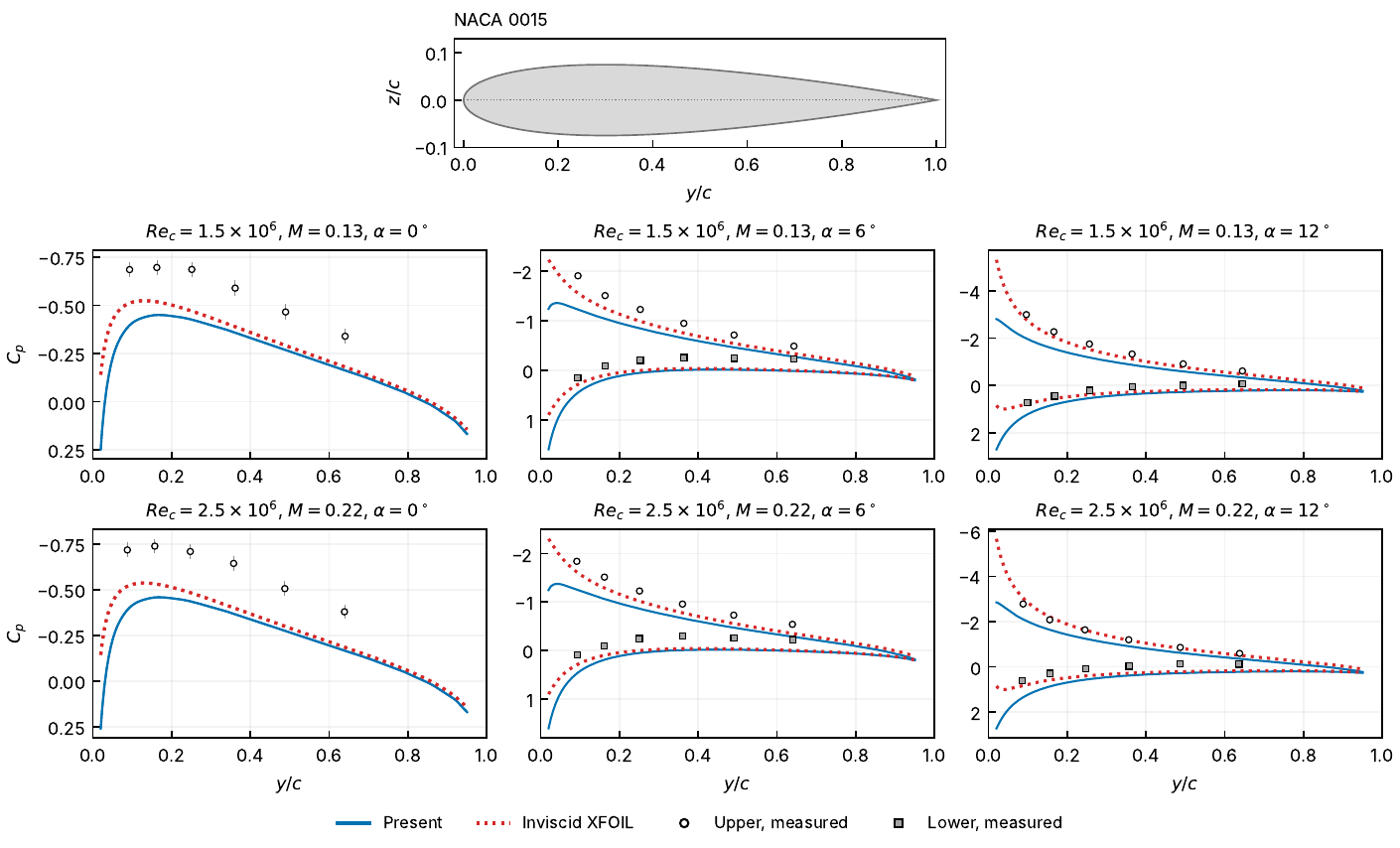}
\caption{NACA 0015. Top: section geometry from the original formula. Below: NASA TP-3151, Figure~18, at span fraction $0.69$; rows $Re_c=1.5\times10^6$ ($M=0.13$) and $2.5\times10^6$ ($M=0.22$), columns $\alpha=0^\circ,6^\circ,12^\circ$. Coincident zero-incidence taps are shown once. Bars are digitization allowances of $0.005c$ and $0.04$ in $C_p$.}\label{fig:val-naca0015}\label{fig:naca0015-pressure}
\end{figure}

\emph{NACA 0015} (Figure~\ref{fig:val-naca0015}). The RMSE is $0.24$--$0.50$ for the present solution, increasing with incidence, and $0.18$--$0.23$ for XFOIL.

\begin{figure}[!htbp]\centering
\includegraphics[width=\linewidth]{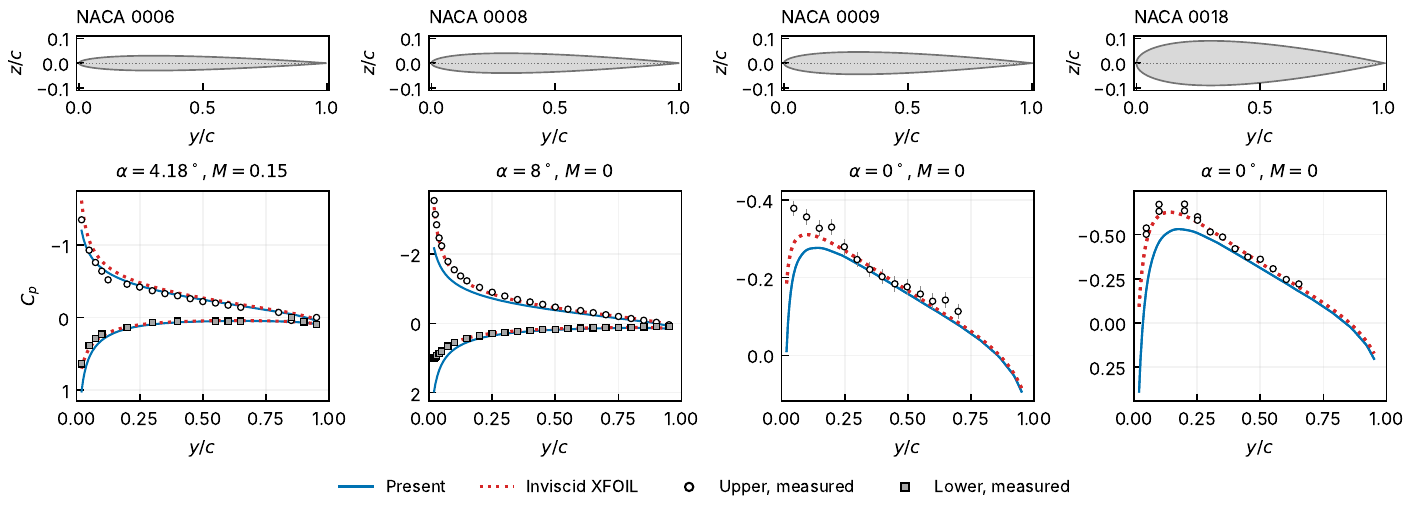}
\caption{NACA 0006, 0008, 0009 and 0018. Top: section geometries from the original formula, at a common scale. Bottom: NACA 0006 at $4.18^\circ$, $Re_c=4.5\times10^6$, $M=0.15$ (TN 3797); NACA 0008 at $8^\circ$, $Re_c=6\times10^6$ (TN 3524); zero-lift NACA 0009 and 0018 upper-surface readings (Report 637). Bars are reading allowances.}\label{fig:val-symmetric-other}
\end{figure}

\emph{NACA 0006, 0008, 0009 and 0018} (Figure~\ref{fig:val-symmetric-other}). The RMSE of the present solution and of XFOIL are $0.089$ and $0.082$ for NACA 0006, $0.410$ and $0.052$ for NACA 0008, $0.058$ and $0.033$ for NACA 0009, and $0.149$ and $0.037$ for NACA 0018.

\subsection{Cambered sections}\label{sec:cambered-results}
\begin{figure}[!htbp]\centering
\includegraphics[width=\linewidth]{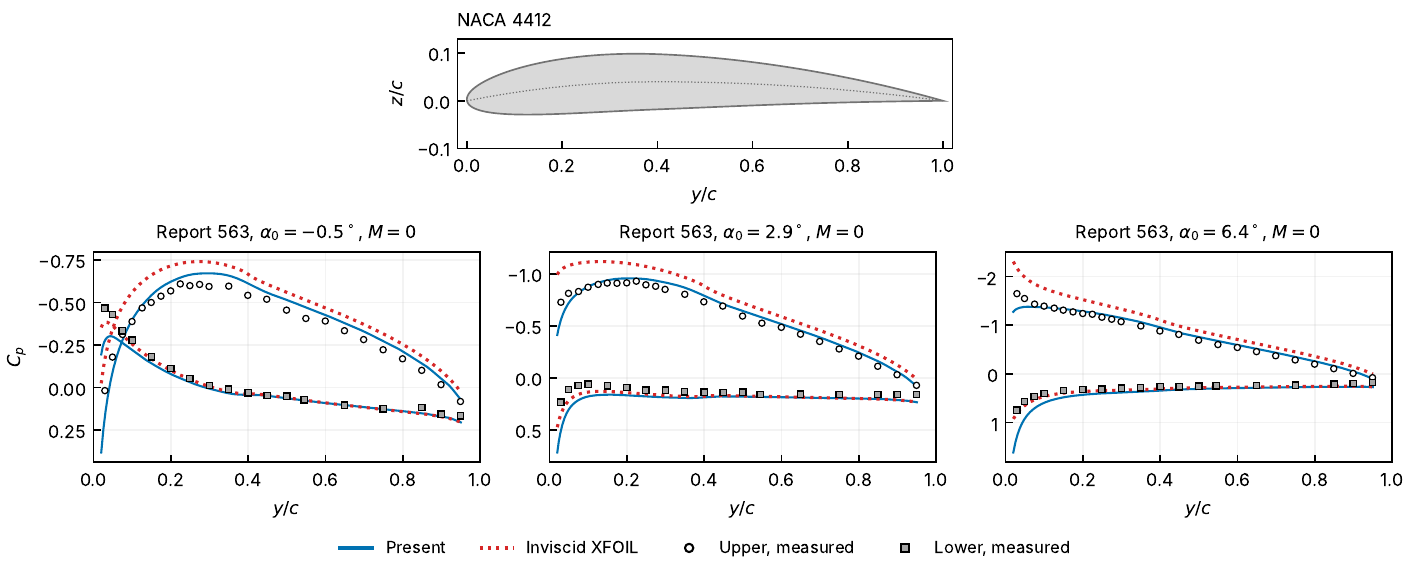}
\caption{NACA 4412. Top: section geometry from the original formulas. Bottom: NACA Report 563 tabulated pressures at mean $Re_c=3.1\times10^6$ and the effective angles $\alpha_0=-0.5^\circ$, $2.9^\circ$, $6.4^\circ$; predictions use $M=0$.}\label{fig:val-naca4412}
\end{figure}

\emph{NACA 4412} (Figure~\ref{fig:val-naca4412}). At the three incidences the RMSE is $0.060$, $0.079$ and $0.137$ for the present solution and $0.098$, $0.138$ and $0.209$ for XFOIL.

\begin{figure}[!htbp]\centering
\includegraphics[width=\linewidth]{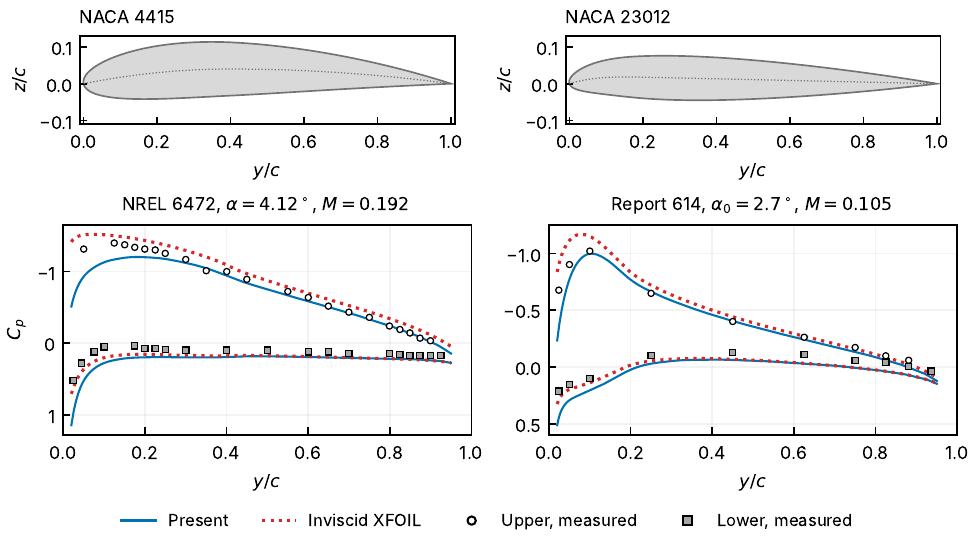}
\caption{NACA 4415 and 23012. Top: section geometries from the original formulas, at a common scale. Bottom: pressures at the source conditions of Table~\ref{tab:validation-sources}; symbols are curve readings with reading allowances.}\label{fig:val-cambered-other}\label{fig:asymmetric-pressure-matrix}
\end{figure}

\emph{NACA 4415 and 23012} (Figure~\ref{fig:val-cambered-other}). The RMSE of the present solution and of XFOIL are $0.143$ and $0.101$ for NACA 4415, and $0.115$ and $0.091$ for NACA 23012.

\begin{figure}[!htbp]\centering
\includegraphics[width=\linewidth]{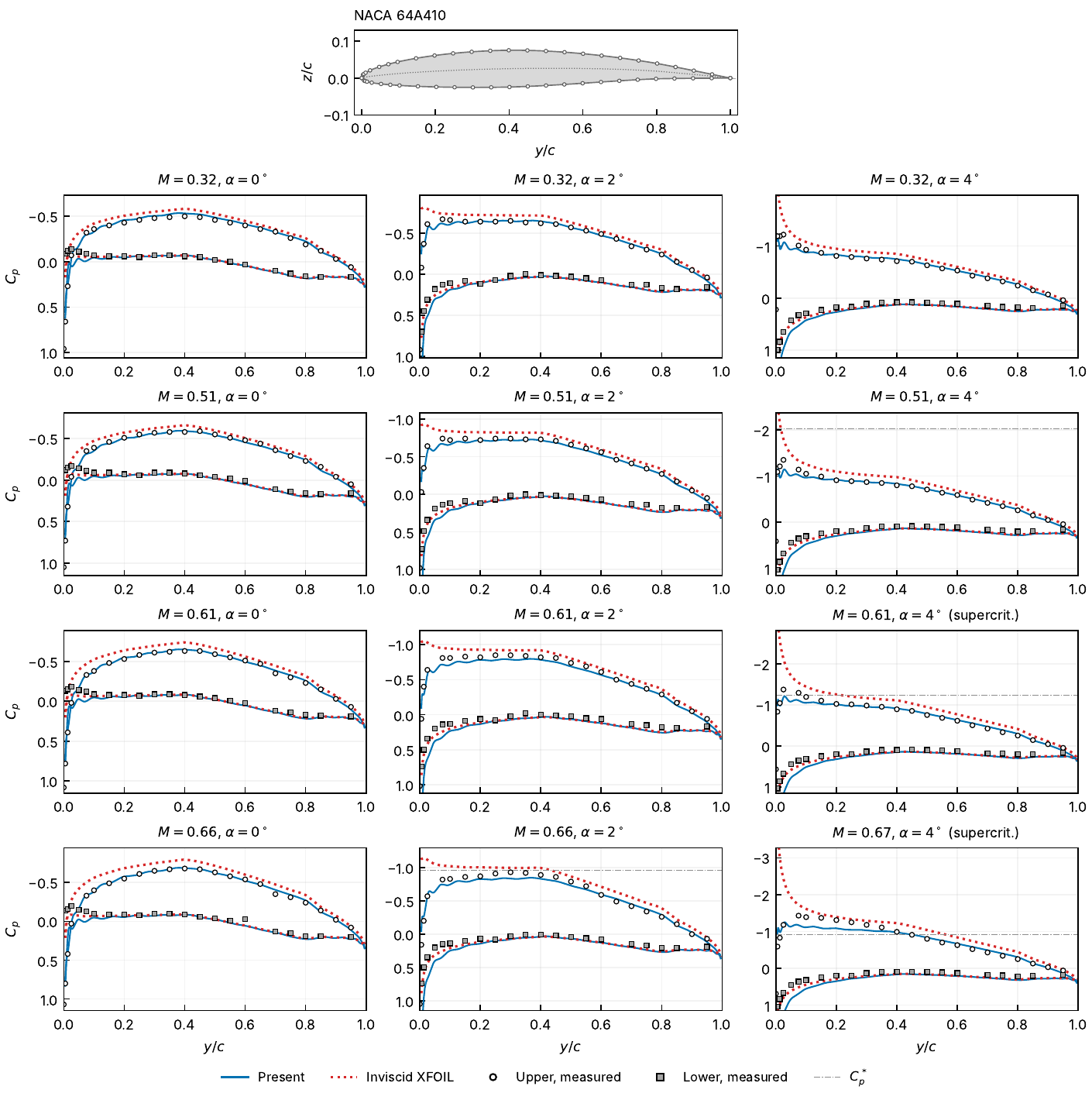}
\caption{NACA 64A410. Top: section geometry, spline through the tabulated ordinates (line) and the ordinates (symbols). Below: NACA TN 3162, $Re_c=1.0$--$1.7\times10^6$; rows group run Mach numbers near $0.32$, $0.51$, $0.61$, $0.66$, each panel title giving the tabulated value; columns $\alpha=0^\circ,2^\circ,4^\circ$. Line styles as in Figure~\ref{fig:val-naca0012}.}\label{fig:val-naca64a410}\label{fig:mach-naca64a410}
\end{figure}

\emph{NACA 64A410} (Figure~\ref{fig:val-naca64a410}). Over the ten subcritical panels the aggregate RMSE is $0.073$ for the present solution and $0.090$ for XFOIL.

\subsection{Summary of the comparisons}\label{sec:validation-summary}\label{sec:mach-comparison}\label{sec:naca0015-comparison}
Neither prediction is uniformly closer to the measurements. XFOIL is closer for the symmetric sections, most clearly at zero incidence, for the thicker sections and at the leading-edge suction peak at high incidence, where the linearized pressure relation underestimates the suction. The present solution is closer over the NACA 4412 interior and for NACA 64A410, and the two are comparable for NACA 0006 and at the higher Mach numbers and incidences of NACA 0012, where the K\'arm\'an--Tsien suction peak of XFOIL exceeds the measured one. XFOIL is closer for NACA 4415 and NACA 23012 and near the leading edge of NACA 4412. These comparisons distinguish an observable's agreement with measurements from accuracy as a solution of the prescribed inviscid contour problem; compensating model and geometry errors, uncorrected tunnel conditions and sparse source geometry can all affect the ranking. They support the analytical solutions and the compressible operator within the subcritical range, and they do not establish general superiority over XFOIL.

\section{Aeroelastic analysis: divergence}\label{sec:aeroelastic-analysis}
The sectional loads are now coupled to an elastic structure, and its static aeroelastic instability, divergence, is determined in closed form: first for a cantilever wing in strip theory, and then for a section with a camber degree of freedom. In both cases the value obtained with inviscid XFOIL loads serves as reference. The derivations are given in Appendix~\ref{app:wing-derivation}.

\subsection{Divergence of a cantilever wing}\label{sec:goland-model}\label{sec:wing-model}\label{sec:symbolic-loads}\label{sec:beam-divergence-condition}\label{sec:cambered-equilibrium}
Consider a uniform cantilever wing of span $S$ and chord $c$, with torsional rigidity $GJ$ and the elastic axis at the chord fraction $e$, at the reference incidence $\alpha_r$. In the typical-section (strip) approximation each spanwise station is an independent two-dimensional section at the incidence $\alpha_r+\theta$, where $\theta$ is the nose-up twist, and the twist satisfies $-GJ\,\theta_{xx}=q_\infty c^2C_m$ with a clamped root and a free tip. Divergence involves the torsion equation only. The divergence dynamic pressure and speed are
\begin{equation}
 q_D=\frac{\pi^2GJ}{4S^2c^2C_{m\alpha}},\qquad
 U_D=\frac{\pi}{2S}\sqrt{\frac{2\,GJ}{\rho_\infty c^2C_{m\alpha}}},\qquad
 C_{m\alpha}=\frac{\partial C_m}{\partial\alpha}(M,\alpha_r),
 \label{beam:divergence-characteristic}
\end{equation}
with the slope of the moment~\eqref{comparison:physical-pressure-force} about the elastic axis; for zero thickness and camber, $C_{m\alpha}=2\pi(e-\tfrac14)/\beta$ at zero incidence. At zero reference incidence the rule~\eqref{sym:pg2} gives the divergence dynamic pressure at a prescribed Mach number in terms of the incompressible slope $C_{m\alpha}(0)=2(\bm Q_m)_{12}$ and $\kappa_m$,
\begin{equation}
 q_D(M)=\frac{\pi^2GJ}{4S^2c^2C_{m\alpha}(0)}\,\frac{\beta}{1+\kappa_mM^2/\beta^2}.
 \label{beam:divergence-explicit-q}
\end{equation}
When the Mach number is that of the divergence speed itself, $M=U_D/a_\infty$, the divergence speed is, to first order in $\kappa_m$,
\begin{equation}
 \begin{gathered}U_D=U_P\Bigl(1-\frac{\kappa_mM_P^2}{2-M_P^2}\Bigr),\qquad
 U_P=U_0\left[\sqrt{1+\frac{M_0^4}{4}}-\frac{M_0^2}{2}\right]^{1/2},\\
 U_0=\frac{\pi}{2S}\sqrt{\frac{2\,GJ}{\rho_\infty c^2C_{m\alpha}(0)}},\qquad M_0=\frac{U_0}{a_\infty},\quad M_P=\frac{U_P}{a_\infty}.\end{gathered}
 \label{beam:divergence-compressible}
\end{equation}
The divergence speed is thus explicit in the structural data, the density and speed of sound of the stream, and two geometry-only numbers of the section; no aerodynamic solution at $M>0$ and no iteration on the Mach number is needed.

\subsection{Dependence of the divergence speed on the section}\label{sec:wing-data}\label{sec:cambered-divergence-bars}\label{sec:section-dependence}
We consider a wing of span $S=6.5$~m, chord $c=0.508$~m, elastic axis at mid-chord, $e=0.50$, and torsional rigidity $GJ=9.539\times10^3$~N\,m$^2$, in air of standard sea-level density, $\rho_\infty=1.225$~kg/m$^3$, and speed of sound $a_\infty=340.3$~m/s, at zero reference incidence. Its sectional properties are those of the high-aspect-ratio wing of Tang and Dowell~\cite{TangDowell2001,TangDowell2016}, scaled geometrically by $10$ and with a longer span, so that the chord Reynolds number at divergence is $1.5$--$1.7\times10^6$ and the Mach number $0.13$--$0.14$. Twenty sections are considered: the symmetric NACA 0006 to 0024; camber series with $2$--$8\%$ camber at $12\%$, $18\%$ and $24\%$ thickness; and NACA 23012. The predictions use the explicit expression~\eqref{beam:divergence-compressible} with the order-30 fits of Section~\ref{sec:xfoil-protocol}. As a reference, the moment slope of inviscid XFOIL on the contours of the original formulas (Appendix~\ref{app:section-formulas}), integrated from its surface pressures and differenced in the incidence, is used in~\eqref{beam:divergence-characteristic} at the same Mach number.

\begin{figure}[!htbp]\centering
\includegraphics[width=\linewidth]{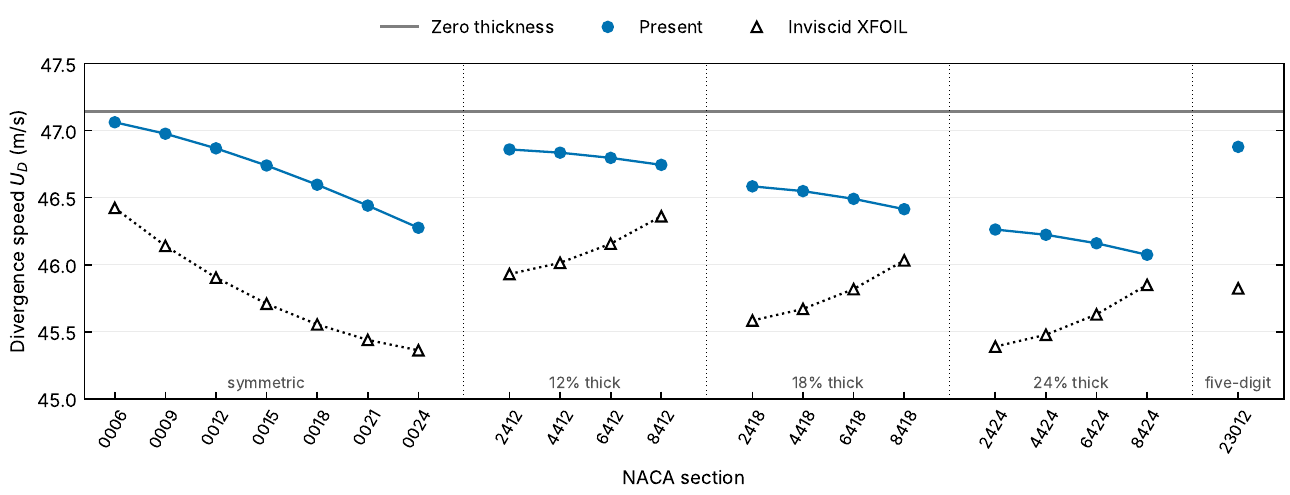}
\caption{Divergence speed of the wing of Section~\ref{sec:wing-data} at zero incidence: explicit expression~\eqref{beam:divergence-compressible} with the present loads, and~\eqref{beam:divergence-characteristic} with the moment of inviscid XFOIL at the same Mach number, for the symmetric sections NACA 0006 to 0024, camber series at $12\%$, $18\%$ and $24\%$ thickness, and NACA 23012. The horizontal line is the zero-thickness value. The Mach number is $0.13$--$0.14$.}\label{tab:aeroelastic-results}\label{fig:section-dependence}
\end{figure}

Figure~\ref{fig:section-dependence} compares the predicted and the XFOIL-based speeds. The predicted speed lies between the zero-thickness value, $47.14$~m/s, and the XFOIL-based speed for all twenty sections; it is within $2.3\%$ of the XFOIL-based speed, $1.7\%$ on average, against $2.9\%$ on average for the zero-thickness theory. It falls with thickness as the XFOIL-based speed does, through a forward displacement of the aerodynamic center, but it does not follow the rise of the XFOIL-based speed with camber: the present loads contain neither the increase of the lift slope with thickness nor the moment of the chordwise pressure force on the faces. These are strip-theory values.

\subsection{Divergence of a section with a camber degree of freedom}\label{sec:camber-divergence}
The divergence of the wing involves its twist alone, and the thickness changes it by a few percent. A second static problem, in which the section itself deforms, shows a larger effect. Murua, Palacios and Peir\'o~\cite{MuruaPalaciosPeiro2010} consider a zero-thickness section in incompressible flow that pitches about its midchord against a torsional spring of stiffness $K_\theta$ per unit span and bends chordwise in a parabolic camber mode against a stiffness $K_\varkappa$. In the notation of Section~\ref{sec:geometry}, the camber mode is the deformation
\begin{equation}
 h_e=\varkappa\,\frac c2\,(1-\eta^2),\qquad h_e'=-2\varkappa\eta,
 \label{camber:mode}
\end{equation}
with dimensionless amplitude $\varkappa$, up to a rigid translation that does not affect the steady loads. The loads that do work on the two coordinates are the pitching moment about the midchord, $m_\theta$, and the camber force $f_\varkappa=\tfrac c2\int_I(p^--p^+)\bigl(\tfrac13-\eta^2\bigr)\,\dd y$. Linearized about the undeformed section at zero incidence,
\begin{equation}
 \begin{pmatrix}m_\theta\\ f_\varkappa\end{pmatrix}
 =q_\infty\frac{c^2}{4}
 \begin{pmatrix}\mathsf m_\theta&\mathsf m_\varkappa\\ \mathsf f_\theta&\mathsf f_\varkappa\end{pmatrix}
 \begin{pmatrix}\theta\\ \varkappa\end{pmatrix},
 \label{camber:matrix}
\end{equation}
and the static equilibrium is $K_\theta\theta=m_\theta$, $K_\varkappa\varkappa=f_\varkappa$. The steady limit of the loads of~\cite{MuruaPalaciosPeiro2010} gives $\mathsf m_\theta=2\pi$, $\mathsf m_\varkappa=0$, $\mathsf f_\theta=-2\pi/3$ and $\mathsf f_\varkappa=\pi/3$. The matrix is triangular, so that the two coordinates diverge independently, the pitch at $q_\infty c^2/4=K_\theta/(2\pi)$ and the camber at $q_\infty c^2/4=3K_\varkappa/\pi$, which is the camber divergence condition of~\cite{MuruaPalaciosPeiro2010}.

For a section with thickness, the four coefficients follow from the analytical solutions at $M=0$; their expressions, the reference computation with inviscid XFOIL and tabulated values are given in Appendix~\ref{app:camber-coefficients}. With $\widehat q=q_\infty c^2/4$, the equilibrium equations with~\eqref{camber:matrix} have a nontrivial solution when
\begin{equation}
 (\mathsf m_\theta\mathsf f_\varkappa-\mathsf m_\varkappa\mathsf f_\theta)\,\widehat q^{\,2}
 -(K_\varkappa\mathsf m_\theta+K_\theta\mathsf f_\varkappa)\,\widehat q+K_\theta K_\varkappa=0,
 \qquad
 \frac{\theta}{\varkappa}=\frac{\widehat q\,\mathsf m_\varkappa}{K_\theta-\widehat q\,\mathsf m_\theta},
 \label{camber:divergence}
\end{equation}
which gives the divergence pressures and the divergence modes in closed form. A coordinate that is free while the other is held diverges at $\widehat q=K_\theta/\mathsf m_\theta$ (pitch) or $\widehat q=K_\varkappa/\mathsf f_\varkappa$ (camber).

\begin{figure}[!htbp]\centering
\includegraphics[width=\linewidth]{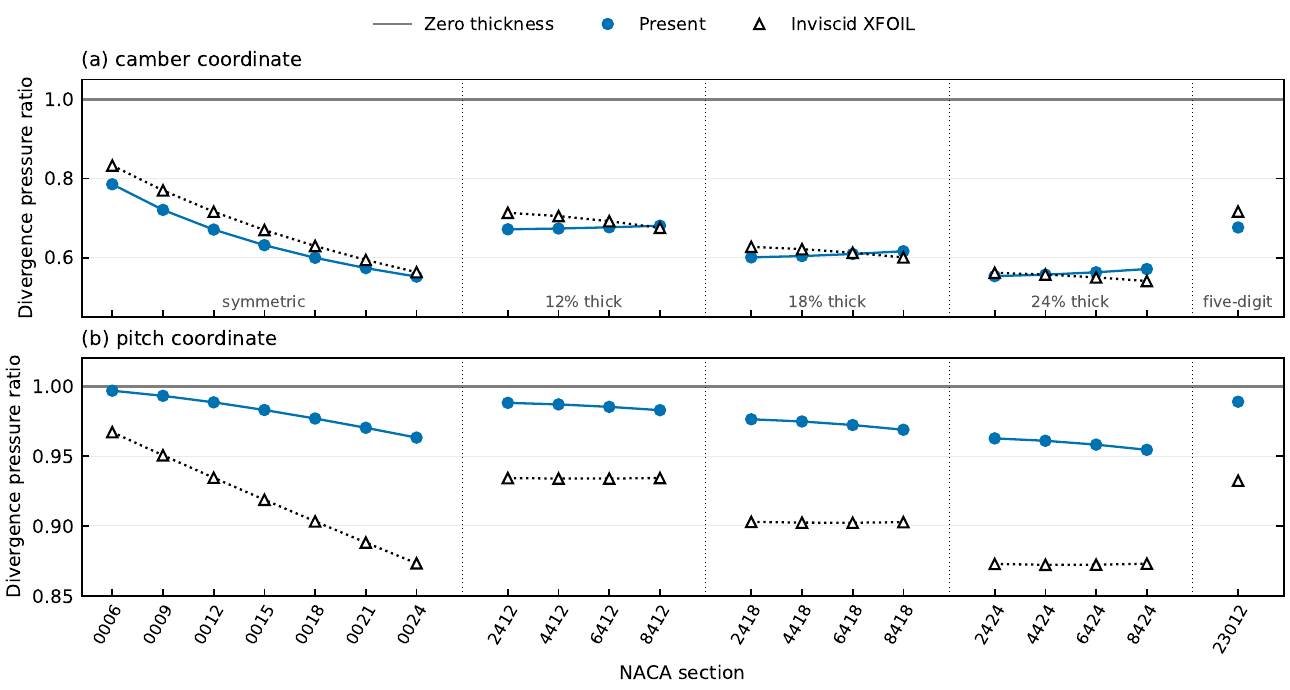}
\caption{Divergence pressure of (a) the camber coordinate, $K_\varkappa/\mathsf f_\varkappa$, and (b) the pitch coordinate about the midchord, $K_\theta/\mathsf m_\theta$, relative to the zero-thickness values of~\cite{MuruaPalaciosPeiro2010}, at $M=0$, for the sections of Figure~\ref{fig:section-dependence}: present loads, from~\eqref{camber:incidence-coefficients} and~\eqref{camber:camber-coefficients}, and inviscid XFOIL on the contours of the original section formulas.}
\label{fig:camber-divergence}
\end{figure}

\begin{figure}[!htbp]\centering
\includegraphics[width=\linewidth]{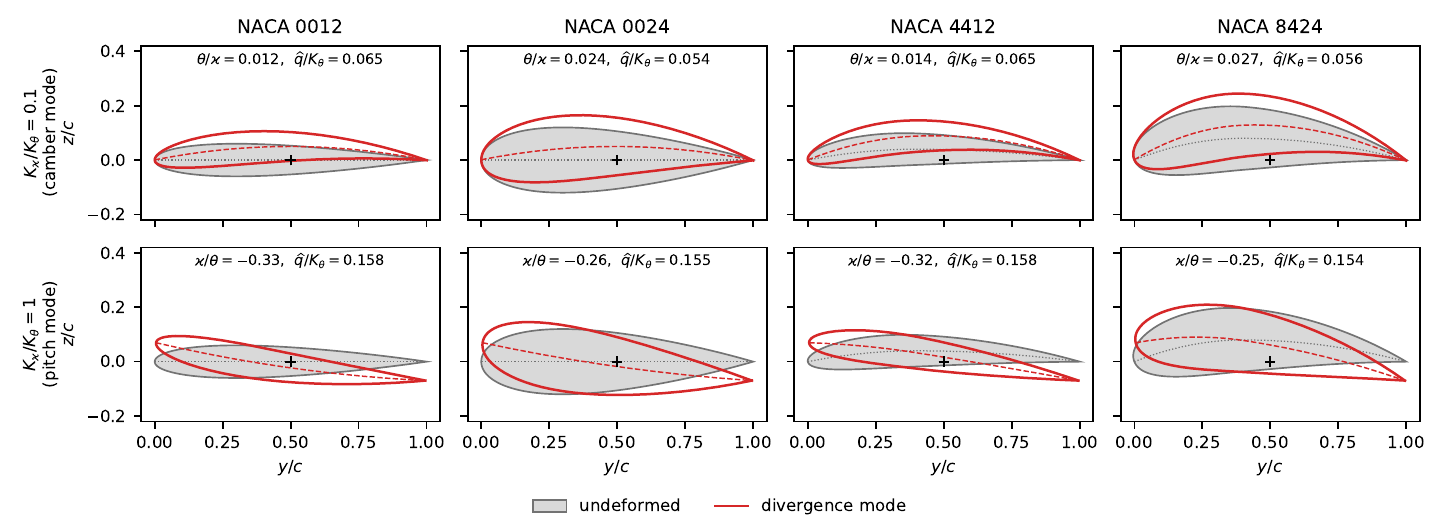}
\caption{Sections with pitch about the midchord (cross) and the camber mode~\eqref{camber:mode}: undeformed section (grey; dotted, its mean line) and divergence mode of the lowest root of~\eqref{camber:divergence} (red; dashed, its mean line), from the present coefficients, for NACA 0012, 0024, 4412 and 8424. Top row: $K_\varkappa/K_\theta=0.1$, shown for $\varkappa=0.1$. Bottom row: $K_\varkappa/K_\theta=1$, shown for $\theta=8^\circ$. Each panel gives the ratio of the two coordinates in the mode and the divergence pressure $\widehat q/K_\theta$. The amplitude of a divergence mode is arbitrary.}
\label{fig:camber-modes}
\end{figure}

Figure~\ref{fig:camber-divergence} shows these two divergence pressures for the twenty sections above. The thickness lowers the divergence pressure of the camber coordinate by $21$--$45\%$; the present value is within $6.4\%$ of that of XFOIL for every section, whereas the zero-thickness value is $20$--$85\%$ above it. For the pitch coordinate the effect is smaller. The thickness also couples the two coordinates, which are uncoupled in the zero-thickness theory, and for stiffness ratios $K_\varkappa/K_\theta$ in an interval containing $\mathsf f_\varkappa/\mathsf m_\theta$ (for NACA 0012, $0.19$--$0.32$) the roots of~\eqref{camber:divergence} are complex: the linear static problem then has no divergence pressure, and the stability is decided by the dynamics, which we do not address. Figure~\ref{fig:camber-modes} shows the divergence modes of four sections for two stiffness ratios outside this interval. The comparison is with inviscid computations; we are not aware of measurements for a section with a camber degree of freedom.

\section{Discussion}\label{sec:discussion}
Existence and uniqueness are known for the full nonlinear problem of subsonic flow past a profile~\cite{Shiffman1952,Bers1954,FinnGilbarg1957}, in a nonconstructive form that does not yield loads for structural coupling, and for the zero-thickness airfoil and Possio equations~\cite{OkadaProssdorf1997,Balakrishnan2003,BalakrishnanShubov,SerryTuffahaGround2017}. Analytical solutions that retain thickness, camber and incidence~\cite{Ayton2017Geometry,BaddooAyton2018} linearize the boundary condition so that these contributions decouple, and the expansion fails near a rounded leading edge. The present system retains the products of surface slopes and disturbance velocities, which couple the thickness and lifting problems, and for the rounded-leading-edge family the velocity remains bounded at the leading edge.

Relative to Mateescu and Abdo~\cite{MateescuAbdo2005}, whose formulation already contains the coupled boundary conditions and the reduction to a finite linear system, the new elements are that the poles of the solution are determined by the operator instead of an assumed polynomial form, that the amplitude equations are polynomial identities on the whole chord instead of collocation conditions, that the amplitude systems are proved to be uniquely solvable, and that repeated poles are included. Relative to geometry-resolved panel and rational solvers and to machine-learning surrogates~\cite{DrelaXfoil,GopalTrefethen2019,SharpeHansman2025NeuralFoil}, the model gives up geometric generality and the exact boundary location in exchange for an explicit dependence of the loads on the geometric coefficients and for a priori solvability and stability statements.

The comparisons of Section~\ref{sec:comparison-pressure-loads} depend on the section, the Mach number and the incidence, and inviscid XFOIL is a geometry-resolved reference, not experimental truth. The pressures and loads reported here use the linearized Bernoulli relation and the pressure jump on the chord. Evaluating the full Bernoulli relation with the same velocities, and integrating the resulting pressures on the surfaces of the section, can potentially improve the accuracy, in particular the suction near a rounded leading edge and the lift slope of thick sections; we leave this to future work.

The scope is steady subsonic potential flow with a collapsed surface boundary; large thickness, high incidence and transonic conditions strain these assumptions, and unsteady phenomena are outside the model. The theorems do not cover sections with two rounded ends. Convergence under fitting requires convergence of the slopes, not only of the profile heights (Corollary~\ref{thm:geometry-stability}), and near-coincident poles and high-degree fits call for conditioning checks. Within this scope, the practical use of the model is repeated, geometry-dependent load evaluation and transparent structural coupling: the loads are quadratic forms in the cosine and sine of the incidence, their slopes follow at every subsonic Mach number from two geometry-only numbers, and the divergence conditions of Section~\ref{sec:aeroelastic-analysis} are available in closed form.

\section{Conclusions}
We revisited the derivation of a thickness-modified airfoil system, a coupled finite Hilbert transform formulation for steady subsonic flow over cambered airfoils of small, nonzero thickness, and treated it analytically. For general geometry the solution was given in weighted $L^2$ spaces in closed form by an integral formula, under conditions on the slope angles that make it unique; for two weighted-polynomial thickness families the solution was obtained as a finite rational expression; and the solution error under geometric approximation was bounded by the slope error. The loads are quadratic forms in the cosine and sine of the incidence, their slopes follow at every subsonic Mach number from two geometry-only numbers, and they give the divergence conditions of an elastic wing and of a section with a camber degree of freedom in closed form.

Compared with published pressure measurements on NACA sections, the accuracy of the model is comparable to that of inviscid XFOIL, and its divergence predictions are closer to XFOIL-based values than those of the zero-thickness theory. Future work may incorporate a finite-span correction, viscosity through a coupled boundary-layer model, and unsteady aerodynamic loading for the study of flutter, where thickness is known to modify Theodorsen-type responses~\cite{MottaGuardoneQuaranta2015,Limacher2025}.

\section*{Acknowledgements}
M.~A.~Serry is funded through his research associate position in the Department of Mechanical and Mechatronics Engineering at the University of Waterloo. The authors thank Dr.~Jun~Liu of the Department of Applied Mathematics, University of Waterloo, for providing access to Claude (Anthropic).

\section*{Declaration of generative AI and AI-assisted technologies}
During the preparation of this work the authors used ChatGPT, including Codex (OpenAI), and Claude (Anthropic) to assist with literature searches, mathematical derivations, the development and debugging of numerical implementations, the extraction and organization of reference data, the preparation of figures, and language editing. The authors defined the research questions, directed each of these tasks, and devoted substantial effort to checking, correcting and refining the results presented here. To the best of their ability, the authors have verified the derivations, data, and references; they take full responsibility for the results, their interpretation, and the content. 


\appendix
\renewcommand{\thesection}{\Alph{section}}
\makeatletter
\newcommand{\prefix@section}{Appendix~}
\renewcommand{\@seccntformat}[1]{\ifcsname prefix@#1\endcsname\csname prefix@#1\endcsname\fi\csname the#1\endcsname.\quad}
\makeatother

\section{Derivation of the trace relations}\label{app:derivation}
The derivation follows the standard route for the airfoil and Possio integral equations: a Fourier transform in the chordwise coordinate reduces the field equation to an ordinary differential equation in the normal coordinate, and its decaying solutions relate the traces of the velocity on the reference plane~\cite{Balakrishnan2003,BalakrishnanShubov,BalakrishnanTuffaha,Tuffaha2009,SerryTuffaha2018,SerryTuffahaGround2017}. Here the transform is applied separately in the upper and lower half-planes, so that the two faces keep their own slopes.

Take the Fourier transform in the entire tangential coordinate $y$, using the convention
\begin{equation}
 \widehat \psi (\xi,z)=\int_{\mathbb R}\psi (y,z)e^{-i\xi y}\,\dd y,
 \qquad
 \psi (y,z)=\frac1{2\pi}\int_{\mathbb R}\widehat \psi (\xi,z)e^{i\xi y}\,\dd\xi.
 \label{air:Fourier-convention}
\end{equation}
Thus $\widehat{\partial_y \psi }=i\xi\widehat \psi $, $\widehat{\partial_{yy}\psi }=-\xi^2\widehat \psi $ and $\widehat{\partial_{yz}\psi }=i\xi\partial_z\widehat \psi $. Applying these identities to the field equation \eqref{main:flow-strip} gives:
\begin{equation}
 D_\alpha\partial_{zz}\widehat\varphi
 -2iC_\alpha\xi\partial_z\widehat\varphi
 -A_\alpha\xi^2\widehat\varphi=0.
 \label{air:fourier-ODE}
\end{equation}

The same reduction can be written directly for the disturbance velocities in either open half-plane. Put $u=\varphi_y$, $v=\varphi_z$. Irrotationality and the field equation give
\begin{equation}
 u_z=v_y,\qquad D_\alpha v_z=2C_\alpha v_y-A_\alpha u_y.
\end{equation}
Consequently, the transformed velocity vector satisfies the first-order system
\begin{equation}
 \partial_z
 \begin{pmatrix}\widehat u\\\widehat v\end{pmatrix}
 =i\xi
 \begin{pmatrix}0&1\\-A_\alpha/D_\alpha&2a\end{pmatrix}
 \begin{pmatrix}\widehat u\\\widehat v\end{pmatrix}.
 \label{air:Fourier-velocity-system}
\end{equation}

\subsection{Transformed potential equation}
For a nonzero frequency, substitute $\widehat\varphi=\widehat\varphi_0(\xi)e^{\varsigma z}$ into~\eqref{air:fourier-ODE}. The characteristic equation and its two roots are
\begin{equation}
 \begin{aligned}
 &D_\alpha \varsigma ^2-2iC_\alpha\xi \varsigma -A_\alpha\xi^2=0,\\
 &\varsigma =\frac{iC_\alpha\xi\pm
 \sqrt{(A_\alpha D_\alpha-C_\alpha^2)\xi^2}}{D_\alpha}
 =ia\xi\pm b|\xi|.
 \end{aligned}
 \label{air:fourier-roots}
\end{equation}
Here $\sqrt{\xi^2}=|\xi|$ for real $\xi$ and $b>0$. In particular, the real parts are $-b|\xi|$ and $+b|\xi|$. Denote the roots selected for the upper and lower domains by
\begin{equation}
 \varsigma _{\rm u}=ia\xi-b|\xi|,\qquad
 \varsigma _{\rm l}=ia\xi+b|\xi|.
 \label{air:upper-lower-roots}
\end{equation}
Before decay is imposed, the general solution in either half-plane is a linear combination of $e^{\varsigma _{\rm u}z}$ and $e^{\varsigma _{\rm l}z}$. Equivalently, the velocity modes of~\eqref{air:Fourier-velocity-system} have the form
\begin{equation}
 \begin{pmatrix}\widehat u\\\widehat v\end{pmatrix}
 =\Pi_{\rm u}(\xi)\begin{pmatrix}i\xi\\\varsigma _{\rm u}\end{pmatrix}e^{\varsigma _{\rm u}z}
 +\Pi_{\rm l}(\xi)\begin{pmatrix}i\xi\\\varsigma _{\rm l}\end{pmatrix}e^{\varsigma _{\rm l}z}.
 \label{air:velocity-general-modes}
\end{equation}
The coefficients in the upper problem are independent of those in the lower problem.

The far-field condition \eqref{air:farfield} requires the disturbance velocity to decay away from the reference plane. For $z>0$, the $\varsigma _{\rm l}$ mode grows as $z\to+\infty$ and is excluded. For $z<0$, the $\varsigma _{\rm u}$ mode grows as $z\to-\infty$ and is excluded. The retained potential modes are therefore
\begin{equation}
 \begin{aligned}
 \widehat\varphi^+(\xi,z)&=\widehat\varphi_0^+(\xi)
 e^{(ia\xi-b|\xi|)z},&&z>0,\\
 \widehat\varphi^-(\xi,z)&=\widehat\varphi_0^-(\xi)
 e^{(ia\xi+b|\xi|)z},&&z<0.
 \end{aligned}
 \label{air:selected-modes}
\end{equation}
The factor $e^{ia\xi z}$ changes only the phase. In both retained modes, the magnitude of the normal factor is $e^{-b|\xi||z|}$. The plus sign before $b|\xi|z$ in the lower domain is thus necessary for decay there. Formally, the inverse transform is
\begin{equation}
 \varphi^\pm(y,z)=\frac1{2\pi}\int_{\mathbb R}
 \widehat\varphi_0^\pm(\xi)e^{i\xi(y+az)}e^{-b|\xi||z|}\,\dd\xi,
 \qquad \pm z>0.
 \label{air:mode-inversion}
\end{equation}

\subsection{Boundary traces and the whole-line Hilbert transform}
Differentiating~\eqref{air:selected-modes}, with $\widehat u=i\xi\widehat\varphi$, and taking the one-sided limits at $z=0$ gives
\begin{equation}
 \begin{aligned}
 \widehat u^\pm&=i\xi\widehat\varphi_0^\pm,\\
 \widehat v^+&=(ia\xi-b|\xi|)\widehat\varphi_0^+,
 &\widehat v^-&=(ia\xi+b|\xi|)\widehat\varphi_0^-.
 \end{aligned}
 \label{air:Fourier-boundary-traces}
\end{equation}
Eliminating the potential amplitudes $\widehat\varphi_0^\pm$, which requires only division by $i\xi$, and using $|\xi|/\xi=\operatorname{sgn}\xi$, yields
\begin{equation}
 \widehat v^+=(a+ib\operatorname{sgn}\xi)\widehat u^+,
 \qquad \widehat v^-=(a-ib\operatorname{sgn}\xi)\widehat u^-.
 \label{air:Fourier-trace-symbols}
\end{equation}
Define the whole-line Hilbert transform by
\begin{equation}
 (\mathcal H \psi )(y)=\frac1\pi\pv\int_{\mathbb R}
 \frac{\psi (\eta)}{y-\eta}\,\dd\eta,\qquad
 \widehat{\mathcal H \psi }(\xi)=-i\operatorname{sgn}\xi\,\widehat \psi (\xi).
 \label{air:whole-line-H}
\end{equation}
Comparing~\eqref{air:Fourier-trace-symbols} with this multiplier and taking inverse transforms gives
\begin{equation}
 v^+=au^+-b\mathcal H u^+,\qquad
 v^-=au^-+b\mathcal H u^-.
 \label{air:trace-pair}
\end{equation}

\subsection{Jumps, averages, and restriction to the finite chord}
With the jumps and averages~\eqref{air:jumps} on the whole line, subtracting the lower relation in~\eqref{air:trace-pair} from the upper relation, and then taking their half-sum, gives
\begin{equation}
 \begin{aligned}
 n&=a(u^+-u^-)-b\mathcal H(u^++u^-)
   =ad-2b\mathcal H\bar u,\\
 \bar v&=a\bar u-\tfrac b2\mathcal H(u^+-u^-)
   =a\bar u-\tfrac b2\mathcal H d.
 \end{aligned}
 \label{air:average-traces}
\end{equation}
The average $\bar u$ is recovered from~\eqref{air:average-traces} directly in the Fourier domain. Subtracting the two symbols in~\eqref{air:Fourier-trace-symbols} gives, for $\xi\ne0$,
\[
 \widehat n-a\widehat d
 =ib\operatorname{sgn}\xi\,(\widehat u^++\widehat u^-)
 =2ib\operatorname{sgn}\xi\,\widehat{\bar u}.
\]
Multiply by $i\operatorname{sgn}\xi$ and use $(\operatorname{sgn}\xi)^2=1$:
\[
 i\operatorname{sgn}\xi\,(\widehat n-a\widehat d)
 =-2b\widehat{\bar u}.
\]
Taking the inverse Fourier transform with the multiplier convention~\eqref{air:whole-line-H} therefore yields
\begin{equation}
 \bar u=\frac1{2b}\mathcal H(n-ad).
 \label{air:average-inversion}
\end{equation}
This calculation uses the selected decaying modes directly, and does not use the identity $\mathcal H^2=-I$ on the whole line. The zero-frequency velocity mode is constant in $z$ and is excluded by the far-field condition, so no additional constant-velocity term appears. Fourier inversion is understood in the same boundary-trace sense as in~\eqref{air:Fourier-boundary-traces}--\eqref{air:trace-pair}.

Extending $d$ and $n$ by zero outside the chord, where they vanish by the assumption stated in Section~\ref{sec:airfoil-derivation}, and restricting to $y\in I$, where the whole-line transform of the zero extension of a chord function $\psi$ is $\HH\psi$, gives
\begin{equation}
 \bar u=\frac1{2b}\HH(n-ad),\qquad
 \bar v=\frac{a}{2b}\HH(n-ad)-\frac b2\HH d,
 \qquad y\in I.
 \label{air:finite-traces}
\end{equation}

\subsection{Tangency on the chord}
The field has now been eliminated in favor of the jumps $d$ and $n$. We impose the collapsed tangency conditions~\eqref{air:tangency} only for $y\in I$, using
\begin{equation}
 u^\pm=\bar u\pm\tfrac12d,\qquad
 v^\pm=\bar v\pm\tfrac12n,\qquad k^\pm=k_c\pm g_t.
 \label{air:face-reconstruction}
\end{equation}
The two face equations become
\begin{equation}
 \begin{aligned}
 \bar v+\tfrac12n-(k_c+g_t)(\bar u+\tfrac12d)
 &=(k_c+g_t)U\cos\alpha-U\sin\alpha,\\
 \bar v-\tfrac12n-(k_c-g_t)(\bar u-\tfrac12d)
 &=(k_c-g_t)U\cos\alpha-U\sin\alpha.
 \end{aligned}
 \label{air:expanded-face-BCs}
\end{equation}
Taking their half-sum and their upper-minus-lower difference, respectively, gives
\begin{equation}
 \begin{aligned}
 \bar v-k_c\bar u-\tfrac12g_t d
 &=k_cU\cos\alpha-U\sin\alpha,\\
 n-k_c d-2g_t\bar u&=2g_tU\cos\alpha.
 \end{aligned}
 \label{air:tangency-averages}
\end{equation}

These relations coincide with the incompressible tangency coupling in Eqs.~(5a,b) of~\cite{MateescuAbdo2005}, after a change of variables and normalization.

Substituting~\eqref{air:finite-traces} gives
\begin{equation}
 \begin{aligned}
 \frac{a-k_c}{2b}\HH(n-ad)-\frac b2\HH d-\frac{g_t}{2}d
 &=k_cU\cos\alpha-U\sin\alpha,\\
 n-k_c d-\frac{g_t}{b}\HH(n-ad)&=2g_tU\cos\alpha.
 \end{aligned}
 \label{air:BC-Hilbert-substitution}
\end{equation}
Multiplying the first equation by $-2$ gives~\eqref{main:total-integral} with the forcings~\eqref{air:total-forcing}.

\section{Loads as functions of the incidence and the Mach number}\label{app:loads-derivation}
The velocities~\eqref{comparison:face-pressure-mapping} follow from~\eqref{air:finite-traces} and~\eqref{air:face-reconstruction}; they are evaluated on the collapsed chord and are not re-evaluated on the finite-height surfaces. The pressure relation~\eqref{comparison:linear-bernoulli} retains terms linear in the disturbance velocity and applies to both the incompressible and the subsonic compressible calculations; compressibility enters the computed velocities through $M$ and the coefficients $a$, $b$.

\subsection{Superposition and quadratic forms}
The incidence enters the solution in two ways: the freestream components $U\cos\alpha$ and $U\sin\alpha$ appear linearly in the right-hand sides of the amplitude systems, while the pole equations and the amplitude matrices contain only the geometry and the flow coefficients $a$ and $b$.

\begin{remark}[Linear dependence on the freestream components]\label{prop:symbolic-linearity}
Fix the geometry, that is, $k_c$ and $Q$ or $R$, and the flow coefficients $a$ and $b$. Under the assumptions of Theorem~\ref{thm:flat-family} or Theorem~\ref{thm:rounded-family}, let $\bm x^{(1)}$, $d^{(1)}$, $n^{(1)}$ denote the amplitudes and jumps obtained when $(U\cos\alpha,U\sin\alpha)$ is replaced by $(1,0)$ in the forcings~\eqref{air:total-forcing} and in the background term~\eqref{total:velocity-background}, with $a$ and $b$ unchanged, and $\bm x^{(2)}$, $d^{(2)}$, $n^{(2)}$ those obtained with $(0,1)$. Then for every $U$ and $\alpha$ with these values of $a$ and $b$
\begin{equation}
 \bm x=\chi_1\bm x^{(1)}+\chi_2\bm x^{(2)},\qquad
 d=\chi_1d^{(1)}+\chi_2d^{(2)},\qquad n=\chi_1n^{(1)}+\chi_2n^{(2)},\qquad
 \bm\chi=U\begin{pmatrix}\cos\alpha\\ \sin\alpha\end{pmatrix},
 \label{sym:linear-superposition}
\end{equation}
and the face velocities~\eqref{comparison:face-pressure-mapping} are the same combinations $u^\pm=\chi_1u^{\pm(1)}+\chi_2u^{\pm(2)}$, $v^\pm=\chi_1v^{\pm(1)}+\chi_2v^{\pm(2)}$ of the two unit fields.
\end{remark}
Two solutions with the same matrix therefore give the field for every freestream speed and, at $M=0$, for every incidence.

The pressures are linear in the velocities and in the freestream components, so the superposition carries over to the loads. With the unit data $\bm\chi^{(1)}=(1,0)$ and $\bm\chi^{(2)}=(0,1)$, define the symmetric pressure kernels
\begin{equation}
 p^\pm_{kl}=-\bigl(\chi^{(k)}_1u^{\pm(l)}+\chi^{(k)}_2v^{\pm(l)}
 +\chi^{(l)}_1u^{\pm(k)}+\chi^{(l)}_2v^{\pm(k)}\bigr),\qquad k,l\in\{1,2\}.
 \label{sym:pressure-kernels}
\end{equation}
By Remark~\ref{prop:symbolic-linearity}, the pressure coefficients normalized as in~\eqref{comparison:pressure-normalization} are $C_p^\pm=U^{-2}\sum_{k,l}\chi_k\chi_l\,p^\pm_{kl}$, at every subsonic Mach number. Inserting the kernels in~\eqref{comparison:physical-pressure-force} gives~\eqref{sym:quadratic-form} with the matrices~\eqref{sym:load-matrices}, since $p^-_{kl}-p^+_{kl}$ is the entry $(k,l)$ of $\bm\Delta$. For zero thickness and zero camber the unit solutions are elementary, $d^{(1)}=0$, $d^{(2)}=2br/(a^2+b^2)$ and $n^{(k)}=0$, which gives~\eqref{sym:thin-matrices}; since $(a^2+b^2)/b=A_\alpha/\beta$, these are $C_N=\pi\beta\sin2\alpha/(1-M^2\cos^2\alpha)$ and $C_m=(e-\tfrac14)C_N$. For the order-30 fit of NACA 0012, $e=\tfrac12$ and $M=0$, the solution gives $(\bm Q_N)_{12}=3.1416$ and $(\bm Q_m)_{12}=0.7947$, against $\pi$ and $\pi/4=0.7854$ for zero thickness: the lift slope remains $2\pi$ for this section, and the thickness moves the aerodynamic center from $0.25c$ to $0.247c$.

\subsection{Slopes with respect to the incidence}
Differentiating~\eqref{sym:quadratic-form}, and recalling that $a$ and $b$ are functions of $M$ and $\alpha$, gives, with $\bm q=(\cos\alpha,\sin\alpha)^T$,
\begin{equation}
 \frac{\partial C_X}{\partial\alpha}=2(\bm Q_X)_{12}\cos2\alpha+\bigl[(\bm Q_X)_{22}-(\bm Q_X)_{11}\bigr]\sin2\alpha
 +\bm q^T\Bigl(\frac{\partial a}{\partial\alpha}\,\partial_a\bm Q_X+\frac{\partial b}{\partial\alpha}\,\partial_b\bm Q_X\Bigr)\bm q,
 \label{sym:slope}
\end{equation}
where, by~\eqref{air:coefficients},
\begin{equation}
 \frac{\partial a}{\partial\alpha}=\frac{M^2\bigl(D_\alpha\cos2\alpha+2M^2\sin^2\alpha\cos^2\alpha\bigr)}{D_\alpha^2},\qquad
 \frac{\partial b}{\partial\alpha}=\frac{2\beta M^2\sin\alpha\cos\alpha}{D_\alpha^2}.
\end{equation}
The first two terms are the slope at frozen flow coefficients. The last term accounts for the change of the operator with the incidence; it vanishes at $M=0$. At zero incidence, $a=0$, $b=\beta$, $\partial a/\partial\alpha=M^2$ and $\partial b/\partial\alpha=0$, which gives~\eqref{sym:slope-zero}. The second term of~\eqref{sym:slope-zero} is of relative size $M^2$: for the moment of NACA 0012 about mid-chord it is $2.2\%$ of the slope at $M=0.5$ and $4.9\%$ at $M=0.7$.

\subsection{Dependence on the Mach number at small incidence}
The dependence on the Mach number can be reduced to the incompressible problem. Dividing~\eqref{stab:system} by $b$ gives
\begin{equation}
 \HH d+\widetilde k\,\HH\ell+\widetilde g\,d=2U\bigl(\widetilde\chi_2-\widetilde k\,\widetilde\chi_1\bigr),\qquad
 \ell-\widetilde k\,d-\widetilde g\,\HH\ell=2U\widetilde g\,\widetilde\chi_1,\qquad
 \widetilde k=\frac{k_c-a}{b},\quad\widetilde g=\frac{g_t}{b},
 \label{sym:similarity}
\end{equation}
with $\widetilde\chi_1=\cos\alpha$ and $\widetilde\chi_2=(\sin\alpha-a\cos\alpha)/b$. This is the system at $M=0$ for the slopes $(\widetilde k,\widetilde g)$ and the freestream components $U(\widetilde\chi_1,\widetilde\chi_2)$. The compressible solution $(d,\ell)$ for a given section is therefore the incompressible solution for the section with slopes $(k_c-a)/b$ and $g_t/b$, which extends the classical similarity rule of linearized compressible flow~\cite{AshleyLandahl1965} to the coupled system; $n=b\ell+ad$ then gives the second jump. At small incidence, $a\simeq M^2\alpha$ and $b\simeq\beta$: the section is stretched by $1/\beta$ normal to the chord.

Thickness and camber enter the slope $\beta\,\partial C_X/\partial\alpha$ at second order in the surface slopes, and by~\eqref{sym:similarity} these are divided by $\beta$. To leading order the departure of this quantity from its incompressible value is therefore proportional to $1/\beta^2-1=M^2/\beta^2$, which gives the rule~\eqref{sym:pg2}. The number $\kappa_X$ is the coefficient of $M^2$ in the expansion of $\beta\,\partial C_X/\partial\alpha$ obtained from~\eqref{sym:slope-zero}. For the moment about mid-chord, $\kappa_m=-0.0017$ for NACA 0012 and $-0.027$ for NACA 8424. With $\kappa_X=0$ the rule departs from the slope of the analytical solution by up to $0.9\%$ at $M=0.5$ and $2.4\%$ at $M=0.7$ for five of the sections of Section~\ref{sec:section-dependence}; with $\kappa_X$ retained the departure is at most $0.03\%$ and $0.2\%$.

\section{Dependence of the poles on the flow coefficients}\label{sec:separation}
The pole equations depend on the Mach number and on the incidence only through the complex number $s$ of~\eqref{total:parameters}, and this dependence can be separated from that on the geometry. Since $b+ia=is$, the functions in~\eqref{total:flat-poles} and~\eqref{total:rounded-poles} are $\Psi_{\rm f}=\Psi^0_{\rm f}-is$ and $\Psi_{\rm r}=-W\,(\Psi^0_{\rm r}-is)$, with
\begin{equation}
 \Psi^0_{\rm f}(\zeta)=Q(\zeta)W(\zeta)+ik_c(\zeta),\qquad
 \Psi^0_{\rm r}(\zeta)=ik_c(\zeta)-\frac{R(\zeta)}{W(\zeta)},
 \label{sep:level-functions}
\end{equation}
so that the pole equations read
\begin{equation}
 \Psi^0_{\rm f}(\zeta_j)=is\quad\text{(flat endpoints)},\qquad\Psi^0_{\rm r}(\zeta_j)=is\quad\text{(rounded leading edge)}.
 \label{sep:level-sets}
\end{equation}
The left-hand sides contain the geometry only and the right-hand side the flow coefficients only. The poles are therefore the points at which a fixed function of the section takes the value $is$; when the Mach number or the incidence changes, they move along the level sets of that function. At a simple pole, differentiation of~\eqref{sep:level-sets} gives
\begin{equation}
 \frac{\dd\zeta_j}{\dd s}=\frac{i}{(\Psi^0)'(\zeta_j)},\qquad
 \frac{\partial\zeta_j}{\partial a}=\frac{\dd\zeta_j}{\dd s},\quad
 \frac{\partial\zeta_j}{\partial b}=-i\,\frac{\dd\zeta_j}{\dd s},
 \label{sep:pole-derivative}
\end{equation}
with $\Psi^0=\Psi^0_{\rm f}$ or $\Psi^0_{\rm r}$. The derivatives of the amplitudes then follow by differentiating the linear systems~\eqref{total:flat-real-system} and~\eqref{total:rounded-real-system}, whose entries depend on $a$ and $b$ explicitly and through the poles, so that no new root finding is needed. For cambered sections of both families at Mach numbers up to $0.7$,~\eqref{sep:pole-derivative} agrees with centered differences of the poles within $10^{-8}$.

\section{Wing model and divergence}\label{app:wing-derivation}
Let $x\in[0,S]$ denote the spanwise position along the elastic axis of a uniform cantilever of chord $c$. The sectional axes $(y,z)$ are those of Section~\ref{sec:geometry} for the undeformed wing: $y$ lies along the secant of the section, positive toward the trailing edge, and $z$ is normal to it. The elastic axis is at the chord fraction $e$. The unknowns are the displacements $h_y(x)$ and $h_z(x)$ of the elastic axis in the $y$ and $z$ directions (chordwise and flapwise bending) and the nose-up twist $\theta(x)$. The linear equations of static equilibrium are~\cite{HodgesDowell1974,HodgesPierce}
\begin{equation}
 EI_y\,h_{y,xxxx}=l_y,\qquad
 EI_z\,h_{z,xxxx}=l_z,\qquad
 -GJ\,\theta_{xx}=m_a,
 \label{beam:dynamic-goland}
\end{equation}
with clamped-root and free-tip conditions for both bending displacements and for the twist. Here $EI_y$ and $EI_z$ are the rigidities for bending displacements in the $y$ and $z$ directions and $GJ$ is the torsional rigidity. The loads per unit span are $l_y$ and $l_z$ in the fixed $y$ and $z$ directions and the nose-up moment $m_a$ about the elastic axis. A section at span position $x$ is at the incidence $\alpha=\alpha_r+\theta$, where $\alpha_r$ is the reference incidence of the undeformed wing. For a straight wing linearized about its undeformed state, the three equations are coupled only through the loads, and the loads depend on the deformation only through the twist.

By~\eqref{sym:quadratic-form} the section-frame coefficients are $C_X(M,\alpha_r+\theta)$ for $X\in\{N,m\}$, with the $2\times2$ matrices $\bm Q_X$ of the section at the flow coefficients $a$, $b$ of the Mach number and of the local incidence. The normal force acts along the normal of the twisted section, so that, with $q_\infty$ from~\eqref{comparison:pressure-normalization},
\begin{equation}
 l_y=q_\infty c\,C_N\sin\theta,\qquad
 l_z=q_\infty c\,C_N\cos\theta,\qquad
 m_a=q_\infty c^2\,C_m.
 \label{beam:qs-loads}
\end{equation}
At $M=0$ these are explicit functions of $\theta$ with coefficients that depend on the geometry alone, and no aerodynamic problem is solved during the structural analysis. For $M>0$ the matrices depend on $\theta$ as well, through $a$ and $b$. The linearized analysis needs only the coefficients and their slopes at the reference incidence, which are those of the analytical solution at $(M,\alpha_r)$. Linearizing about the undeformed wing, with the steady coefficients $C_{X0}=C_X(M,\alpha_r)$ and the slopes $C_{X\alpha}=\partial C_X/\partial\alpha$ at $(M,\alpha_r)$ from~\eqref{sym:slope},
\begin{equation}
 \frac{l_y}{q_\infty c}=C_{N0}\,\theta,\qquad
 \frac{l_z}{q_\infty c}=C_{N0}+C_{N\alpha}\,\theta,\qquad
 \frac{m_a}{q_\infty c^2}=C_{m0}+C_{m\alpha}\,\theta.
 \label{beam:dynamic-aero-matrices}
\end{equation}

The three equations~\eqref{beam:dynamic-goland} decouple in the following order. The twist satisfies $-GJ\theta_{xx}=q_\infty c^2(C_{m0}+C_{m\alpha}\theta)$, and the two bending displacements then follow from $\theta$ by quadrature. With $\kappa^2=q_\infty c^2C_{m\alpha}/GJ$ and $C_{m\alpha}>0$, the twist of the cantilever is
\begin{equation}
 \theta_0(x)=\frac{C_{m0}}{C_{m\alpha}}\left[\frac{\cos\kappa(S-x)}{\cos\kappa S}-1\right],
 \label{beam:cambered-equilibrium-solution}
\end{equation}
which is nonzero for a cambered section and becomes unbounded as $\kappa S\to\pi/2$. The homogeneous problem $\theta_{xx}+\kappa^2\theta=0$, $\theta(0)=\theta_x(S)=0$, has a nontrivial solution at the same value, which gives~\eqref{beam:divergence-characteristic}. The moment slope is given by~\eqref{sym:slope} in terms of the matrix $\bm Q_m$; at zero reference incidence it is, by~\eqref{sym:slope-zero}, $C_{m\alpha}=2(\bm Q_m)_{12}+M^2\partial_a(\bm Q_m)_{11}$ at $(a,b)=(0,\beta)$. The bending rigidities and the mass distribution do not enter, and the camber enters through the dependence of $\bm Q_m$ on the geometry, not through the size of the steady moment $C_{m0}$.

When the Mach number is not prescribed but is that of the divergence speed itself, $M=U_D/a_\infty$, the divergence speed is the solution of the scalar equation
\begin{equation}
 U_D=\frac{\pi}{2S}\sqrt{\frac{2\,GJ}{\rho_\infty c^2\,C_{m\alpha}(U_D/a_\infty)}}.
 \label{beam:divergence-fixed-point}
\end{equation}
Let $U_0$ and $M_0=U_0/a_\infty$ denote the divergence speed and its Mach number evaluated with the incompressible slope. With~\eqref{sym:pg2}, equation~\eqref{beam:divergence-fixed-point} reads $U_D^2\,(1+\kappa_mM^2/\beta^2)=U_0^2\,\beta$. For $\kappa_m=0$ this is a quadratic equation for $U_D^2$,
\begin{equation}
 U_D^4+M_0^2U_0^2\,U_D^2-U_0^4=0,
 \label{beam:divergence-quadratic}
\end{equation} with the positive root $U_P$ of~\eqref{beam:divergence-compressible}, and the first-order correction in $\kappa_m$ gives~\eqref{beam:divergence-compressible}. For the sections of Section~\ref{sec:section-dependence} the explicit speed agrees with the solution of~\eqref{beam:divergence-fixed-point} within $10^{-4}$~m/s.

\section{Pitch--camber coefficients}\label{app:camber-coefficients}
For a section with thickness, the four coefficients follow from the analytical solutions at $M=0$. With the pressure jump~\eqref{comparison:linear-bernoulli} and the unit solutions of Remark~\ref{prop:symbolic-linearity}, the derivative of $\Delta C_p$ with respect to the incidence at zero incidence is $2(d^{(2)}+n^{(1)})$, so that
\begin{equation}
 \mathsf m_\theta=8\int_0^1\bigl(\tfrac12-t\bigr)\bigl(d^{(2)}+n^{(1)}\bigr)\,\dd t,\qquad
 \mathsf f_\theta=4\int_0^1\bigl(\tfrac13-\eta^2\bigr)\bigl(d^{(2)}+n^{(1)}\bigr)\,\dd t,
 \label{camber:incidence-coefficients}
\end{equation}
which are twice the off-diagonal entries of the matrices~\eqref{sym:load-matrices} with the corresponding weights; no differentiation is needed. The camber mode~\eqref{camber:mode} enters the system through the slope, $k_c\mapsto k_c-2\varkappa\eta$, with both faces displaced by $h_e$ and the thickness distribution unchanged, and
\begin{equation}
 \mathsf m_\varkappa=8\int_0^1\bigl(\tfrac12-t\bigr)\,\partial_\varkappa d^{(1)}\,\dd t,\qquad
 \mathsf f_\varkappa=4\int_0^1\bigl(\tfrac13-\eta^2\bigr)\,\partial_\varkappa d^{(1)}\,\dd t,
 \label{camber:camber-coefficients}
\end{equation}
where $\partial_\varkappa d^{(1)}$ is the derivative of the first unit solution with respect to the camber amplitude at $\varkappa=0$. Because the camber changes the geometry, and hence the poles and the amplitude matrix, this derivative is evaluated as a centered difference of the analytical solutions at $\varkappa=\pm0.003$.

With the thickness of NACA 0012 scaled to $10^{-5}c$, the present loads return $6.2832$, $0.0000$, $-2.0943$ and $1.0473$, the zero-thickness values to four digits. We evaluated the coefficients for the twenty sections of Section~\ref{sec:section-dependence}, with the order-30 fits. As a numerical reference, the same coefficients were computed with inviscid XFOIL on the contours of the original section formulas, with both faces displaced by~\eqref{camber:mode}, from the difference of the surface pressures of the two faces at the same chordwise station; the differences were taken over $\pm0.5^\circ$ in $\theta$ and $\pm0.005$ in $\varkappa$, and doubling these steps changes the coefficients by less than $0.003$. For the cambered sections the loads at zero incidence deform the section below the divergence pressure, and the divergence pressures are those at which this linear static response becomes unbounded. Table~\ref{tab:camber-divergence} lists the coefficients for three symmetric sections.

\begin{table}[!htbp]\centering\small
\caption{Pitch--camber coefficients of~\eqref{camber:matrix} for symmetric NACA four-digit sections at $M=0$, and the divergence pressures of the camber and of the pitch coordinate alone relative to their zero-thickness values, $\pi/(3\mathsf f_\varkappa)$ and $2\pi/\mathsf m_\theta$. XFOIL is inviscid, on the contour of the original section formula.}
\label{tab:camber-divergence}
\begin{tabular}{llrrrrrr}
\toprule
$t/c$ & Loads & $\mathsf m_\theta$ & $\mathsf m_\varkappa$ & $\mathsf f_\theta$ & $\mathsf f_\varkappa$ & camber & pitch\\
\midrule
$0$ & Zero thickness~\cite{MuruaPalaciosPeiro2010} & $6.283$ & $0$ & $-2.094$ & $1.047$ & $1$ & $1$\\
\midrule
$0.06$ & Present & $6.304$ & $0.035$ & $-1.799$ & $1.333$ & $0.79$ & $1.00$\\
 & XFOIL & $6.497$ & $0.214$ & $-1.879$ & $1.258$ & $0.83$ & $0.97$\\
\midrule
$0.12$ & Present & $6.357$ & $0.111$ & $-1.554$ & $1.560$ & $0.67$ & $0.99$\\
 & XFOIL & $6.724$ & $0.428$ & $-1.683$ & $1.462$ & $0.72$ & $0.93$\\
\midrule
$0.18$ & Present & $6.432$ & $0.204$ & $-1.346$ & $1.745$ & $0.60$ & $0.98$\\
 & XFOIL & $6.956$ & $0.636$ & $-1.498$ & $1.663$ & $0.63$ & $0.90$\\
\bottomrule
\end{tabular}
\end{table}

\section{Transform identities and coefficient matching}\label{sec:transform-identities}
\subsection{Transform identities and polynomial data}
For polynomial camber and deformation, define the total mean-slope degree by
\begin{equation}
 {\mathsf p}=\begin{cases}\deg k_c,&k_c\not\equiv0,\\0,&k_c\equiv0.\end{cases}
 \label{total:deformation-degree}
\end{equation}
For a complex pole parameter $\zeta \notin[-1,1]$, define
\begin{equation}
 W(\zeta )=\sqrt{\zeta -1}\sqrt{\zeta +1},\quad W(\zeta )\sim \zeta ,\qquad
 K(\zeta )=\frac{\zeta -1}{W(\zeta )}.
 \label{total:branch-functions}
\end{equation}
The exact transform identities are
\begin{equation}
 \HH\frac{r}{\zeta -\eta}=\frac{K(\zeta )}{\zeta -\eta},\qquad
 \HH\frac{w}{\zeta -\eta}=-1+\frac{W(\zeta )}{\zeta -\eta},\qquad
 \HH r=1.
 \label{total:transform-identities}
\end{equation}
They follow by partial fractions from the endpoint-weight Cauchy integrals; see Appendix \ref{app:weightedCauchyIntegrals}. They apply to traces on the collapsed chord and do not require a smooth extension joining the two half-plane fields.

The solution constructions take $U,M,\alpha$ and the polynomial coefficients of $k_c$ as inputs. The flow coefficients $a,b$ are defined once in~\eqref{air:coefficients}; the forcings $F,G$ are defined in~\eqref{air:total-forcing}; and $\lambda$ is defined in~\eqref{total:parameters}. For rigid pitch~\eqref{air:rigid-perturbation}, $h_e=0$ and $k_c=(2/c)\partial_\eta f_c$; pitch changes $\alpha$, and therefore $a,b$, rather than adding a constant to the local camber slope.

The dependence on incidence is carried by $a,b$, the pole equations and the right-hand sides of the amplitude systems below. Arguments are suppressed in the solution notation. The prescribed camber and deformation enter through $k_c$ and $\lambda$ of~\eqref{total:parameters}.

\paragraph{Background term}
The thickness forcing $G$ is accommodated by a constant background term in the normal-velocity jump. Since $\HH r=1$ by~\eqref{total:transform-identities},
\begin{equation}
 N_0:=-2b\,U\cos\alpha,\qquad\HH(rN_0)=-2b\,U\cos\alpha.
 \label{total:velocity-background}
\end{equation}
The flat-endpoint construction uses $N_0$; the rounded-leading-edge construction needs no background term.

\begin{remark}[Camber]\label{rem:camber-transfer}
The camber $f_c$ and the deformation $h_e$ enter the retained equations only through their combined slope $k_c$, since the face slopes are $k_c\pm g_t$ and both functions vanish at the endpoints, so that the secant and the incidence are unchanged. A cambered section with camber $f_c$ and thickness $f_t$ therefore gives exactly the same system, and hence the same poles, amplitude systems and solution, as the symmetric section $\pm f_t$ with the deformation $f_c+h_e$. Theorems~\ref{thm:flat-family} and~\ref{thm:rounded-family} are accordingly stated for an arbitrary polynomial $k_c$. This is an exact reparameterization within the model, not a linearization in camber: adding the thin-airfoil camber load to a symmetric thick solution does not solve the retained equations, because $k_c$ also enters the operator through $\lambda$ and so changes the poles.
\end{remark}

\subsection{Polynomial coefficient matching and repeated poles}\label{sec:coefficient-matching}
The amplitude systems below are obtained by matching polynomial coefficients. Choose a real polynomial basis
\begin{equation}
 \{\mathsf b_k\}_{k\ge0},\qquad \mathsf b_0=1,\qquad \deg\mathsf b_k=k.
 \label{total:poly-basis}
\end{equation}
For a polynomial $\Pi$ of degree at most ${\mathsf q}$, its coefficient vector in this basis, padded with zeros, is
\begin{equation}
 [\Pi]_{\mathsf q}=(\mathsf c_0,\mathsf c_1,\ldots,\mathsf c_{\mathsf q})^T\in\mathbb C^{{\mathsf q}+1},\qquad
 \Pi=\sum_{k=0}^{\mathsf q} \mathsf c_k\mathsf b_k .
 \label{total:coefficient-vector}
\end{equation}
The first unit vector is
\begin{equation}
 \bm e_0^{({\mathsf q})}=(1,0,\ldots,0)^T\in\mathbb R^{{\mathsf q}+1},\qquad
 [1]_{\mathsf q}=\bm e_0^{({\mathsf q})}.
 \label{total:unit-vector}
\end{equation}
Two polynomials of degree at most ${\mathsf q}$ are equal if and only if their coefficient vectors~\eqref{total:coefficient-vector} are equal; this is the sense in which the amplitude systems enforce the integral equations. Monomials give a direct construction; the supplied implementation uses Chebyshev polynomials, including for polynomial deformation. All vectors in a given system must use the same basis.

For any polynomial $\Pi$, define its polynomial divided difference by
\begin{equation}
 \mathcal D_\zeta \Pi(\eta)=\frac{\Pi(\eta)-\Pi(\zeta )}{\eta-\zeta }.
 \label{total:divided-difference}
\end{equation}
In particular, $\mathcal D_\zeta k_c=0$ when $k_c$ is constant. This operation is polynomial division, not differentiation of the unknown jumps.

For pole orders $\nu\ge1$, introduce the normalized derivative in the complex pole parameter,
\begin{equation}
 \mathcal L_{\zeta ,\nu}=\frac{(-1)^{\nu-1}}{(\nu-1)!}\partial_\zeta ^{\nu-1},
 \qquad
 \mathcal L_{\zeta ,\nu}\frac1{\zeta -\eta}=\frac1{(\zeta -\eta)^\nu}.
 \label{total:confluent-operator}
\end{equation}
Define the explicit transform kernels
\begin{equation}
 \begin{aligned}
 \Phi_\nu(\zeta ,\eta)&=\mathcal L_{\zeta ,\nu}\frac{K(\zeta )}{\zeta -\eta}
 =\sum_{k=0}^{\nu-1}\frac{(-1)^k K^{(k)}(\zeta )}{k!\,(\zeta -\eta)^{\nu-k}},\\
 \Omega_\nu(\zeta ,\eta)&=\mathcal L_{\zeta ,\nu}\left[-1+\frac{W(\zeta )}{\zeta -\eta}\right]
 =-\delta_{\nu1}+\sum_{k=0}^{\nu-1}\frac{(-1)^k W^{(k)}(\zeta )}{k!\,(\zeta -\eta)^{\nu-k}}.
 \end{aligned}
 \label{total:confluent-kernels}
\end{equation}
Here $\delta_{\nu1}=1$ for $\nu=1$ and is zero otherwise. Differentiating~\eqref{total:transform-identities} at off-chord parameters gives
\begin{equation}
 \HH\!\left(\frac{r}{(\zeta -\eta)^\nu}\right)=\Phi_\nu(\zeta ,\eta),\qquad
 \HH\!\left(\frac{w}{(\zeta -\eta)^\nu}\right)=\Omega_\nu(\zeta ,\eta).
 \label{total:confluent-transforms}
\end{equation}
Thus a pole $\zeta _j$ of multiplicity $\mathsf r_j$ is represented by its full principal part, with orders $\nu=1,\ldots,\mathsf r_j$. All differentiations are with respect to the pole parameter; no derivatives of the unknown jumps are introduced.

The flat-endpoint construction below uses every admissible pole with its multiplicity. Its pole count and matrix invertibility are proved in Appendix~\ref{app:proof-flat-family}. For rounded-leading-edge profiles with $P(-1)>0$ and $P(1)=0$, the complete confluent system is also uniquely solvable after one redundant endpoint equation is removed. In both families, compatibility with the prescribed square-root branch is essential, and no completeness statement in larger endpoint spaces is made.

\section{Worked examples}\label{app:examples}
The two examples below illustrate Theorems~\ref{thm:flat-family} and~\ref{thm:rounded-family} for low-degree geometries. Both are symmetric and admit explicit poles and amplitudes.

\begin{example}[Flat endpoints: two simple poles]\label{ex:flat-poles-amplitudes}
Take $M=0$, $k_c=0$ and $P=\varepsilon/3$, with $\varepsilon>0$. Then
\[
 f^\pm=\pm\frac{c\varepsilon}{6}(1-\eta^2)^{3/2},\qquad
 Q(\eta)=-\varepsilon\eta,\qquad a=\lambda=0,\quad b=1,\quad s=-i,\quad {\mathsf q}=1.
\]
The incidence $\alpha$ may be nonzero. The candidate equation and the required unsquared equation are
\[
 \varepsilon^2\zeta ^2(\zeta ^2-1)=1,\qquad -\varepsilon \zeta W(\zeta )=1.
\]
Define
\[
 \Upsilon=\left(\frac{\sqrt{1+4/\varepsilon^2}-1}{2}\right)^{1/2}.
\]
The candidate roots $\zeta =\pm\sqrt{1+\Upsilon^2}$ fail the unsquared equation, since $\zeta W(\zeta )>0$ there. The retained roots are $\zeta _1=i\Upsilon$ and $\zeta _2=-i\Upsilon$, where $W(\pm i\Upsilon)=\pm i\sqrt{1+\Upsilon^2}$ and $\varepsilon\Upsilon\sqrt{1+\Upsilon^2}=1$. Both are simple zeros of $\Psi_{\rm f}$ and both are retained. Put $K_0=K(i\Upsilon)=(\Upsilon+i)/\sqrt{1+\Upsilon^2}$, so that $|K_0|=1$ and $K(-i\Upsilon)=\overline K_0$. Direct polynomial division gives
\begin{gather*}
 E_{1,1}=\varepsilon(1-\eta-i\Upsilon),\qquad
 E_{1,2}=\varepsilon(1-\eta+i\Upsilon),\\
 E_{2,1}=i\varepsilon K_0(1+\eta+i\Upsilon),\qquad
 E_{2,2}=i\varepsilon\overline K_0(1+\eta-i\Upsilon).
\end{gather*}
Matching the constant and linear coefficients in~\eqref{total:flat-amplitudes} yields
\begin{gather*}
 \Re(A_1+A_2)=0,\qquad \Im(A_1-A_2)=\frac{U\sin\alpha}{\varepsilon\Upsilon},\\
 \Im(K_0A_1+\overline K_0A_2)=0,\qquad \Re(K_0A_1-\overline K_0A_2)=-\frac{U\cos\alpha}{\varepsilon\Upsilon}.
\end{gather*}
Thus the amplitudes are explicitly
\begin{equation}
 A_1=\frac{U}{2\varepsilon\Upsilon}
 \bigl(i\sin\alpha-\overline K_0\cos\alpha\bigr),\qquad
 A_2=\frac{U}{2\varepsilon\Upsilon}
 \bigl(-i\sin\alpha+K_0\cos\alpha\bigr).
 \label{example:flat-amplitudes}
\end{equation}
Substitution into~\eqref{total:flat-ansatz}, with $N_0=-2U\cos\alpha$, supplies the jumps; their transforms follow from~\eqref{app:flat-solution-transforms}.
\end{example}

\begin{example}[Rounded leading edge: a cubic pole equation]\label{ex:rounded-poles-amplitudes}
Take $M=0$, $k_c=0$ and $P(\eta)=\varepsilon(1-\eta)$, so that
\[
 f^\pm=\pm\frac{c\varepsilon}{2}(1-\eta)w(\eta),\qquad
 R(\eta)=\varepsilon(2\eta^2-\eta-1),\qquad
 a=\lambda=0,\quad b=1,\quad s=-i.
\]
This profile has a rounded leading edge and satisfies the sharp-trailing-edge restriction $P(1)=0$ used in the comparisons of Section~\ref{sec:comparison-pressure-loads}. Here ${\mathsf q}_1=1$, ${\mathsf q}_2=2$. The candidate equation factors as
\[
 (\zeta -1)\{\varepsilon^2(\zeta -1)(2\zeta +1)^2-(\zeta +1)\}=0.
\]
Discard the endpoint $\zeta =1$, solve the remaining cubic, and test each root in $R(\zeta )+W(\zeta )=0$. For $\varepsilon=0.1$, the retained simple poles and branch values are
\[
 \begin{array}{c|rr}
 j&\zeta _j&W_j:=W(\zeta _j)\\\hline
 1&-1.022042564&-0.211118457\\
 2&-4.485634775&-4.372747344
 \end{array}
\]
The third cubic root, $\zeta \simeq5.507677339$, fails the unsquared test. Both retained poles are real and less than $-1$; each is included once on the selected branch.

To compute the amplitudes, set $A_j=A^{\rm r}_j+iA^{\rm i}_j$ and define the real quantities
\[
 E_{j,0}=\varepsilon(1-2\zeta _j)-1,\quad
 \mathcal T_{j,0}=-\zeta _j+\varepsilon+\varepsilon W_j(1-2\zeta _j),\quad
 \mathcal T_{j,1}=\varepsilon-1-2\varepsilon W_j.
\]
The theorem's polynomials reduce to
\[
 E_j=E_{j,0}-2\varepsilon\eta,\qquad
 \mathcal T_j=i(\mathcal T_{j,0}+\mathcal T_{j,1}\eta-2\varepsilon\eta^2).
\]
Consequently coefficient matching, in the monomial basis, is exactly
\begin{equation}
 \begin{gathered}
 E_{1,0}A^{\rm r}_1+E_{2,0}A^{\rm r}_2=U\sin\alpha,\qquad A^{\rm r}_1+A^{\rm r}_2=0,\\
 \mathcal T_{1,0}A^{\rm i}_1+\mathcal T_{2,0}A^{\rm i}_2=\varepsilon U\cos\alpha,\qquad
 \mathcal T_{1,1}A^{\rm i}_1+\mathcal T_{2,1}A^{\rm i}_2=\varepsilon U\cos\alpha,\qquad
 A^{\rm i}_1+A^{\rm i}_2=U\cos\alpha.
 \end{gathered}
 \label{example:rounded-coefficients}
\end{equation}
For these poles the five-equation, four-unknown system is consistent and has rank four. Its solution can be written
\begin{equation}
 \begin{aligned}
 A^{\rm r}_1&=-\frac{U\sin\alpha}{2\varepsilon(\zeta _1-\zeta _2)},&A^{\rm r}_2&=-A^{\rm r}_1,\\
 A^{\rm i}_1&=\frac{U\cos\alpha[-1/(2\varepsilon)-W_2]}{W_1-W_2},&
 A^{\rm i}_2&=U\cos\alpha-A^{\rm i}_1.
 \end{aligned}
 \label{example:rounded-amplitudes}
\end{equation}
The remaining constant-coefficient equation in~\eqref{example:rounded-coefficients} is verified using the unsquared pole relations. For $U=1$ and $\alpha=5^\circ$,
\[
 A_1\simeq-0.125817-0.150149i,\qquad
 A_2\simeq0.125817+1.146344i.
\]
Using these amplitudes in~\eqref{total:rounded-ansatz} completes the solution. This calculation illustrates the rounded theorem's solvability result: its one redundant equation corresponds to the trailing-edge compatibility relation. If a retained root has unsquared multiplicity two, the same procedure adds an independent amplitude $A_{j2}$ multiplying $(\zeta _j-\eta)^{-2}$ and the coefficient columns $-\partial_\zeta E$ and $-\partial_\zeta\mathcal{\mathcal T}$ evaluated at $\zeta _j$. The simple-pole column is retained as well, as prescribed by the confluent construction.
\end{example}

\section{Section geometries and data sources}\label{app:section-formulas}
This subsection gives the formulas of all sections used in the paper. Here $t=y/c$ is the chord fraction, $t_{\rm m}$ the maximum thickness as a fraction of the chord, $z_t$ the half-thickness and $z_c$ the ordinate of the mean line, both as fractions of the chord.

\emph{Four-digit and five-digit thickness distribution.} All four- and five-digit sections use~\cite{Abbott1959}
\begin{equation}
 z_t(t)=5\,t_{\rm m}\bigl(0.2969\sqrt t-0.1260\,t-0.3516\,t^2+0.2843\,t^3-0.1036\,t^4\bigr),
 \label{app:naca-thickness}
\end{equation}
in which the last coefficient is $-0.1036$ instead of the standard $-0.1015$, so that $z_t(1)=0$ and the trailing edge is closed. The sections of this form are NACA 0006, 0008, 0009, 0012, 0015, 0018, 0021 and 0024, with $t_{\rm m}$ given by the last two digits, and the cambered sections below.

\emph{Four-digit mean line.} For a section with maximum camber $m_c$ at the chord fraction $p_c$, given by the first and second digits as $m_c=(\text{first digit})/100$ and $p_c=(\text{second digit})/10$,
\begin{equation}
 z_c(t)=\begin{cases}\dfrac{m_c}{p_c^2}\bigl(2p_ct-t^2\bigr),&0\le t<p_c,\\[8pt]\dfrac{m_c}{(1-p_c)^2}\bigl(1-2p_c+2p_ct-t^2\bigr),&p_c\le t\le1.\end{cases}
 \label{app:mean-line-4}
\end{equation}
The sections of this form are NACA 2412, 4412, 6412, 8412, 4415, 2418, 4418, 6418, 8418, 2424, 4424, 6424 and 8424, all with $p_c=0.4$.

\emph{Five-digit mean line.} NACA 23012 uses the standard non-reflexed mean line
\begin{equation}
 z_c(t)=\frac{k_1}{6}
 \begin{cases}
 t^3-3r_ct^2+r_c^2(3-r_c)t,&0\le t<r_c,\\
 r_c^3(1-t),&r_c\le t\le1,
 \end{cases}
 \qquad r_c=0.2025,\quad k_1=15.957,
 \label{app:mean-line-230}
\end{equation}
with the thickness~\eqref{app:naca-thickness} for $t_{\rm m}=0.12$.

\emph{Reference contour and graph used for the fits.} With $\theta_c=\arctan z_c'(t)$, the reference contour, used by XFOIL and in the geometry panels, adds the thickness along the normal to the mean line: the upper and lower faces are the curves
\begin{equation}
 \bigl(t\mp z_t\sin\theta_c,\ z_c\pm z_t\cos\theta_c\bigr),\qquad0\le t\le1,
 \label{app:normal-construction}
\end{equation}
in units of the chord. The analytical model requires faces that are graphs over the chord with a common mean line, and its fits are made to the graph in which the thickness is added vertically,
\begin{equation}
 \frac{f^\pm}{c}=z_c(t)\pm z_t(t),\qquad\text{that is,}\quad f_c=c\,z_c,\quad f_t=c\,z_t .
 \label{app:vertical-construction}
\end{equation}
For symmetric sections the two constructions coincide.

\emph{NACA 64A410.} This section has no closed-form ordinates. The ordinates tabulated in~\cite{Stivers1954} are used: the trailing-edge gap of $0.00042c$ is closed linearly, the section is rotated to its secant frame, and each face is interpolated by a cubic spline in $\sqrt t$, as described in Section~\ref{sec:cambered-results}.

\subsection{XFOIL settings and fitting procedure}
XFOIL is a widely used numerical tool for airfoil analysis that combines a panel method with boundary-layer and transition models~\cite{DrelaXfoil,KatzPlotkin2001}. In a panel method, singularity distributions are assigned to panels on the actual contour and their strengths are determined by enforcing impermeability and a trailing-edge condition, so that the inviscid solution involves neither the collapse of the boundary onto the chord nor the linearization of the flow equation. We use it as the numerical reference for the analytical results. XFOIL 6.99 is run in inviscid mode for every validation case. Its contour is generated from the original formula of each section (Appendix~\ref{app:section-formulas}), not from the polynomial fit used by the analytical model. The standard calculation uses 481 input contour points and 320 redistributed panel points. XFOIL supplies its own surface pressures. At nonzero Mach number its K\'arm\'an--Tsien correction differs from the compressible operator derived here~\cite{XfoilPrimer}. Its pressures are interpolated on each face at the positions of the pressure taps. Changing from 240 to 320 panel points changes the interior pressures by less than $0.0060$.

Two representations of each section are used. The reference contour, used by XFOIL and drawn in the geometry panels of the figures, is that of the original formulas in Appendix~\ref{app:section-formulas}: the standard NACA thickness distribution~\cite{Abbott1959}, with the trailing-edge coefficient chosen to close the trailing edge, and, for cambered sections, the thickness added along the normal to the mean line. The analytical model uses a member of the rounded-leading-edge family fitted to the section at order 30. Its camber and thickness factors are fitted to samples of the heights and slopes of the graph obtained by adding the thickness vertically to the mean line, with an endpoint-constrained camber factor of degree 30 and a thickness factor of degree 31; the fit is a least-squares fit in a Chebyshev basis at cosine-spaced points, with a relative slope weight of $0.02$. Symmetric sections of other thickness ratios are obtained by scaling the order-30 fit of NACA 0012. The pole equations are solved as polynomial equations with a check of the branch of each root, and the largest residual of the two integral equations is below $3\times10^{-11}$. NACA 64A410 has no formula; its reference contour is a spline through the tabulated ordinates, and its fit is described in Section~\ref{sec:cambered-results}. The differences between XFOIL and the present predictions therefore contain three contributions: the error of the fit, the difference between the vertical and the normal construction for cambered sections, and the error of the collapsed-boundary model. Section~\ref{sec:fit-order} shows for NACA 0012 that the first is small at order 30. The nominal sections have a rounded leading edge and a finite trailing-edge angle; they fall under part~(iii) of Theorem~\ref{thm:existence}, and the fitted sections, which have zero thickness slope at the trailing edge, approximate them in the sense of Theorem~\ref{thm:te-angle}. Neither representation reconstructs the individual tested specimens.

The pressures are evaluated with the linearized relation~\eqref{comparison:linear-bernoulli}. The fitting order of the rounded family denotes $\deg\widetilde P$ in $P=(1-\eta)\widetilde P$; all comparisons below use order 30, that is, a profile factor of degree 31. Zero thickness is evaluated analytically. All validation calculations have zero imposed deformation.

The plotting conventions are common to all comparisons: blue solid curves denote the present solution, and red dotted curves denote inviscid XFOIL. Experimental markers identify faces or data sets as stated in each caption. Figure-reading allowances are distinguished from experimental confidence intervals. Errors are evaluated at the same observations for each model and are not pooled across different observables.

\subsection{Remarks on the pressure comparisons}
\emph{NACA 0012.} Figure~\ref{fig:val-naca0012} combines the low-speed Gregory--O'Reilly upper-surface data with the Mach-number sweep of NASA TM-100526. At low speed the pressure RMSE of the present solution is $0.081$ at $0^\circ$ and $0.252$ at $10^\circ$, against $0.012$ and $0.103$ for XFOIL. Over the nine subcritical TM-100526 panels the aggregate RMSE is $0.108$ for the present solution and $0.079$ for XFOIL. At $\alpha=0^\circ$ the present solution underestimates the suction, with errors of $0.07$--$0.11$ against $0.02$--$0.06$ for XFOIL. As $M$ and $\alpha$ increase, the K\'arm\'an--Tsien XFOIL suction peak grows faster than the measured one and the two errors approach each other: at $M=0.6$, $\alpha=4^\circ$ they are $0.169$ and $0.184$, and at $M=0.7$, $\alpha=2^\circ$ they are $0.141$ and $0.152$. On the lower surface both predictions lie below the measured suction at nonzero incidence, a common offset consistent with the absence of wall-interference corrections. At $M=0.7$, $\alpha=4^\circ$ the measured shock near $y/c\simeq0.25$ is not represented by either prediction.

\emph{NACA 0015.} Figure~\ref{fig:val-naca0015} uses six clearly resolved interior taps per face at span fraction $0.69$ of the image-endplate model in NASA TP-3151~\cite{McAlisterTakahashi1991}; reconstructed endpoint symbols are excluded. The two Reynolds-number cases are computed at their separate nominal Mach numbers (Appendix~\ref{app:nasa-source-checks}). The errors of the present solution are $0.24$--$0.50$, increasing with incidence, and those of XFOIL $0.18$--$0.23$. Both predictions underestimate the measured zero-incidence suction. The data are uncorrected tunnel data from a specimen with a finite trailing-edge gap, which is absent from the profile class; these shared discrepancies prevent reading the errors as a matched free-air validation.

\emph{NACA 0006, 0008, 0009 and 0018.} Figure~\ref{fig:val-symmetric-other} collects single-condition data for four further thickness ratios. For NACA 0006 at $4.18^\circ$ the two predictions are close, with errors of $0.089$ for the present solution and $0.082$ for XFOIL. For NACA 0008 at $8^\circ$ the errors are $0.410$ and $0.052$: at this incidence the linearized pressure relation misses most of the leading-edge suction peak. For the zero-lift NACA 0009 and 0018 readings the error of the present solution grows with thickness, from $0.058$ to $0.149$, against $0.033$ and $0.037$ for XFOIL.

\emph{NACA 4412.} Figure~\ref{fig:val-naca4412} uses Pinkerton's tabulated both-face pressures at mean $Re_c=3.1\times10^6$~\cite{Pinkerton1936}. The geometric angles $0^\circ$, $4^\circ$ and $8^\circ$ correspond to the tabulated effective angles $\alpha_0=-0.5^\circ$, $2.9^\circ$ and $6.4^\circ$, which are used unchanged; 43 readings on $0.02\le y/c\le0.95$ are retained per angle. The graph surrogate differs from the listed orifice ordinates by RMSE $0.00105c$. Here the present solution gives the smaller interior errors, $0.060$, $0.079$ and $0.137$, against $0.098$, $0.138$ and $0.209$ for XFOIL. This ordering does not extend to the six additional nose readings per angle, where XFOIL has the smaller errors at the two larger angles.

\emph{NACA 4415 and 23012.} Figure~\ref{fig:val-cambered-other} extends the cambered comparison to a further four-digit section and the five-digit NACA 23012. The NACA 4415 data are the clean-surface circles at the recorded incidence $4.12^\circ$, and the NACA 23012 data are the main airfoil tested without the flap, with the Reynolds number converted to the main chord. The RMSE values of the present solution and of XFOIL are $0.143$ and $0.101$ for NACA 4415, and $0.115$ and $0.091$ for NACA 23012. For NACA 4415 the present solution misses part of the suction plateau. The nominal-Mach sensitivities are small compared with these errors.

\emph{NACA 64A410.} Figure~\ref{fig:val-naca64a410} compares with Stivers' Mach-number sweep~\cite{Stivers1954}, whose run Mach numbers are grouped near $0.32$, $0.51$, $0.61$ and $0.66$; each panel uses its tabulated run Mach number. This section has no closed-form ordinates, so the table itself is fitted. The $0.00042c$ trailing-edge gap is closed linearly, each face is interpolated by a cubic spline in $\sqrt{y/c}$, and the camber and thickness factors are fitted at order 30 to dense height and slope samples, as for the formula-based NACA sections. The fit reproduces the tabulated ordinates with RMSE $0.0002c$ and maximum deviation $0.001c$. The small ripples near the leading edge come from interpolating the sparse table and are shared by both predictions. Over the ten subcritical panels the aggregate RMSE is $0.073$ for the present solution and $0.090$ for XFOIL. The larger XFOIL error at $M\ge0.61$ again comes mainly from the leading-edge suction peak. Fits of order 24, 32 and 48 give subcritical RMSE values within $0.001$ of each other, so that the result is insensitive to the fit and the larger XFOIL error persists. A direct low-order least-squares fit of the sparse ordinates, which does not resample the slopes, gives a larger error, $0.081$, and does not converge as its degree is raised.

\subsection{Source details of the measurements}\label{app:pressure-source-details}\label{app:nasa-source-checks}
\emph{NACA 4412.} The pressures are read from Table~I of Report~563 and the effective angles from its Table~II; the sign convention and the staggered surface positions of Table~I are preserved.

\emph{NACA 0015.} The largest model of NASA TP-3151, the only one instrumented for pressure, has a chord of $0.518$~m; Figure~18 of the report is its image-endplate configuration. The report associates $Re\simeq1.5\times10^6$ on this chord with $U\simeq46$~m/s and $M\simeq0.13$, and scaling to $Re\simeq2.5\times10^6$ gives about $77$~m/s and $M\simeq0.22$. Each nominal Mach number is used in both the analytical operator and XFOIL. The pressure orifices lie at the span fraction $y/s=0.69$. Estimated endpoint pressures, shown by open symbols in the source, are excluded, and coincident zero-incidence observations are counted once. The data are used as published, without wall-interference corrections.

\emph{NACA 4415 and 23012.} The NACA 4415 readings are the clean-airfoil symbols of Figure~B149 of NREL/TP-442-6472, 21 on the upper and 20 on the lower surface. The model chord is $0.4572$~m; with air at $20^\circ$C and one atmosphere, $U=Re_c\nu/c$ gives the nominal Mach number $0.192$. Repeating the predictions at $M=0.17$ and $0.21$ changes the interior $C_p$ by at most $0.011$. The NACA 23012 readings are from Figure~3 of Report~614, for the main airfoil without the flap. The main and flap chords are 20 and 4~in and the Reynolds number of the source refers to their sum, so the main-chord Reynolds number is $1.46\times10^6/1.2$; the nominal condition of 80~mph at sea level gives $M\simeq0.105$. Ten readings per face are used on $0.02\le t\le0.95$. The transcription of this figure in~\cite{ASPIREPressureData} was checked against the report, from which the conditions and the surface ordering are taken. The reading allowances $(\Delta t,\Delta C_p)$ are $(0.005,0.05)$ and $(0.005,0.025)$ for the two sections; they describe the resolution of the source figures and are not statistical uncertainties.

\section{MATLAB implementation for symmetric NACA sections}\label{app:matlab}
The function below computes the surface pressures on a symmetric NACA 00XX section from the closed-form solution of Theorem~\ref{thm:rounded-family}. Its inputs are the section number, the freestream speed $U$ and the incidence $\alpha$ in degrees, and optionally the speed of sound $a_\infty$, the density $\rho_\infty$, the static pressure $p_\infty$ and the fit order; the speed of sound, density and pressure default to sea-level values. The speed of sound enters only through the Mach number $M=U/a_\infty$, and the density and the static pressure only through the dimensional face pressures $p^\pm=p_\infty+q_\infty C_p^\pm$.

The function follows the steps of the paper. The thickness~\eqref{app:naca-thickness} and its slope are sampled at cosine-spaced points of the chord. The rounded-leading-edge family~\eqref{total:rounded-profile-slope} with $P=(1-\eta)\widetilde P$ is fitted to these samples by least squares, with $\widetilde P$ a Chebyshev series of degree 30 and a relative slope weight of $0.02$, as in Appendix~\ref{app:section-formulas}; each section is fitted directly. Algorithm~\ref{alg:rounded} is then applied with $k_c=0$, so that $\lambda=a$. The poles are found from the map $\zeta=(\omega+\omega^{-1})/2$, which takes the slit plane onto the unit disk: $\omega^{\mathsf m}\Psi_{\rm r}$, with ${\mathsf m}=\deg R$, is a polynomial in $\omega$ of degree $2{\mathsf m}$, its coefficients are computed from samples on the unit circle, and its roots inside the disk are the poles, so that no separate branch test is needed. The polynomial identities~\eqref{total:rounded-amplitudes}, of degrees ${\mathsf q}_1$ and ${\mathsf q}_2$, are imposed at ${\mathsf q}_1+1$ and ${\mathsf q}_2+1$ Chebyshev points, which is equivalent to matching their coefficients, and the redundant endpoint equation is absorbed by a least-squares solve of the consistent system. The jumps follow from~\eqref{total:rounded-ansatz}, their finite Hilbert transforms from~\eqref{total:transform-identities}, the face velocities from~\eqref{comparison:face-pressure-mapping} and the pressure coefficients from~\eqref{comparison:linear-bernoulli}. The implementation assumes simple poles, which is the generic case, and issues a warning when two poles nearly coincide. For NACA 0012 at $M=0.147$, $\alpha=4^\circ$ with a fit of order 30 and at $M=0.588$, $\alpha=2^\circ$ with a fit of order 20, its pressure coefficients agree with those of the implementation used for the figures of this paper to $3\times10^{-12}$. A call such as \texttt{naca00xx\_cp('0012', 50, 4)} without an output argument plots the two pressure distributions.

\begin{lstlisting}
function out = naca00xx_cp(naca, U, alphaDeg, aInf, rhoInf, pInf, N)
%NACA00XX_CP  Surface pressures on a symmetric NACA 00XX section from the
%  closed-form solution of the thickness-modified airfoil system for the
%  rounded-leading-edge family (Algorithm 2), with the linearized Bernoulli
%  relation for the face pressures.
%
%  out = naca00xx_cp(naca, U, alphaDeg)                      % sea-level air
%  out = naca00xx_cp(naca, U, alphaDeg, aInf, rhoInf, pInf, N)
%
%  naca      thickness digits, e.g. 12 or '0012' (only the last two are used)
%  U         freestream speed [m/s];  alphaDeg  incidence [deg]
%  aInf      speed of sound [m/s]  (default 340.29)
%  rhoInf    density [kg/m^3]      (default 1.225)
%  pInf      static pressure [Pa]  (default 101325)
%  N         fit order: degree of Pt in P = (1-eta)*Pt (default 30)
%  out       struct: x (= y/c), CpU, CpL, pU, pL, CL, Cm (about c/4), M,
%            poles, amplitudes, fitError
%  Without an output argument the pressure distributions are plotted.

if ischar(naca) || isstring(naca), naca = str2double(char(naca)); end
if nargin < 4 || isempty(aInf),   aInf   = 340.29; end
if nargin < 5 || isempty(rhoInf), rhoInf = 1.225;  end
if nargin < 6 || isempty(pInf),   pInf   = 101325; end
if nargin < 7 || isempty(N),      N      = 30;     end
tm = mod(naca, 100)/100;                       % thickness ratio t_m
M  = U/aInf;  al = alphaDeg*pi/180;
if M >= 1, error('Subsonic flow (U < aInf) is required.'); end

% Flow coefficients a, b; no camber, so k_c = 0 and lambda = a
D = 1 - M^2*sin(al)^2;  a = M^2*sin(al)*cos(al)/D;  b = sqrt(1 - M^2)/D;
s = a - 1i*b;  lam = a;

% Steps 1-3: sample the section and fit the rounded-leading-edge family
[eta, T, dT] = sample_section(tm, 400);
[cR, fitError] = fit_rounded(eta, T, dT, N, 0.02);

% Step 4: Algorithm 2 (poles and amplitudes)
[zj, A] = algorithm2(cR, b, lam, U, al);

% Step 5: jumps, face velocities and pressure coefficients
K = 201;  x = -cos(pi*((1:K)' - 0.5)/K);  w = sqrt(1 - x.^2);
Z  = (1./(zj.' - x))*A;                        % sum_j A_j/(zeta_j - eta)
HZ = (-1 + Wb(zj).'./(zj.' - x))*A;            % H_I[w Z]
d  = 2*w.*real(Z);   n = 2*w.*real(s*Z);       % velocity jumps
Hd = 2*real(HZ);     Hm = 2*b*imag(HZ);        % H_I d and H_I(n - a d)
u  = Hm/(2*b);       v = a*Hm/(2*b) - b*Hd/2;  % mean face velocities
CpU = -2/U*(cos(al)*(u + d/2) + sin(al)*(v + n/2));
CpL = -2/U*(cos(al)*(u - d/2) + sin(al)*(v - n/2));

% Normal force and moment by Gauss-Chebyshev quadrature
dCp = CpL - CpU;  t = (1 + x)/2;
CN  = pi/K*sum(dCp.*w)/2;  Cm = pi/K*sum((0.25 - t).*dCp.*w)/2;

q = 0.5*rhoInf*U^2;
out = struct('x', t, 'CpU', CpU, 'CpL', CpL, 'pU', pInf + q*CpU, ...
    'pL', pInf + q*CpL, 'CL', CN*cos(al), 'Cm', Cm, 'M', M, ...
    'poles', zj, 'amplitudes', A, 'fitError', fitError);
if nargout == 0
    plot(t, -CpU, 'b-', t, -CpL, 'r--'); grid on
    xlabel('y/c'); ylabel('-C_p'); legend('upper', 'lower');
    title(sprintf('NACA 00%02d, M = %.3f, \\alpha = %g^\\circ', round(100*tm), M, alphaDeg));
    clear out
end
end

% ------------------------------------------------------------------------
function [eta, T, dT] = sample_section(tm, Ns)
% Cosine-spaced interior points eta = 2t-1 and the NACA thickness
% (closed trailing edge): T = 2 z_t (full thickness /
% chord) and its slope dT/deta = dz_t/dt.
eta = -cos(pi*((1:Ns)' - 0.5)/Ns);  t = (1 + eta)/2;
T  = 10*tm*(0.2969*sqrt(t) - 0.1260*t - 0.3516*t.^2 + 0.2843*t.^3 - 0.1036*t.^4);
dT =  5*tm*(0.2969/2./sqrt(t) - 0.1260 - 0.7032*t + 0.8529*t.^2 - 0.4144*t.^3);
end

function [cR, err] = fit_rounded(eta, T, dT, N, wt)
% Least-squares fit of the rounded family T = w P, P = (1-eta)Pt, with Pt a
% Chebyshev series of degree N. The slope rows
% match R = w g_t with w dT/deta (weight wt), which removes the singularity
% of the slope at the nose. Returns the Chebyshev coefficients of R.
th = acos(eta);  k = 0:N;  w = sqrt(1 - eta.^2);
Tk = cos(th*k);  dTk = sin(th*k).*k./sin(th);  % T_k and T_k' at the samples
P  = (1 - eta).*Tk;  dP = -Tk + (1 - eta).*dTk;
c  = [w.*P; wt*((1 - eta.^2).*dP - eta.*P)] \ [T; wt*w.*dT];
err = max(abs(w.*P*c - T));                    % largest thickness error
cP = [c; 0] - chebx(c);                        % P = (1 - eta) Pt
dP = chebder(cP);                              % R = (1-eta^2)P' - eta P
cR = [dP; 0; 0] - chebx(chebx(dP)) - chebx(cP);
end

function [zj, A] = algorithm2(cR, b, lam, U, al)
% Algorithm 2 for k_c = 0 (constant lambda) with simple poles.
m = numel(cR) - 1;                             % deg R; there are m poles
% Step 1 (poles): zeros of Psi_r = R + (b + i lam) W off the chord. With
% zeta = (om + 1/om)/2 the slit plane maps onto |om| < 1, and om^m Psi_r is
% a polynomial of degree 2m in om, recovered from samples on |om| = 1.
Kc = 4*m;  om = exp(2i*pi*(0:Kc-1)'/Kc);
f  = om.^m.*(chebval(cR, real((om + 1./om)/2)) + (b + 1i*lam)*(1 - om.^2)./(2*om));
C  = fft(f)/Kc;  r = roots(flipud(C(1:2*m+1)));
r  = r(abs(r) < 1 - 1e-8);  zj = (r + 1./r)/2;
if numel(zj) ~= m, warning('Found %d poles instead of %d.', numel(zj), m); end
cdR = chebder(cR);                             % Newton polish on Psi_r
for it = 1:3
    zj = zj - (chebval(cR, zj) + (b + 1i*lam)*Wb(zj)) ./ ...
              (chebval(cdR, zj) + (b + 1i*lam)*zj./Wb(zj));
end
if min(min(abs(zj - zj.') + eye(m)*1e9)) < 1e-6
    warning('Nearly repeated poles: the confluent form is needed.');
end
% Step 2 (amplitudes): the polynomial identities 2Re sum A_j E_j = F_r and
% 2Re sum A_j T_j = G_r, of degrees m-1 and m, are imposed at m and m+1
% Chebyshev points, which is equivalent to matching their coefficients.
eE = cos(pi*((1:m)' - 0.5)/m);  eT = cos(pi*((1:m+1)' - 0.5)/(m+1));
DR = @(e) (chebval(cR, e) - chebval(cR, zj.'))./(e - zj.');  % D_zeta R
E  = -DR(eE) - b - 1i*lam;
Tm = (zj.' + eT)*(lam - 1i*b) - 1i*chebval(cR, eT) - 1i*Wb(zj.').*DR(eT);
Mr = 2*[real(E), -imag(E); real(Tm), -imag(Tm)];
h  = [2*U*sin(al)*ones(m, 1); 2*U*cos(al)*chebval(cR, eT)];
xA = Mr\h;  A = xA(1:m) + 1i*xA(m+1:end);    % one redundant row: T(1) = 0
end

function y = Wb(z)                             % W = sqrt(z-1)sqrt(z+1) ~ z
y = sqrt(z - 1).*sqrt(z + 1);
end

function y = chebval(c, z)                     % Clenshaw: sum c_k T_k(z)
b1 = 0*z;  b2 = b1;
for k = numel(c):-1:2, [b1, b2] = deal(c(k) + 2*z.*b1 - b2, b1); end
y = c(1) + z.*b1 - b2;
end

function y = chebx(c)                          % coefficients of x*sum c_k T_k
c = c(:);  n = numel(c);  y = zeros(n + 1, 1);
y(2) = c(1);  y(3:n+1) = y(3:n+1) + c(2:n)/2;  y(1:n-1) = y(1:n-1) + c(2:n)/2;
end

function d = chebder(c)                        % coefficients of the derivative
c = c(:);  n = numel(c) - 1;  d = zeros(n + 2, 1);
for k = n:-1:1, d(k) = d(k + 2) + 2*k*c(k + 1); end
d(1) = d(1)/2;  d = d(1:max(n, 1));
end
\end{lstlisting}

\section{Proofs}\label{app:family-proofs}
\subsection{Proof of Theorem~\ref{thm:existence}}\label{app:proof-geometry-stability}
Part~(i) is proved from the factorization~\eqref{stab:factorization} by the Neumann series, using the properties of $\HH$ on $X$ in Lemma~\ref{lem:H-unitary}. Parts~(ii) and~(iii) are proved by reducing the system to that of a flat plate in a weighted space.

\begin{lemma}[The finite Hilbert transform on $X$]\label{lem:H-unitary}
The finite Hilbert transform $\HH$ is a unitary operator on $X$, and
\begin{equation}
 \HH^{-1}g=-r\,\HH\Bigl[\frac gr\Bigr],\qquad g\in X.
 \label{app:H-inverse}
\end{equation}
\end{lemma}
\begin{proof}
The space $X$ is $L^2$ on the chord with the weight $1/r=(1+\eta)^{1/2}(1-\eta)^{-1/2}$. This is a power weight whose exponents at the two endpoints lie in $(-1,1)$, so that it satisfies the Muckenhoupt condition $A_2$ on the interval, and $\HH$ is bounded on $X$~\cite[Proposition~2.2]{AstalaPaivarintaSaksman1996}. Moreover,
\[
 \frac{1-\eta}{1+\eta}\cdot\frac1{r(\eta)}=r(\eta)=(1-\eta)^{1/2}(1+\eta)^{-1/2}
\]
is again a power weight with exponents in $(-1,1)$ and satisfies $A_2$. By~\cite[Theorem~3.1]{AstalaPaivarintaSaksman1996}, applied with $p=2$, the weight $1/r$ and the function $h=r$, the operator $\HH$ is therefore an isomorphism of $X$ onto itself, with the inverse $-r\,\HH[\,\cdot\,/r]$, which is~\eqref{app:H-inverse}.

It remains to show that $\HH$ preserves the norm. Let $f,g\in X$. Then $g/r$ belongs to $L^2((-1,1);r\,\dd\eta)$, the dual of $X$ in the unweighted pairing, and the antisymmetry of the finite Hilbert transform in this pairing~\cite[Eq.~(3.1)]{AstalaPaivarintaSaksman1996} gives
\[
 \int_{-1}^1(\HH f)\,g\,\frac{\dd\eta}{r}=-\int_{-1}^1f\,\HH\Bigl[\frac gr\Bigr]\dd\eta
 =\int_{-1}^1f\,\Bigl(-r\,\HH\Bigl[\frac gr\Bigr]\Bigr)\frac{\dd\eta}{r}.
\]
The adjoint of $\HH$ in $X$ is thus $-r\,\HH[\,\cdot\,/r]=\HH^{-1}$, so that $\HH$ is unitary.
\end{proof}

\begin{recalled}[Existence and uniqueness, part~(i)]{Theorem~\ref{thm:existence}}
Fix $U>0$, $0\le M<1$ and $\alpha$, and hence $a,b$.

\emph{(i) Small bounded slopes.} Let $\lambda,g_t\in L^\infty(-1,1)$ with $\mu>0$. Then $b\mathsf J+\mathsf K$ has a bounded inverse on $X\times X$, given by the Neumann series
\begin{equation}
 (b\mathsf J+\mathsf K)^{-1}=-\frac1b\sum_{n\ge0}\Bigl(\frac{\mathsf J\mathsf K}{b}\Bigr)^{n}\mathsf J,
 \tag{\ref{stab:neumann}}
\end{equation}
which converges in the operator norm, and $\mathsf A^{-1}=\mathsf D^{-1}(b\mathsf J+\mathsf K)^{-1}\mathsf E^{-1}$. In particular,~\eqref{stab:system} has exactly one solution $(d,\ell)\in X\times X$, and $\norm{d}_X+\norm{\ell}_X\le\mu^{-1}\bigl(\norm{F}_X+\norm{G}_X\bigr)$.
\end{recalled}

\begin{proof}[Proof of Theorem~\ref{thm:existence}, part~(i)]
\emph{Step 1: the factors of $\mathsf A$.} Use the norm $\norm{d}_X+\norm{\ell}_X$ on $X\times X$. By Lemma~\ref{lem:H-unitary}, the operators $\mathsf E$, $\mathsf J$ and $\mathsf D$ preserve this norm, and $\mathsf J^{-1}=-\mathsf J$. The multiplication operator $\mathsf K$ satisfies $\norm{\mathsf K}\le\Lambda+\Gamma$.

\emph{Step 2: inversion by the Neumann series.} Write
\[
 b\mathsf J+\mathsf K=b\mathsf J\Bigl(\mathsf I-\frac{\mathsf J\mathsf K}{b}\Bigr),\qquad\Bigl\lVert\frac{\mathsf J\mathsf K}{b}\Bigr\rVert\le\frac{\Lambda+\Gamma}{b}<1.
\]
By the Neumann series theorem~\cite[Theorem~2.14]{Kress2014}, $\mathsf I-\mathsf J\mathsf K/b$ has the bounded inverse $\sum_{n\ge0}(\mathsf J\mathsf K/b)^n$, of norm at most $b/\mu$. Multiplication by $(b\mathsf J)^{-1}=-\mathsf J/b$ gives~\eqref{stab:neumann} and $\norm{(b\mathsf J+\mathsf K)^{-1}}\le1/\mu$, and~\eqref{stab:factorization} gives $\mathsf A^{-1}$ with $\norm{\mathsf A^{-1}}\le1/\mu$.

\emph{Step 3: bound on the solution.} With $\norm{1}_X^2=\pi$, $\norm{F}_X\le2\sqrt\pi(U|\sin\alpha|+\norm{k_c}_\infty U|\cos\alpha|)$ and $\norm{G}_X\le2\sqrt\pi\,\Gamma U|\cos\alpha|$, the solution satisfies
\begin{equation}
 \norm{d}_X+\norm{\ell}_X\le\frac{2\sqrt\pi}{\mu}\Bigl[U|\sin\alpha|+\bigl(\norm{k_c}_\infty+\Gamma\bigr)U|\cos\alpha|\Bigr].
 \label{stab:apriori}
\end{equation}
\end{proof}

\begin{recalled}[Existence and uniqueness, parts~(ii) and~(iii)]{Theorem~\ref{thm:existence}}
Fix $U>0$, $0\le M<1$ and $\alpha$, and hence $a,b$. A function on $[-1,1]$ is called piecewise H\"older continuous if there are finitely many points $\eta_k$ of the open chord such that the function is H\"older continuous on each of the closed intervals into which these points divide $[-1,1]$, with one-sided limits at the points $\eta_k$. At these points the angles~\eqref{stab:angles} are required to satisfy
\begin{equation}
 \bigl|\phi^\pm(\eta_k+0)-\phi^\pm(\eta_k-0)\bigr|<\frac\pi2\qquad\text{for every }k\text{ and for both faces.}
 \tag{\ref{stab:corner-angles}}
\end{equation}

\emph{(ii) Bounded slopes.} Let $\lambda$ and $g_t$ be piecewise H\"older continuous with~\eqref{stab:corner-angles}, and let
\begin{equation}
 \bigl|\phi^+(-1)-\phi^-(-1)\bigr|<\frac\pi2,\qquad\bigl|\phi^+(1)-\phi^-(1)\bigr|<\frac\pi2 .
 \tag{\ref{stab:edge-angles}}
\end{equation}
Then $\mathsf A$ has a bounded inverse on $X\times X$. In particular,~\eqref{stab:system} has exactly one solution $(d,\ell)\in X\times X$.

\emph{(iii) Rounded leading edge.} Let
\begin{equation}
 g_t=\frac{S}{\sqrt{1+\eta}},\qquad S(-1)>0,
 \tag{\ref{stab:wedge-class}}
\end{equation}
with piecewise H\"older-continuous $\lambda$ and $S$ and with~\eqref{stab:corner-angles}, and let
\begin{equation}
 \varDelta_{\rm te}:=\bigl|\phi^+(1)-\phi^-(1)\bigr|<\frac\pi2 .
 \tag{\ref{stab:te-angle}}
\end{equation}
Then~\eqref{stab:system} has exactly one solution $(d,\ell)\in Y\times Y$.
\end{recalled}

The operator of the system is written as a product in which the geometry enters only through multiplications by functions, and the inversion of the system is reduced to the inversion of the finite Hilbert transform $\HH$ on weighted $L^2$ spaces, for which the result of Astala, P\"aiv\"arinta and Saksman~\cite{AstalaPaivarintaSaksman1996} is invoked. The construction adapts two ideas from the theory of systems of singular integral equations. Following Peters~\cite{Peters1972}, the system is written for the boundary values on the two sides of the chord of the Cauchy integral of the unknowns, here combined into the single complex unknown $f=d-i\ell$. Following Porter~\cite{Porter1986}, this Cauchy integral is written as the product of an analytic function and the Cauchy integral of a new unknown; with the canonical function~\eqref{wp:canonical} as the analytic factor, the system becomes that of a flat plate, and the relation between the two unknowns is the formula~\eqref{wp:G} for $\mathsf G^\sharp$, which has the form of~\cite[Eq.~(2.7)]{Porter1986}. Both works assume H\"older-continuous coefficients and solutions with integrable endpoint singularities; here the coefficients may have corners and the unbounded slopes of a rounded leading edge, and the solutions are sought in the weighted spaces $X$ and $Y$. Parts~(ii) and~(iii) can also be obtained by another route. The map $\omega=\zeta-W(\zeta)$ takes the slit plane onto the unit disk and the two banks of the chord onto the unit circle, and the face conditions~\eqref{wp:rows} become a Riemann--Hilbert problem on the circle, with a piecewise continuous unimodular coefficient formed from $c_\pm/|c_\pm|$. By Muskhelishvili's symmetrization~\cite{Muskhelishvili1953} this is a problem of linear conjugation, for which existence, uniqueness and solution formulas in weighted classes of Cauchy-type integrals are available, including coefficients with jumps~\cite{KhuskivadzeKokilashviliPaatashvili1998,KokilashviliPaatashviliSamko2005,KokilashviliPaatashvili2007}. In that setting each corner and each edge contributes a jump of the coefficient, and the conditions~\eqref{stab:corner-angles},~\eqref{stab:edge-angles} and~\eqref{stab:te-angle} correspond to the conditions on these jumps under which the problem is uniquely solvable. These results give no explicit bound on the inverse. The reduction on the chord is used here because it gives the solution formula~\eqref{wp:solution} on the chord directly, and because its canonical function is also used for the explicit bounds of Section~\ref{app:rh-bounds}.

\subsubsection{Step 1: the system in complex form}
Throughout, $\sigma_0=r$ in part~(ii) and $\sigma_0=w$ in part~(iii). For a positive weight $\sigma$ write
\begin{equation}
 L^2_{\sigma_0}=L^2\bigl((-1,1);\sigma_0^{-1}\dd\eta\bigr),\qquad
 L^2(\sigma)=L^2\bigl((-1,1);\sigma\,\dd\eta\bigr),
 \label{wp:spaces}
\end{equation}
so that $X=L^2_r$ and $Y=L^2_w$ in~\eqref{stab:space} and~\eqref{stab:space-Y}. The same symbols denote the spaces of complex functions, and a pair $(d,\ell)$ of real functions is identified with $f=d-i\ell$, so that $\norm{f}^2=\norm{d}^2+\norm{\ell}^2$. With the angles $\phi^\pm$ of~\eqref{stab:angles} put
\begin{equation}
 \phi_{\rm d}=\phi^+-\phi^-,\qquad\phi_{\rm s}=\phi^++\phi^-,\qquad
 c_\pm=\lambda\mp g_t+ib=\frac{i\,b\,e^{i\phi^\pm}}{\cos\phi^\pm},\qquad
 [\phi^\pm]_k=\phi^\pm(\eta_k+0)-\phi^\pm(\eta_k-0).
 \label{wp:angles}
\end{equation}
In part~(ii) the angles $\phi^\pm=\arctan((\pm g_t-\lambda)/b)$ are piecewise H\"older continuous with values in a compact subset of $(-\pi/2,\pi/2)$. In part~(iii), $\phi^\pm=\pm\pi/2\mp\arctan\bigl(b\sqrt{1+\eta}/(S\mp\lambda\sqrt{1+\eta})\bigr)$ near the leading edge, so that $\phi^\pm$ are again piecewise H\"older continuous on $[-1,1]$, with $\phi^\pm(-1)=\pm\pi/2$, $\phi_{\rm d}(-1)=\pi$ and $\phi_{\rm s}(-1)=0$. In both parts $|[\phi^\pm]_k|<\pi/2$ by~\eqref{stab:corner-angles} and $|\phi_{\rm d}(1)|<\pi/2$.

For real $d,\ell$ in $X$ or in $Y$ put
\begin{equation}
 u^\pm=\tfrac12\HH\ell\pm\tfrac12d,\qquad
 v^\pm-au^\pm=-\tfrac b2\HH d\pm\tfrac b2\ell,
 \label{app:face-velocities}
\end{equation}
For $f=d-i\ell$ define
\begin{equation}
 \mathcal V(\zeta )=\frac{i}{2\pi}\int_{-1}^1\frac{f(\eta)}{\zeta -\eta}\,\dd\eta ,
 \label{app:cauchy-integral}
\end{equation}
which is analytic off the chord and $O(\zeta ^{-1})$ at infinity. By the Plemelj--Sokhotski formulas~\cite{Muskhelishvili1953}, its boundary values on the upper and lower banks of the chord are
\begin{equation}
 \mathcal V_\pm=\pm\tfrac12f+\tfrac i2\HH f=u^\pm-\frac ib\,(v^\pm-au^\pm).
 \label{app:plemelj}
\end{equation}
For a complex function $f$ on the chord put
\begin{equation}
 \mathsf P_\pm f=\pm\tfrac12f+\tfrac i2\HH f ,
 \label{wp:projections}
\end{equation}
so that $\mathsf P_\pm f=\mathcal V_\pm$ by~\eqref{app:plemelj}. The half-sum and the difference of the two rows of~\eqref{stab:matrix-form} give
\begin{equation}
 \mathsf A=\mathsf S\,\mathsf N,\qquad
 \mathsf S=\begin{pmatrix}-1&-1\\1&-1\end{pmatrix},\qquad
 \mathsf N\begin{pmatrix}d\\ \ell\end{pmatrix}=\begin{pmatrix}h^+\\ h^-\end{pmatrix},\qquad
 h^\pm=v^\pm-(k_c\pm g_t)u^\pm=\Re\bigl[c_\pm\,\mathsf P_\pm f\bigr],
 \label{wp:rows}
\end{equation}
with $f=d-i\ell$. The last expression defines $\mathsf N$ also when $g_t$ is unbounded, and~\eqref{stab:system} is the equation $\mathsf N(d,\ell)^{\mathsf T}=(h^+,h^-)^{\mathsf T}$ with the tangency data
\begin{equation}
 h^\pm=(k_c\pm g_t)\,U\cos\alpha-U\sin\alpha .
 \label{wp:data}
\end{equation}
Parts~(ii) and~(iii) follow from the next statement, which is proved in the remainder of this subsection.

\begin{proposition}[The operator $\mathsf N$]\label{prop:N-isomorphism}
Under the hypotheses of part~(ii) or of part~(iii) of Theorem~\ref{thm:existence}, $\mathsf N$ is an isomorphism
\begin{equation}
 \mathsf N:\ L^2_{\sigma_0}\times L^2_{\sigma_0}\ \longrightarrow\ L^2\Bigl(\frac{\cos^2\phi^+}{\sigma_0}\Bigr)\times L^2\Bigl(\frac{\cos^2\phi^-}{\sigma_0}\Bigr),\qquad
 \cos^2\phi^\pm=\frac{b^2}{(\lambda\mp g_t)^2+b^2},
 \label{wp:isomorphism}
\end{equation}
and its inverse is given by~\eqref{wp:solution}.
\end{proposition}

\subsubsection{Step 2: a function that removes the slope angles}
With $W$ from~\eqref{total:branch-functions}, define for $\zeta\notin[-1,1]$
\begin{equation}
 \Theta(\zeta)=-\frac1{2\pi}\int_{-1}^1\frac{\phi_{\rm d}(t)}{t-\zeta}\,\dd t
 -\frac{W(\zeta)}{2\pi i}\int_{-1}^1\frac{\phi_{\rm s}(t)}{w(t)\,(t-\zeta)}\,\dd t,\qquad
 \Xi(\zeta)=e^{\Theta(\zeta)} .
 \label{wp:canonical}
\end{equation}
The function $\Xi$ is analytic and has no zeros off the chord, and $|\Xi(\infty)|=1$, because the first term of $\Theta$ vanishes at infinity and the second tends to the purely imaginary number $-\tfrac{i}{2\pi}\int_{-1}^1\phi_{\rm s}/w\,\dd t$. The points $\pm1$ and $\eta_k$ are called the special points. The following three properties are used; they follow from the Plemelj--Sokhotski formulas and from the behaviour of Cauchy integrals near the ends of the line of integration and near points of discontinuity of the density~\cite[Chapter~4]{Muskhelishvili1953}.

\emph{(a) Boundary values.} On the two sides of the chord $W=\pm iw$. Hence, at every point of the chord that is not special, $\Theta_\pm=-i\phi^\pm+\tfrac12\HH\phi_{\rm d}\pm\tfrac12w\,\HH[\phi_{\rm s}/w]$, that is,
\begin{equation}
 \arg\Xi_\pm=-\phi^\pm,\qquad
 |\Xi_\pm|=\exp\Bigl\{\tfrac12\HH\phi_{\rm d}\pm\tfrac12\,w\,\HH\Bigl[\frac{\phi_{\rm s}}{w}\Bigr]\Bigr\} .
 \label{wp:boundary-values}
\end{equation}
The argument of $\Xi$ on each side of the chord is thus minus the slope angle of the corresponding face, so that
\begin{equation}
 c_\pm\,\Xi_\pm=\frac{i\,b\,|\Xi_\pm|}{\cos\phi^\pm}
 \label{wp:imaginary}
\end{equation}
is purely imaginary. The functions $\Xi_\pm$ are H\"older continuous on every closed interval of the chord that contains no special point.

\emph{(b) Moduli on the chord.} A jump $[\psi]_c$ of a H\"older-continuous density $\psi$ at a point $c$ of the open chord contributes $\tfrac{[\psi]_c}\pi\ln|\eta-c|$ to its finite Hilbert transform, and the end values contribute logarithms at the two ends, as $\HH[1]=\tfrac1\pi\ln\tfrac{1+\eta}{1-\eta}$ shows. With the jumps $[\phi^+]_k-[\phi^-]_k$ of $\phi_{\rm d}$ and $[\phi^+]_k+[\phi^-]_k$ of $\phi_{\rm s}$,
\begin{equation}
 \begin{aligned}
 \HH\phi_{\rm d}&=-\frac{\phi_{\rm d}(1)}\pi\ln(1-\eta)+\frac{\phi_{\rm d}(-1)}\pi\ln(1+\eta)+\sum_k\frac{[\phi^+]_k-[\phi^-]_k}\pi\ln|\eta-\eta_k|+\text{a bounded function},\\
 w\,\HH\Bigl[\frac{\phi_{\rm s}}{w}\Bigr]&=\sum_k\frac{[\phi^+]_k+[\phi^-]_k}\pi\ln|\eta-\eta_k|+\text{a bounded function},
 \end{aligned}
 \label{wp:log-behaviour}
\end{equation}
where the bounded functions are H\"older continuous on each closed interval between consecutive special points. Hence there is a constant $C\ge1$ with
\begin{equation}
 C^{-1}\le\frac{|\Xi_\pm|}{(1-\eta)^{-\phi_{\rm d}(1)/(2\pi)}\,(1+\eta)^{\phi_{\rm d}(-1)/(2\pi)}\prod_k|\eta-\eta_k|^{\pm[\phi^\pm]_k/\pi}}\le C .
 \label{wp:moduli}
\end{equation}
At the two edges the moduli on the two faces are comparable. At a corner they are comparable only if the two faces turn by opposite angles; in general the two faces carry different powers there.

\emph{(c) Behaviour off the chord.} Near each special point $c$, both $\Xi$ and $1/\Xi$ are $O(|\zeta-c|^{-1/2})$. Indeed, the exponents in~\eqref{wp:moduli} have modulus at most $\tfrac12$ at the edges, since $|\phi_{\rm d}|\le\pi$, and modulus less than $\tfrac12$ at the corners.

\subsubsection{Step 3: the weights left by the slope angles}
Define the two weights
\begin{equation}
 \sigma_\pm=\frac{|\Xi_\pm|^2}{\sigma_0},\qquad\sigma_{\min}=\min\{\sigma_+,\sigma_-\} .
 \label{wp:weight}
\end{equation}
By~\eqref{wp:moduli}, each of $\sigma_+$, $\sigma_-$ and $\sigma_{\min}$ is bounded above and below by constant multiples of a weight of the form
\begin{equation}
 \begin{gathered}
 (1-\eta)^{e_{\rm te}}(1+\eta)^{e_{\rm le}}\prod_k|\eta-\eta_k|^{e_k},\qquad
 e_{\rm te}=-\frac{\phi_{\rm d}(1)}\pi-\frac12,\\
 e_{\rm le}=\frac{\phi_{\rm d}(-1)}\pi+\frac12\ \ (\sigma_0=r),\qquad
 e_{\rm le}=\frac{\phi_{\rm d}(-1)}\pi-\frac12\ \ (\sigma_0=w),
 \end{gathered}
 \label{wp:exponents}
\end{equation}
with the same $e_{\rm te}$ and $e_{\rm le}$ for the three weights, and with $e_k=2[\phi^+]_k/\pi$ for $\sigma_+$, $e_k=-2[\phi^-]_k/\pi$ for $\sigma_-$ and the larger of these two numbers for $\sigma_{\min}$. A weight of this form satisfies the Muckenhoupt condition $A_2$ on the interval if and only if all its exponents lie in $(-1,1)$, and the condition is not affected by a factor that is bounded above and below. Under the hypotheses of the theorem,
\begin{equation}
 \begin{array}{llll}
 \text{part (ii):}&-1<e_{\rm te}<0,&0<e_{\rm le}<1,&-1<e_k<1,\\[2pt]
 \text{part (iii):}&-1<e_{\rm te}<0,&e_{\rm le}=\tfrac12,&-1<e_k<1 .
 \end{array}
 \label{wp:ranges}
\end{equation}
The following four properties hold for $\sigma\in\{\sigma_+,\sigma_-,\sigma_{\min}\}$.
\begin{enumerate}
 \item[(W1)] $\sigma\in A_2$ and $r^2\sigma=\dfrac{1-\eta}{1+\eta}\,\sigma\in A_2$; the second weight has the exponents $e_{\rm te}+1\in(0,1)$, $e_{\rm le}-1\in(-1,0]$ and $e_k$.
 \item[(W2)] Outside any neighbourhood of the corners $\eta_k$, the weights $\sigma_+$ and $\sigma_-$ are comparable.
 \item[(W3)] $1/\sigma$ and $(1+1/r)^2/\sigma$ are integrable on $(-1,1)$: at the trailing edge the second function behaves like $(1-\eta)^{-1-e_{\rm te}}$ with $e_{\rm te}<0$, at the leading edge both are bounded by a multiple of $(1+\eta)^{-e_{\rm le}}$ with $e_{\rm le}<1$, and at the corners like $|\eta-\eta_k|^{-e_k}$ with $e_k<1$.
 \item[(W4)] $\HH$ and $r\,\HH[\,\cdot\,/r]$ are bounded on $L^2(\sigma)$, by~(W1) and~\cite[Proposition~2.2]{AstalaPaivarintaSaksman1996}; $\HH$ is also bounded on $L^2_{\sigma_0}$, since $1/r$ and $1/w$ are power weights with exponents $\pm\tfrac12$.
\end{enumerate}
These are the places where the conditions of the theorem are used: $|\phi_{\rm d}(1)|<\pi/2$ is equivalent to $-1<e_{\rm te}<0$; in part~(ii), $|\phi_{\rm d}(-1)|<\pi/2$ is equivalent to $0<e_{\rm le}<1$; and~\eqref{stab:corner-angles} is equivalent to $-1<e_k<1$ for both faces. In part~(iii), $\phi_{\rm d}(-1)=\pi$ would give $e_{\rm le}=\tfrac32$ with $\sigma_0=r$, which is why the space $Y$ is used for a rounded leading edge.

\subsubsection{Step 4: solution of the flat-plate system}
The operator $\mathsf N$ of a flat plate, $\lambda=g_t=0$, is $\mathsf N_0=\tfrac b2\bigl(\begin{smallmatrix}-\HH&1\\-\HH&-1\end{smallmatrix}\bigr)$; in the notation~\eqref{wp:projections} its two components are $-b\,\Im[\mathsf P_\pm\widetilde f]$.

\begin{lemma}[Flat-plate system]\label{lem:flat-plate}
Let $\widetilde h^\pm\in L^2(\sigma_\pm)$ be real. Then there is exactly one $\widetilde f=\widetilde d-i\widetilde\ell\in L^2(\sigma_{\min})$ with
\begin{equation}
 -b\,\Im\bigl[\mathsf P_\pm\widetilde f\bigr]=-\tfrac b2\,\HH\widetilde d\pm\tfrac b2\,\widetilde\ell=\widetilde h^\pm ,
 \label{wp:flat-system}
\end{equation}
namely
\begin{equation}
 \widetilde\ell=\frac{\widetilde h^+-\widetilde h^-}{b},\qquad
 \widetilde d=r\,\HH\Bigl[\frac{\widetilde h^++\widetilde h^-}{b\,r}\Bigr],
 \label{wp:flat-inverse}
\end{equation}
and it satisfies $\mathsf P_\pm\widetilde f\in L^2(\sigma_\pm)$ with
\begin{equation}
 \norm{\mathsf P_+\widetilde f}_{L^2(\sigma_+)}+\norm{\mathsf P_-\widetilde f}_{L^2(\sigma_-)}\le C\Bigl(\norm{\widetilde h^+}_{L^2(\sigma_+)}+\norm{\widetilde h^-}_{L^2(\sigma_-)}\Bigr).
 \label{wp:flat-bound}
\end{equation}
\end{lemma}
\begin{proof}
\emph{Uniqueness.} Since $\sigma_{\min}\le\sigma_\pm$, the data belong to $L^2(\sigma_{\min})$. The difference and the sum of the two equations~\eqref{wp:flat-system} are $\widetilde\ell=(\widetilde h^+-\widetilde h^-)/b$ and $\HH\widetilde d=-(\widetilde h^++\widetilde h^-)/b$. By~(W1) and~\cite[Theorem~3.1]{AstalaPaivarintaSaksman1996}, applied with $p=2$ and the weight $\sigma_{\min}$, the finite Hilbert transform is an isomorphism of $L^2(\sigma_{\min})$ onto itself, with the inverse $\HH^{-1}=-r\,\HH[\,\cdot\,/r]$. Hence~\eqref{wp:flat-inverse} is the only solution in $L^2(\sigma_{\min})$.

\emph{The bound.} The imaginary parts of $\mathsf P_\pm\widetilde f$ are $-\widetilde h^\pm/b$. For the real parts, $\Re[\mathsf P_\pm\widetilde f]=\pm\tfrac12\widetilde d+\tfrac12\HH\widetilde\ell$, and~\eqref{wp:flat-inverse} gives
\[
 \Re\bigl[\mathsf P_\pm\widetilde f\bigr]=\pm\frac1{2b}\Bigl(\mathsf Q_1\widetilde h^\pm+\mathsf Q_2\widetilde h^\mp\Bigr),\qquad
 \mathsf Q_1=r\,\HH\Bigl[\frac{\,\cdot\,}{r}\Bigr]+\HH,\qquad
 \mathsf Q_2=r\,\HH\Bigl[\frac{\,\cdot\,}{r}\Bigr]-\HH .
\]
The operator $\mathsf Q_1$ is bounded on $L^2(\sigma_\pm)$ by~(W4). The operator $\mathsf Q_2$ couples the two faces. It is an integral operator whose kernel has no singularity on the diagonal,
\[
 (\mathsf Q_2h)(y)=\frac1\pi\int_{-1}^1\frac{r(y)-r(t)}{r(t)\,(y-t)}\,h(t)\,\dd t ,
\]
and it is bounded from $L^2(\sigma_-)$ to $L^2(\sigma_+)$ and from $L^2(\sigma_+)$ to $L^2(\sigma_-)$. To see this for $h\in L^2(\sigma_-)$, let $\mathcal U$ be the union of the intervals $|y-\eta_k|<\varepsilon$, with $\varepsilon$ so small that the intervals $|y-\eta_k|<2\varepsilon$ are disjoint and at a positive distance from $\pm1$. Outside $\mathcal U$, $\sigma_+\le C\sigma_-$ by~(W2), and $\mathsf Q_2$ is bounded on $L^2(\sigma_-)$ by~(W4), so that $\int_{(-1,1)\setminus\mathcal U}|\mathsf Q_2h|^2\sigma_+\,\dd y\le C\norm{h}_{L^2(\sigma_-)}^2$. For $y\in\mathcal U$ the kernel is bounded by $C(1+1/r(t))$ for all $t$: for $|y-t|\ge\varepsilon$ because $r(y)$ is bounded on $\mathcal U$, and for $|y-t|<\varepsilon$ because $r$ is smooth and $1/r$ is bounded on the intervals $|t-\eta_k|<2\varepsilon$. By the Cauchy--Schwarz inequality and~(W3),
\[
 |(\mathsf Q_2h)(y)|\le C\int_{-1}^1\Bigl(1+\frac1r\Bigr)|h|\,\dd t\le C\Bigl(\int_{-1}^1\Bigl(1+\frac1r\Bigr)^2\frac{\dd t}{\sigma_-}\Bigr)^{1/2}\norm{h}_{L^2(\sigma_-)},\qquad y\in\mathcal U,
\]
and $\sigma_+$ is integrable on $\mathcal U$. The other direction is the same. This proves~\eqref{wp:flat-bound}.
\end{proof}

\subsubsection{Step 5: division and multiplication by the canonical function}
Define, for complex functions on the chord,
\begin{equation}
 \mathsf G f=\frac{\mathsf P_+f}{\Xi_+}-\frac{\mathsf P_-f}{\Xi_-},\qquad
 \mathsf G^\sharp\widetilde f=\Xi_+\,\mathsf P_+\widetilde f-\Xi_-\,\mathsf P_-\widetilde f
 =\tfrac12(\Xi_++\Xi_-)\,\widetilde f+\tfrac i2(\Xi_+-\Xi_-)\,\HH\widetilde f .
 \label{wp:G}
\end{equation}
These are the jumps across the chord of the Cauchy integral of $f$ divided by $\Xi$, and of the Cauchy integral of $\widetilde f$ multiplied by $\Xi$.

\begin{lemma}[Division by $\Xi$]\label{lem:division}
\emph{(1)} For every $f\in L^2_{\sigma_0}$, the function $\mathsf Gf$ belongs to $L^2(\sigma_{\min})$, and
\begin{equation}
 \mathsf P_\pm(\mathsf G f)=\frac{\mathsf P_\pm f}{\Xi_\pm}\in L^2(\sigma_\pm),\qquad
 \norm{\mathsf P_\pm f/\Xi_\pm}_{L^2(\sigma_\pm)}=\norm{\mathsf P_\pm f}_{L^2_{\sigma_0}}\le C\norm{f}_{L^2_{\sigma_0}} .
 \label{wp:division}
\end{equation}
\emph{(2)} For every $\widetilde f$ given by Lemma~\ref{lem:flat-plate}, the function $\mathsf G^\sharp\widetilde f$ belongs to $L^2_{\sigma_0}$, and
\begin{equation}
 \mathsf P_\pm(\mathsf G^\sharp\widetilde f)=\Xi_\pm\,\mathsf P_\pm\widetilde f,\qquad
 \norm{\mathsf G^\sharp\widetilde f}_{L^2_{\sigma_0}}\le\norm{\mathsf P_+\widetilde f}_{L^2(\sigma_+)}+\norm{\mathsf P_-\widetilde f}_{L^2(\sigma_-)} .
 \label{wp:multiplication}
\end{equation}
\end{lemma}
\begin{proof}
The norm identities follow from $|\Xi_\pm|^2=\sigma_\pm\,\sigma_0$, and the last inequality in~\eqref{wp:division} from~(W4). It remains to prove the two relations for $\mathsf P_\pm$.

\emph{(1) for smooth functions.} Let $f$ be infinitely differentiable with compact support in the chord, vanishing near every special point, and let $\mathcal V$ be its Cauchy integral~\eqref{app:cauchy-integral}. Then $\mathcal V$ is bounded, it is analytic outside the support of $f$, it vanishes at infinity, and its boundary values $\mathsf P_\pm f$ are H\"older continuous. The function $\mathcal V/\Xi$ is analytic off the chord and vanishes at infinity; its boundary values $\mathsf P_\pm f/\Xi_\pm$ are H\"older continuous on closed intervals without special points; and it is $O(|\zeta-c|^{-1/2})$ near each special point $c$, by property~(c). Its jump across the chord is $\mathsf Gf$. Near each special point $f=0$, so that $\mathsf Gf=\tfrac i2\HH f\,(1/\Xi_+-1/\Xi_-)$ there, which by~\eqref{wp:boundary-values} and~\eqref{wp:log-behaviour} is, on each side of the special point, a power of the distance to it, with exponent of modulus at most $\tfrac12$, multiplied by a H\"older-continuous function. The Cauchy integral of $\mathsf Gf$ therefore exists, has the jump $\mathsf Gf$, and is $o(|\zeta-c|^{-1})$ near each special point~\cite[Chapter~4]{Muskhelishvili1953}. The difference between $\mathcal V/\Xi$ and this Cauchy integral has no jump across the chord away from the special points, so that it is analytic there; its singularities at the special points are isolated and of order less than one, hence removable; and it vanishes at infinity. By Liouville's theorem it is zero. Thus $\mathcal V/\Xi$ is the Cauchy integral of $\mathsf Gf$, and its boundary values give $\mathsf P_\pm(\mathsf Gf)=\mathsf P_\pm f/\Xi_\pm$.

\emph{(1) in general.} The functions used above are dense in $L^2_{\sigma_0}$. Since $\sigma_{\min}\le\sigma_\pm$, the maps $f\mapsto\mathsf P_\pm f/\Xi_\pm$ are bounded from $L^2_{\sigma_0}$ to $L^2(\sigma_{\min})$; hence so is $\mathsf G$, and so are $f\mapsto\mathsf P_\pm(\mathsf Gf)$, because $\HH$ is bounded on $L^2(\sigma_{\min})$ by~(W4). Both sides of the relation are thus bounded operators from $L^2_{\sigma_0}$ to $L^2(\sigma_{\min})$ that agree on a dense subspace.

\emph{(2) for smooth data.} Let $\widetilde h^\pm$ be infinitely differentiable with compact support in the chord, vanishing near every special point, and let $\widetilde f$ be given by~\eqref{wp:flat-inverse}. Then $\widetilde\ell$ has the same properties as the data, and $\widetilde d=r$ times an infinitely differentiable function on $(-1,1)$ that is analytic near $\pm1$. The Cauchy integral $\widetilde{\mathcal V}$ of $\widetilde f$ is bounded near the trailing edge and near the corners, and $O(|\zeta+1|^{-1/2})$ near the leading edge. By property~(c) and~\eqref{wp:moduli}, the product $\widetilde{\mathcal V}\,\Xi$ is $o(|\zeta-c|^{-1})$ near each special point: at the leading edge the exponent is $-\tfrac12+\phi_{\rm d}(-1)/(2\pi)$, which is larger than $-\tfrac34$ in part~(ii) and equal to zero in part~(iii). Its jump across the chord is $\mathsf G^\sharp\widetilde f$, which near each special point is a power of the distance multiplied by a H\"older-continuous function. The argument of~(1) shows that $\widetilde{\mathcal V}\,\Xi$ is the Cauchy integral of $\mathsf G^\sharp\widetilde f$, so that $\mathsf P_\pm(\mathsf G^\sharp\widetilde f)=\Xi_\pm\mathsf P_\pm\widetilde f$.

\emph{(2) in general.} The data used above are dense in $L^2(\sigma_+)\times L^2(\sigma_-)$. By~\eqref{wp:flat-bound} and the norm identity, the maps $(\widetilde h^+,\widetilde h^-)\mapsto\Xi_\pm\mathsf P_\pm\widetilde f$ are bounded from this space to $L^2_{\sigma_0}$; hence so are $(\widetilde h^+,\widetilde h^-)\mapsto\mathsf G^\sharp\widetilde f$ and, by~(W4), $(\widetilde h^+,\widetilde h^-)\mapsto\mathsf P_\pm(\mathsf G^\sharp\widetilde f)$. Both sides of the relation are bounded operators that agree on a dense subspace.
\end{proof}

\subsubsection{Step 6: inversion of the system}
\begin{proof}[Proof of Proposition~\ref{prop:N-isomorphism}]
\emph{(a) Reduction to the flat plate.} Let $f=d-i\ell\in L^2_{\sigma_0}$ and $\widetilde f=\mathsf Gf=\widetilde d-i\widetilde\ell$. By~\eqref{wp:rows},~\eqref{wp:division} and~\eqref{wp:imaginary},
\begin{equation}
 h^\pm=\Re\bigl[c_\pm\,\Xi_\pm\,\mathsf P_\pm\widetilde f\bigr]
 =-\frac{b\,|\Xi_\pm|}{\cos\phi^\pm}\,\Im\bigl[\mathsf P_\pm\widetilde f\bigr]
 =\frac{|\Xi_\pm|}{\cos\phi^\pm}\Bigl(-\tfrac b2\,\HH\widetilde d\pm\tfrac b2\,\widetilde\ell\Bigr),
 \label{wp:reduction}
\end{equation}
that is,
\begin{equation}
 \mathsf N=\begin{pmatrix}\dfrac{|\Xi_+|}{\cos\phi^+}&0\\[8pt]0&\dfrac{|\Xi_-|}{\cos\phi^-}\end{pmatrix}\,\mathsf N_0\,\mathsf G,\qquad
 \mathsf N_0=\frac b2\begin{pmatrix}-\HH&1\\-\HH&-1\end{pmatrix},
 \label{wp:factorization}
\end{equation}
where $\mathsf G$ acts on $d-i\ell$ and $\mathsf N_0$ on the pair $(\widetilde d,\widetilde\ell)$. The geometry enters through the multiplications by $|\Xi_\pm|/\cos\phi^\pm$ and through $\mathsf G$, which contains only $\Xi_\pm$ and $\HH$.

\emph{(b) Boundedness.} By~\eqref{wp:rows}, $|h^\pm|\le|c_\pm|\,|\mathsf P_\pm f|=b\,|\mathsf P_\pm f|/\cos\phi^\pm$, so that $\norm{h^\pm}_{L^2(\cos^2\phi^\pm/\sigma_0)}\le b\norm{\mathsf P_\pm f}_{L^2_{\sigma_0}}\le C\norm{f}_{L^2_{\sigma_0}}$ by~(W4).

\emph{(c) Injectivity.} Let $\mathsf N(d,\ell)^{\mathsf T}=0$. By~\eqref{wp:reduction}, $\widetilde f=\mathsf Gf\in L^2(\sigma_{\min})$ satisfies~\eqref{wp:flat-system} with zero data, so that $\widetilde f=0$ by Lemma~\ref{lem:flat-plate}. Then $\mathsf P_\pm f=\Xi_\pm\mathsf P_\pm\widetilde f=0$ by~\eqref{wp:division}, and $f=\mathsf P_+f-\mathsf P_-f=0$.

\emph{(d) Surjectivity and the inverse.} Let $h^\pm\in L^2(\cos^2\phi^\pm/\sigma_0)$ and put
\begin{equation}
 \widetilde h^\pm=\frac{\cos\phi^\pm}{|\Xi_\pm|}\,h^\pm,\qquad\text{so that}\quad
 \norm{\widetilde h^\pm}_{L^2(\sigma_\pm)}=\norm{h^\pm}_{L^2(\cos^2\phi^\pm/\sigma_0)} .
 \label{wp:data-tilde}
\end{equation}
Let $\widetilde f=\widetilde d-i\widetilde\ell$ be given by Lemma~\ref{lem:flat-plate}, and define
\begin{equation}
 d-i\ell=\mathsf G^\sharp\bigl(\widetilde d-i\widetilde\ell\bigr),
 \label{wp:solution}
\end{equation}
with $\widetilde d$, $\widetilde\ell$ from~\eqref{wp:flat-inverse} and $\mathsf G^\sharp$ from~\eqref{wp:G}. By Lemma~\ref{lem:division} and~\eqref{wp:flat-bound}, $f=d-i\ell\in L^2_{\sigma_0}$ with $\norm{f}_{L^2_{\sigma_0}}\le C(\norm{h^+}_{L^2(\cos^2\phi^+/\sigma_0)}+\norm{h^-}_{L^2(\cos^2\phi^-/\sigma_0)})$, and $\mathsf P_\pm f=\Xi_\pm\mathsf P_\pm\widetilde f$. Therefore, by~\eqref{wp:imaginary} and~\eqref{wp:flat-system},
\[
 \Re\bigl[c_\pm\,\mathsf P_\pm f\bigr]=-\frac{b\,|\Xi_\pm|}{\cos\phi^\pm}\,\Im\bigl[\mathsf P_\pm\widetilde f\bigr]=\frac{|\Xi_\pm|}{\cos\phi^\pm}\,\widetilde h^\pm=h^\pm ,
\]
that is, $\mathsf N(d,\ell)^{\mathsf T}=(h^+,h^-)^{\mathsf T}$. Thus $\mathsf N$ is a bounded bijection with the bounded inverse~\eqref{wp:solution}.
\end{proof}

\begin{proof}[Proof of Theorem~\ref{thm:existence}, parts~(ii) and~(iii)]
\emph{Part~(ii).} The slopes are bounded, so that $\cos\phi^\pm$ is bounded below by a positive constant and $L^2(\cos^2\phi^\pm/\sigma_0)=X$ with equivalent norms. By Proposition~\ref{prop:N-isomorphism}, $\mathsf N$ is an isomorphism of $X\times X$ onto itself, and so is $\mathsf A=\mathsf S\,\mathsf N$.

\emph{Part~(iii).} Here $\cos\phi^\pm=b/\sqrt{(\lambda\mp g_t)^2+b^2}$, and the data~\eqref{wp:data} grow like $g_t$ at the leading edge, so that $\cos\phi^\pm\,h^\pm$ is bounded on the chord. Since $1/w$ is integrable, $h^\pm\in L^2(\cos^2\phi^\pm/\sigma_0)$, and Proposition~\ref{prop:N-isomorphism} gives exactly one solution in $Y\times Y$.
\end{proof}

\emph{The Kutta condition.} The solution~\eqref{wp:solution} belongs to $L^2_{\sigma_0}$, whose weight excludes an inverse-square-root singularity at the trailing edge. In the construction this is the choice of the inverse $-r\,\HH[\,\cdot\,/r]$ of the finite Hilbert transform in Lemma~\ref{lem:flat-plate}, the inverse that vanishes at the trailing edge.

\emph{Scope.} The proof gives no explicit constants, because the weights $\sigma_\pm$ are general power weights. For rounded sections without slope discontinuities the weights are comparable to that of $X$, on which $\HH$ is unitary, and explicit bounds follow from the same canonical function in Section~\ref{app:rh-bounds}; they are used in the error bounds of Section~\ref{sec:geometry-stability}. For slopes that are only bounded, the moduli $|\Xi_\pm|$ are no longer comparable to power weights, and the properties~(W1) to~(W4) would have to be established in another way; small bounded slopes are covered by part~(i).

\subsection{Explicit bounds for rounded sections}\label{app:rh-bounds}
The error bounds of Section~\ref{sec:geometry-stability} use explicit constants for rounded sections without slope discontinuities. They are obtained here from three facts that are already available: the canonical function of Section~\ref{app:proof-geometry-stability}, the unitarity of $\HH$ on $X$ (Lemma~\ref{lem:H-unitary}), and an isometry property of $\HH$ on $Y$. Existence and uniqueness are taken from Theorem~\ref{thm:existence}; this subsection only bounds the solution.

\begin{lemma}[The finite Hilbert transform on $Y$]\label{lem:H-Y}
For every complex $f\in Y$,
\begin{equation}
 \norm{\HH f}_Y=\norm{f}_Y,\qquad
 \norm{\mathsf P_+f}_Y^2+\norm{\mathsf P_-f}_Y^2=\norm{f}_Y^2 .
 \label{app:Y-isometry}
\end{equation}
\end{lemma}
\begin{proof}
Let $\mathsf C_j$ and $\mathsf U_j$ be the Chebyshev polynomials of the first and of the second kind. The functions $w\,\mathsf U_{j-1}$, $j\ge1$, form an orthogonal basis of $Y$ with $\norm{w\,\mathsf U_{j-1}}_Y^2=\int_{-1}^1w\,\mathsf U_{j-1}^2\,\dd\eta=\pi/2$; the functions $\mathsf C_j$, $j\ge1$, are orthogonal in $Y$ with $\norm{\mathsf C_j}_Y^2=\int_{-1}^1\mathsf C_j^2\,\dd\eta/w=\pi/2$; and $\HH[w\,\mathsf U_{j-1}]=\mathsf C_j$~\cite{MasonHandscomb2003}. Hence $\norm{\HH f}_Y=\norm{f}_Y$ for finite linear combinations of the functions $w\,\mathsf U_{j-1}$, and for all $f\in Y$ because $\HH$ is bounded on $Y$ by~(W4). The second identity follows from the first and from the pointwise identity $|\mathsf P_+f|^2+|\mathsf P_-f|^2=\tfrac12|f|^2+\tfrac12|\HH f|^2$, which holds by~\eqref{wp:projections}.
\end{proof}

By~\eqref{wp:rows} and~\eqref{wp:angles}, $c_\pm=i\,b\,e^{i\phi^\pm}/\cos\phi^\pm$, so that the two rows of the system are
\begin{equation}
 \Im\bigl[e^{i\phi^\pm}\,\mathsf P_\pm f\bigr]=-\frac{\cos\phi^\pm}{b}\,h^\pm,\qquad
 \frac{\cos\phi^\pm}{b}=\frac1{\sqrt{(\lambda\mp g_t)^2+b^2}} ,
 \label{app:normalized-rows}
\end{equation}
with $f=d-i\ell$. The next lemma bounds $f$ by the left-hand sides. It involves two pairs of angles: the angles $\phi^\pm$ of the system, which need not be continuous, and reference angles $\phi_\circ^\pm$, from which the canonical function is built.

\begin{lemma}[A priori bound for rounded sections]\label{lem:rounded-bounds}
Let $\phi_\circ^\pm$ be H\"older continuous on $[-1,1]$ with
\begin{equation}
 \phi_\circ^\pm(-1)=\pm\frac\pi2,\qquad\phi_\circ^+(1)=\phi_\circ^-(1),
 \label{app:reference-conditions}
\end{equation}
let $\Xi_\circ$ be the canonical function~\eqref{wp:canonical} formed with $\phi_\circ^\pm$ in place of $\phi^\pm$, and let
\begin{equation}
 \sigma_{\circ\pm}=\frac{|\Xi_{\circ\pm}|^2}{w}
 \label{app:modulus-ratio}
\end{equation}
be the corresponding weights~\eqref{wp:weight} with $\sigma_0=w$. Then $r\sigma_{\circ\pm}$ is bounded above and below by positive constants on the chord. Put
\begin{equation}
 \mathcal N_\circ=2\,\Bigl(\frac{\sup r\sigma_\circ}{\inf r\sigma_\circ}\Bigr)^{1/2},
 \label{app:RH-bound}
\end{equation}
with the supremum and the infimum taken over the chord and over both faces. Let $\phi^\pm$ be measurable functions on the chord with
\begin{equation}
 \varDelta_\circ=\max_\pm\,\sup_{-1<\eta<1}\bigl|2\bigl(\phi^\pm-\phi_\circ^\pm\bigr)\bigr|<\frac\pi2 .
 \label{app:reference-distance}
\end{equation}
Then every complex $f\in Y$ satisfies
\begin{equation}
 \norm{f}_Y\le\frac{\mathcal N_\circ}{\cos\varDelta_\circ}\Bigl(\sum_\pm\bigl\|\Im\bigl[e^{i\phi^\pm}\,\mathsf P_\pm f\bigr]\bigr\|_Y^2\Bigr)^{1/2}.
 \label{app:rounded-stability}
\end{equation}
\end{lemma}
\begin{proof}
\emph{Step 1: the weights are those of $X$.} By~\eqref{app:reference-conditions}, $\phi_{\circ\rm d}=\phi_\circ^+-\phi_\circ^-$ equals $\pi$ at the leading edge and $0$ at the trailing edge, and there are no corners. By~\eqref{wp:log-behaviour}, $\HH\phi_{\circ\rm d}=\ln(1+\eta)$ plus a bounded function, and $w\,\HH[(\phi_\circ^++\phi_\circ^-)/w]$ is bounded. Hence $r\sigma_{\circ\pm}=|\Xi_{\circ\pm}|^2/(1+\eta)$ is bounded above and below, and
\begin{equation}
 \sigma_{\circ\pm}=\frac{r\sigma_{\circ\pm}}{r}
 \label{app:weights-X}
\end{equation}
is comparable to $1/r$, the weight of $X$.

\emph{Step 2: division by the canonical function.} The angles $\phi_\circ^\pm$ have the properties of the angles of part~(iii) of Theorem~\ref{thm:existence} that are used in the construction of the canonical function and in Lemma~\ref{lem:division}: they are H\"older continuous, $\phi_{\circ\rm d}(-1)=\pi$, $\phi_\circ^+(-1)+\phi_\circ^-(-1)=0$ and $|\phi_{\circ\rm d}(1)|<\pi/2$. Lemma~\ref{lem:division} therefore holds for $\Xi_\circ$. Let $f\in Y$, and let $\widetilde f=\mathsf Gf$ be formed with $\Xi_\circ$. By~\eqref{wp:division} and~\eqref{app:weights-X}, $\widetilde f\in X$ and
\begin{equation}
 \widetilde f_\pm:=\mathsf P_\pm\widetilde f=\frac{\mathsf P_\pm f}{\Xi_{\circ\pm}}\in X,\qquad
 \frac{|\mathsf P_\pm f|^2}{w}=\sigma_{\circ\pm}\,|\widetilde f_\pm|^2 .
 \label{app:divided}
\end{equation}

\emph{Step 3: an identity from the unitarity of $\HH$ on $X$.} By Lemma~\ref{lem:H-unitary}, $\int_{-1}^1(\HH u)(\HH v)\,\dd\eta/r=\int_{-1}^1u\,v\,\dd\eta/r$ for real $u,v\in X$, and therefore $\int_{-1}^1(\HH\widetilde f)^2\,\dd\eta/r=\int_{-1}^1\widetilde f^{\,2}\,\dd\eta/r$ for complex $\widetilde f\in X$; here the squares are those of complex numbers, without moduli. Since $\widetilde f_+^2+\widetilde f_-^2=\tfrac12\widetilde f^{\,2}-\tfrac12(\HH\widetilde f)^2$ by~\eqref{wp:projections},
\begin{equation}
 \int_{-1}^1\bigl(\widetilde f_+^2+\widetilde f_-^2\bigr)\frac{\dd\eta}{r}=0 .
 \label{app:energy-identity}
\end{equation}

\emph{Step 4: the estimate.} Put $\varepsilon_\pm=2(\phi^\pm-\phi_\circ^\pm)$. Since $\Xi_{\circ\pm}=|\Xi_{\circ\pm}|\,e^{-i\phi_\circ^\pm}$,
\[
 \Im\bigl[e^{i\varepsilon_\pm/2}\,\widetilde f_\pm\bigr]=\frac{\Im\bigl[e^{i\phi^\pm}\,\mathsf P_\pm f\bigr]}{|\Xi_{\circ\pm}|} .
\]
Every complex number $\zeta$ satisfies $\zeta^2=|\zeta|^2+2i\,(\Im\zeta)\,\zeta$. With $\zeta=e^{i\varepsilon_\pm/2}\widetilde f_\pm$ this gives
\[
 \widetilde f_\pm^{\,2}=e^{-i\varepsilon_\pm}|\widetilde f_\pm|^2+2i\,e^{-i\varepsilon_\pm/2}\,\widetilde f_\pm\,\frac{\Im\bigl[e^{i\phi^\pm}\,\mathsf P_\pm f\bigr]}{|\Xi_{\circ\pm}|},
\]
and the real part of~\eqref{app:energy-identity} is
\[
 \sum_\pm\int_{-1}^1\cos\varepsilon_\pm\,|\widetilde f_\pm|^2\,\frac{\dd\eta}{r}
 =-\Re\sum_\pm\int_{-1}^1 2i\,e^{-i\varepsilon_\pm/2}\,\widetilde f_\pm\,\frac{\Im\bigl[e^{i\phi^\pm}\,\mathsf P_\pm f\bigr]}{|\Xi_{\circ\pm}|}\,\frac{\dd\eta}{r} .
\]
On the left $\cos\varepsilon_\pm\ge\cos\varDelta_\circ>0$, and the Cauchy--Schwarz inequality on the right gives
\[
 \Bigl(\sum_\pm\int_{-1}^1|\widetilde f_\pm|^2\,\frac{\dd\eta}{r}\Bigr)^{1/2}
 \le\frac{2}{\cos\varDelta_\circ}\Bigl(\sum_\pm\int_{-1}^1\frac{\bigl(\Im\bigl[e^{i\phi^\pm}\,\mathsf P_\pm f\bigr]\bigr)^2}{|\Xi_{\circ\pm}|^2}\,\frac{\dd\eta}{r}\Bigr)^{1/2}.
\]
By~\eqref{app:Y-isometry} and~\eqref{app:divided}, $\norm{f}_Y^2=\sum_\pm\int\sigma_{\circ\pm}|\widetilde f_\pm|^2\,\dd\eta\le\sup r\sigma_\circ\,\sum_\pm\int|\widetilde f_\pm|^2\,\dd\eta/r$. Since $|\Xi_{\circ\pm}|^2\,r=(r\sigma_{\circ\pm})\,w$, the right-hand side is at most $(2/\cos\varDelta_\circ)\bigl(\sum_\pm\norm{\Im[e^{i\phi^\pm}\mathsf P_\pm f]}_Y^2/\inf r\sigma_\circ\bigr)^{1/2}$. The three inequalities give~\eqref{app:rounded-stability}.
\end{proof}

The lemma separates the two ways in which the angles enter. The H\"older-continuous reference angles are removed by the canonical function, at the price of the ratio of the largest to the smallest value of its modulus on the chord, once the factor $\sqrt{1+\eta}$ of the rounded leading edge has been taken out. What remains of the angles only has to be smaller than $\pi/2$, and it costs the factor $1/\cos\varDelta_\circ$; it may be discontinuous.

\emph{Constants of a target with a common tangent at the trailing edge.} Let the target be of the form~\eqref{stab:rounded-class} with H\"older-continuous $k_c^*$ and $R_1^*$. Its angles $\phi^{*\pm}$ are H\"older continuous and satisfy~\eqref{app:reference-conditions}, so that Lemma~\ref{lem:rounded-bounds} applies with $\phi_\circ^\pm=\phi^\pm=\phi^{*\pm}$ and $\varDelta_\circ=0$. The constants of Corollary~\ref{thm:geometry-stability} are
\begin{equation}
 \mathcal N^*=2\,\Bigl(\frac{\sup r\sigma^*}{\inf r\sigma^*}\Bigr)^{1/2},\qquad
 \underline\varpi^*=\inf_\pm\,\inf_{-1<\eta<1}\sqrt{\frac{1+\eta}2}\,\sqrt{(\lambda^*\mp g_t^*)^2+b^2}>0,
 \label{stab:rounded-constants}
\end{equation}
where $\sigma^*_\pm$ are the weights~\eqref{app:modulus-ratio} formed with the target angles. The second constant is positive because $\sqrt{1+\eta}\,|g_t^*|$ tends to a positive limit at the leading edge and the second square root is at least $b$. Since $\cos\phi^{*\pm}/b=1/|c^*_\pm|$, the definition of $\underline\varpi^*$ gives
\begin{equation}
 \frac{\cos\phi^{*\pm}}{b}\le\frac1{\underline\varpi^*}\sqrt{\frac{1+\eta}2}.
 \label{app:two-lower-bounds}
\end{equation}
For pairs of functions put $\norm{(F,G)}_X=\bigl(\norm{F}_X^2+\norm{G}_X^2\bigr)^{1/2}$. Since $1/r=(1+\eta)/w\le2/w$, for every function $v$ and every $u\in Y$,
\begin{equation}
 \norm{u}_X\le\sqrt2\,\norm{u}_Y,\qquad
 \norm{v\,u}_X\le\norm{\sqrt{1+\eta}\,v}_\infty\norm{u}_Y .
 \label{app:weighted-product}
\end{equation}
Let $\mathsf A^*$ be the operator~\eqref{stab:matrix-form} of the target.

\begin{lemma}[Inverse of the target operator]\label{lem:target-inverse}
For every $(d,\ell)\in Y\times Y$, $\mathsf A^*(d,\ell)^{\mathsf T}\in X\times X$ and
\begin{equation}
 \norm{(d,\ell)}_Y\le\frac{\mathcal N^*}{2\,\underline\varpi^*}\,\bigl\|\mathsf A^*(d,\ell)^{\mathsf T}\bigr\|_X .
 \label{app:target-inverse}
\end{equation}
In particular, the solution of the target satisfies
\begin{equation}
 \norm{(d^*,\ell^*)}_Y\le\frac{\sqrt\pi\,\mathcal N^*}{\underline\varpi^*}\Bigl[U|\sin\alpha|+\bigl(\norm{k_c^*}_\infty+\norm{\sqrt{1+\eta}\,g_t^*}_\infty\bigr)U|\cos\alpha|\Bigr].
 \label{stab:rounded-apriori-solution}
\end{equation}
\end{lemma}
\begin{proof}
The entries of $\mathsf A^*(d,\ell)^{\mathsf T}$ are $b\HH d-\lambda^*\HH\ell+g_t^*d$ and $\lambda^*d+b\ell-g_t^*\HH\ell$. By Lemma~\ref{lem:H-Y} and~\eqref{app:weighted-product}, each term belongs to $X$, since $\lambda^*$ and $\sqrt{1+\eta}\,g_t^*$ are bounded. Let $f=d-i\ell$ and $h^\pm=\Re[c^*_\pm\mathsf P_\pm f]$. By~\eqref{wp:rows}, $\mathsf A^*(d,\ell)^{\mathsf T}=\mathsf S(h^+,h^-)^{\mathsf T}$, and $\mathsf S^{\mathsf T}\mathsf S=2\mathsf I$, so that $\norm{h^+}_X^2+\norm{h^-}_X^2=\tfrac12\norm{\mathsf A^*(d,\ell)^{\mathsf T}}_X^2$. The identity $c^*_\pm=i\,b\,e^{i\phi^{*\pm}}/\cos\phi^{*\pm}$ gives, for every $f$, $\Im[e^{i\phi^{*\pm}}\mathsf P_\pm f]=-(\cos\phi^{*\pm}/b)\,h^\pm$, as in~\eqref{app:normalized-rows}. By~\eqref{app:two-lower-bounds},
\[
 \Bigl\|\frac{\cos\phi^{*\pm}}{b}\,h^\pm\Bigr\|_Y^2\le\frac1{2(\underline\varpi^*)^2}\int_{-1}^1(1+\eta)\,(h^\pm)^2\,\frac{\dd\eta}{w}=\frac{\norm{h^\pm}_X^2}{2(\underline\varpi^*)^2}.
\]
Lemma~\ref{lem:rounded-bounds}, with $\phi_\circ^\pm=\phi^\pm=\phi^{*\pm}$ and $\varDelta_\circ=0$, gives
\[
 \norm{f}_Y\le\mathcal N^*\Bigl(\sum_\pm\Bigl\|\frac{\cos\phi^{*\pm}}{b}\,h^\pm\Bigr\|_Y^2\Bigr)^{1/2}
 \le\frac{\mathcal N^*}{\sqrt2\,\underline\varpi^*}\bigl(\norm{h^+}_X^2+\norm{h^-}_X^2\bigr)^{1/2}
 =\frac{\mathcal N^*}{2\,\underline\varpi^*}\bigl\|\mathsf A^*(d,\ell)^{\mathsf T}\bigr\|_X ,
\]
which is~\eqref{app:target-inverse}. For the target solution, $\mathsf A^*(d^*,\ell^*)^{\mathsf T}=(F^*,G^*)$ with $F^*=2U(\sin\alpha-k_c^*\cos\alpha)$ and $G^*=2g_t^*U\cos\alpha$. Since $\norm{1}_X^2=\norm{1}_Y^2=\pi$, $\norm{F^*}_X\le2\sqrt\pi\,(U|\sin\alpha|+\norm{k_c^*}_\infty U|\cos\alpha|)$ and, by~\eqref{app:weighted-product}, $\norm{G^*}_X\le2\sqrt\pi\,\norm{\sqrt{1+\eta}\,g_t^*}_\infty U|\cos\alpha|$. With $\norm{(F^*,G^*)}_X\le\norm{F^*}_X+\norm{G^*}_X$, \eqref{app:target-inverse} gives~\eqref{stab:rounded-apriori-solution}.
\end{proof}

\subsection{Proof of Corollary~\ref{thm:geometry-stability}}\label{app:proof-corollary}
\begin{recalled}[Geometry-error bounds]{Corollary~\ref{thm:geometry-stability}}
Fix $U>0$, $0\le M<1$ and $\alpha$, and hence $a,b$. The target and the approximating geometry share these flow data.

\emph{(i) Bounded slopes.} Let the target slopes $k_c^*,g_t^*\in L^\infty(-1,1)$ satisfy $\mu^*>0$, with solution $(d^*,\ell^*)$ from part~(i) of Theorem~\ref{thm:existence}. Let $(d,\ell)\in X\times X$ solve~\eqref{stab:system} for other bounded slopes $(k_c,g_t)$, and set $\epsilon_g=\norm{\delta k}_\infty+\norm{\delta g}_\infty$. Then
\begin{equation}
 \norm{d-d^*}_X+\norm{\ell-\ell^*}_X
 \le\frac{\epsilon_g}{\mu^*}
 \Bigl(2\sqrt\pi\,U|\cos\alpha|+\norm{d}_X+\norm{\ell}_X\Bigr).
 \tag{\ref{stab:aposteriori}}
\end{equation}
Moreover $\norm{n-n^*}_X\le|a|\norm{d-d^*}_X+b\norm{\ell-\ell^*}_X$. In particular, the flat-endpoint solutions of Theorem~\ref{thm:flat-family} converge in $X$, at the rate of the slope error, whenever their slopes $(k_c,g_t)$ converge uniformly to a target with $\mu^*>0$.

\emph{(ii) Rounded leading edges.} Let the target be of the form~\eqref{stab:rounded-class} with H\"older-continuous $k_c^*$ and $R_1^*$ on $[-1,1]$, and let the solution be $(d^*,\ell^*)\in Y\times Y$. Then there exist constants $\mathcal N^*>0$ and $\underline\varpi^*>0$, depending only on the target geometry and the flow data, with the following property. Let $(d,\ell)\in Y\times Y$ solve~\eqref{stab:system} for another geometry with bounded $k_c$ and bounded $\sqrt{1+\eta}\,g_t$, and set
\begin{equation}
 \epsilon_r=\norm{\delta k}_\infty+\norm{\sqrt{\tfrac{1+\eta}2}\,\delta g}_\infty.
 \tag{\ref{stab:rounded-error-measure}}
\end{equation}
Then
\begin{equation}
 \norm{(d-d^*,\ell-\ell^*)}_Y\le \frac{\mathcal N^*}{\underline\varpi^*}\,\epsilon_r
 \Bigl(2\sqrt\pi\,U|\cos\alpha|+\norm{d}_Y+\norm{\ell}_Y\Bigr).
 \tag{\ref{stab:rounded-aposteriori}}
\end{equation}
In particular, the rounded-family solutions of Theorem~\ref{thm:rounded-family} converge in $Y$ to the target solution, at the rate of $\epsilon_r$, whenever $\norm{\delta k}_\infty+\norm{\sqrt{1+\eta}\,\delta g}_\infty\to0$.
\end{recalled}

\begin{proof}
Both parts follow from the perturbation theorem for bounded linear operators: if $\mathsf A_0$ has a bounded inverse and $\norm{\mathsf A_0^{-1}\mathsf B}<1$, then $\mathsf A=\mathsf A_0-\mathsf B$ has a bounded inverse with $\norm{\mathsf A^{-1}}\le\norm{\mathsf A_0^{-1}}/(1-\norm{\mathsf A_0^{-1}\mathsf B})$~\cite[Theorem~10.1]{Kress2014}; and, since $\mathsf A^*(\mathsf x-\mathsf x^*)=(\mathsf A^*-\mathsf A)\mathsf x+\mathsf f-\mathsf f^*$, the solutions of $\mathsf A\mathsf x=\mathsf f$ and $\mathsf A^*\mathsf x^*=\mathsf f^*$ satisfy $\norm{\mathsf x-\mathsf x^*}\le\norm{\mathsf A^{*-1}}\,\norm{(\mathsf A^*-\mathsf A)\mathsf x+\mathsf f-\mathsf f^*}$.

\emph{Part~(i).} Let $\mathsf A^*$ and $(F^*,G^*)$ belong to the target, so that $\norm{\mathsf A^{*-1}}\le1/\mu^*$. With $\lambda=\lambda^*-\delta k$, $g_t=g_t^*+\delta g$, $F=F^*-2\delta k\,U\cos\alpha$ and $G=G^*+2\delta g\,U\cos\alpha$,
\[
 (\mathsf A^*-\mathsf A)(d,\ell)+(F,G)-(F^*,G^*)=\bigl(-\delta k\,(2U\cos\alpha+\HH\ell)-\delta g\,d,\;\delta g\,(2U\cos\alpha+\HH\ell)+\delta k\,d\bigr),
\]
whose norm is at most $\epsilon_g(2\sqrt\pi U|\cos\alpha|+\norm{d}_X+\norm{\ell}_X)$. The perturbation estimate gives~\eqref{stab:aposteriori}. With $\mathcal E=\norm{d-d^*}_X+\norm{\ell-\ell^*}_X$ and $\norm{d}_X+\norm{\ell}_X\le\norm{d^*}_X+\norm{\ell^*}_X+\mathcal E$, \eqref{stab:aposteriori} and~\eqref{stab:apriori} for the target show that, for $\epsilon_g<\mu^*$, $\mathcal E$ is bounded by a multiple of $\epsilon_g$ that depends on the target only. The estimate for $n-n^*$ follows from $n=b\ell+ad$. For flat-endpoint approximants, Theorem~\ref{thm:flat-family} supplies $(d,\ell)\in X\times X$ for every admissible polynomial geometry, and uniform slope convergence means $\epsilon_g\to0$; no margin condition on the approximants themselves is needed.

\emph{Part~(ii).} This is the same estimate, with Lemma~\ref{lem:target-inverse} in place of the bound for $\mathsf A^{*-1}$: the difference of the two systems is measured in $X$ and the solution in $Y$. As in part~(i),
\[
 \mathsf A^*\begin{pmatrix}d-d^*\\ \ell-\ell^*\end{pmatrix}=\begin{pmatrix}-\delta k\,(2U\cos\alpha+\HH\ell)-\delta g\,d\\ \delta g\,(2U\cos\alpha+\HH\ell)+\delta k\,d\end{pmatrix}.
\]
Put $\epsilon_k=\norm{\delta k}_\infty$ and $\epsilon_\delta=\norm{\sqrt{(1+\eta)/2}\,\delta g}_\infty$, so that $\epsilon_r=\epsilon_k+\epsilon_\delta$. By~\eqref{app:weighted-product}, $\norm{\delta k\,v}_X\le\sqrt2\,\epsilon_k\norm{v}_Y$ and $\norm{\delta g\,v}_X\le\sqrt2\,\epsilon_\delta\norm{v}_Y$ for $v\in Y$, and $\norm{2U\cos\alpha+\HH\ell}_Y\le2\sqrt\pi\,U|\cos\alpha|+\norm{\ell}_Y$ by Lemma~\ref{lem:H-Y}. The two entries on the right therefore have norms in $X$ at most $\sqrt2(\epsilon_k\norm{2U\cos\alpha+\HH\ell}_Y+\epsilon_\delta\norm{d}_Y)$ and $\sqrt2(\epsilon_\delta\norm{2U\cos\alpha+\HH\ell}_Y+\epsilon_k\norm{d}_Y)$, and the Cauchy--Schwarz inequality bounds the norm of the pair by
\[
 2\bigl(\epsilon_k^2+\epsilon_\delta^2\bigr)^{1/2}\bigl(\norm{2U\cos\alpha+\HH\ell}_Y^2+\norm{d}_Y^2\bigr)^{1/2}
 \le2\,\epsilon_r\bigl(2\sqrt\pi\,U|\cos\alpha|+\norm{d}_Y+\norm{\ell}_Y\bigr).
\]
Lemma~\ref{lem:target-inverse} gives~\eqref{stab:rounded-aposteriori}. With $\mathcal E=\norm{(d-d^*,\ell-\ell^*)}_Y$ and $\norm{d}_Y+\norm{\ell}_Y\le\sqrt2\,(\norm{(d^*,\ell^*)}_Y+\mathcal E)$,
\[
 \Bigl(1-\frac{\sqrt2\,\mathcal N^*\epsilon_r}{\underline\varpi^*}\Bigr)\mathcal E\le\frac{\mathcal N^*\epsilon_r}{\underline\varpi^*}\Bigl(2\sqrt\pi\,U|\cos\alpha|+\sqrt2\,\norm{(d^*,\ell^*)}_Y\Bigr),
\]
and~\eqref{stab:rounded-apriori-solution} bounds the last norm by the target data. Hence, once $\sqrt2\,\mathcal N^*\epsilon_r<\underline\varpi^*$, the error is bounded by a multiple of $\epsilon_r$ that depends on the target only, which gives the convergence statement of part~(ii).

Finally, let $T=wP^*$, where $P^*\in C^1[-1,1]$ has a H\"older-continuous derivative, $P^*(1)=0$ and $P^*(-1)>0$. Then $R^*=wT'=(1-\eta)[(1+\eta)P^{*\prime}-\eta P^*/(1-\eta)]$, and $P^*/(1-\eta)$ is H\"older continuous because $P^*(1)=0$; also $R_1^*(-1)=P^*(-1)/2>0$, so~\eqref{stab:rounded-class} holds. For the approximants $T_N=wP_N$ of Theorem~\ref{thm:rounded-approximation}, with $\delta P_N=P_N-P^*$ and $\delta P_N(1)=0$,
\[
 \sqrt{1+\eta}\,(T_N'-T')=(1+\eta)\sqrt{1-\eta}\,\delta P_N'-\frac{\eta\,\delta P_N}{\sqrt{1-\eta}},
 \qquad|\delta P_N(\eta)|\le(1-\eta)\norm{\delta P_N'}_\infty,
\]
so that $\norm{\sqrt{1+\eta}\,(T_N'-T')}_\infty\le3\sqrt2\,\norm{\delta P_N'}_\infty\to0$.
\end{proof}

\paragraph{Remarks on Corollary~\ref{thm:geometry-stability}.}
In words, Corollary~\ref{thm:geometry-stability} says that a small error in the surface slopes produces a proportionally small error in the solution. This matters in practice because a measured, tabulated or deformed section is never represented exactly: it is fitted, and the fit is what the solver sees. The corollary identifies what the fit must capture. The slopes, not the ordinates, enter the equations, so a fit that matches the ordinates closely but has oscillating slopes can still give a poor solution.

For sections with flat endpoints the approximation results of Section~\ref{sec:families} supply such fits. If the thickness is continuously differentiable with zero slope at the ends, Theorem~\ref{thm:flat-approximation} gives polynomial approximants whose thickness slopes converge uniformly, Remark~\ref{rem:asymmetric-density} does the same for the camber, and the corresponding solutions then converge to the solution for the true section.

For a rounded leading edge the slope is infinite at the nose, so the error is measured with the weight $\sqrt{1+\eta}$, which removes exactly that singularity. The approximation results again supply the fits: if the regular factor $P^*=T/w$ of the target thickness $T$ is smooth enough, with $P^*(1)=0$ and $P^*(-1)>0$, the approximants of Theorem~\ref{thm:rounded-approximation} satisfy $\norm{\sqrt{1+\eta}\,\delta g}_\infty\le3\sqrt2\,\norm{P_N'-P^{*\prime}}_\infty$, which tends to zero. Only the target section needs this regularity; the approximating section enters only through the size of its slope error and of its own solution.

The constants in part~(ii) should not be read as error estimates. They are given explicitly in~\eqref{stab:rounded-constants} and can be computed from the target geometry, but they are large, and they grow as the nose gets sharper, since $\underline\varpi^*\le P^*(-1)/2$. What the corollary guarantees is that the solution depends continuously, indeed in a Lipschitz manner, on the geometry; it does not say how small the error is.

Finally, both parts compare two solutions of the same model equations. They state how much the solution changes when the geometry changes, not how close the model is to the real flow.

\subsection{Proof of Theorem~\ref{thm:te-angle}}\label{app:proof-te-angle}
\begin{recalled}[Approximation of sections with a finite trailing-edge angle]{Theorem~\ref{thm:te-angle}}
Fix $U>0$, $0\le M<1$ and $\alpha$. Let the target geometry $(k_c^*,g_t^*)$ be of the form~\eqref{stab:wedge-class} with H\"older-continuous $k_c^*$ and $S^*$ on $[-1,1]$ and with~\eqref{stab:te-angle}, with slope angles $\phi^{*\pm}$ from~\eqref{stab:angles}, trailing-edge jump $\varDelta_{\rm te}$, and solution $(d^*,\ell^*)\in Y\times Y$, whose chordwise face velocities are $u^{*\pm}$ as in~\eqref{comparison:face-pressure-mapping}. For a second geometry $(k_c,g_t)$ with slope angles $\phi^\pm$, and with $\delta k$ and $\delta g$ as above, define
\begin{equation}
 \norm{\delta}_*=\Biggl(\sum_\pm\int_{-1}^1\frac{|\delta k\pm\delta g|^2}{(a-k_c\mp g_t)^2+b^2}\,\bigl(U\cos\alpha+u^{*\pm}\bigr)^2\,\frac{\dd\eta}{w}\Biggr)^{1/2},
 \tag{\ref{stab:target-weighted-norm}}
\end{equation}
\begin{equation}
 \varDelta_{\rm ang}=\max_\pm\,\sup_{-1<\eta<1}\Bigl|2\bigl(\phi^\pm-\phi^{*\pm}\bigr)\pm\bigl(\phi^{*+}(1)-\phi^{*-}(1)\bigr)\Bigl(1-\frac{\arccos\eta}{\pi}\Bigr)\Bigr| .
 \tag{\ref{stab:angle-distance}}
\end{equation}

\emph{(i) Distance between the solutions.} Let the second geometry have a bounded mean slope $k_c$ and a bounded $\sqrt{1+\eta}\,g_t$, let $(d,\ell)\in Y\times Y$ be a solution of~\eqref{stab:system} for it, and let $\varDelta_{\rm ang}<\pi/2$. Then
\begin{equation}
 \norm{(d-d^*,\ell-\ell^*)}_Y\le\frac{\mathcal N_{\rm c}^*}{\cos\varDelta_{\rm ang}}\,\norm{\delta}_*,
 \tag{\ref{stab:integral-bound}}
\end{equation}
where the constant $\mathcal N_{\rm c}^*$, given by~\eqref{app:target-constant}, depends on the target geometry and the flow data only.

\emph{(ii) Approximation by the rounded-leading-edge family.} Write the target thickness as $T^*=(f^+-f^-)/c=wP^*$, and suppose that $P^*\in C^1(-1,1)$ is continuous on $[-1,1]$ with $P^*(1)=0$ and $P^*(-1)>0$, and that $wP^{*\prime}$ extends continuously to $[-1,1]$. Then there are polynomials $P_N$, with $P_N(1)=0$ and, for all sufficiently large $N$, $P_N(-1)>0$, and polynomials $k_N$ with $\int_{-1}^1k_N\,\dd\eta=0$, such that the sections of the rounded-leading-edge family with thickness $T_N=wP_N$ and mean slope $k_N$ satisfy
\begin{equation}
 \norm{T_N-T^*}_\infty\longrightarrow0,\qquad
 \norm{\delta_N}_*\longrightarrow0,\qquad
 \limsup_{N\to\infty}\varDelta_{{\rm ang},N}\le\varDelta_{\rm te}<\frac\pi2 .
 \tag{\ref{stab:integral-approximation}}
\end{equation}

\emph{(iii) Convergence.} The solutions $(d_N,\ell_N)$ of Theorem~\ref{thm:rounded-family} for these sections converge to the solution for the target, $\norm{(d_N-d^*,\ell_N-\ell^*)}_Y\to0$, and the distance is bounded by~\eqref{stab:integral-bound}.
\end{recalled}

\begin{proof}
Throughout, the two faces are parametrized by $\eta=\cos\tau$, with the lower face on $0<\tau<\pi$ and the upper face on $-\pi<\tau<0$, so that $\dd\tau=\dd\eta/w$ on each face and a function on the two faces is a function of $\tau$. In particular $\phi$ and $\phi^*$ denote the slope angles on the face that corresponds to $\tau$, and
\begin{equation}
 \mathcal B=a-k_c\pm g_t,\qquad\mathcal B^*=a-k_c^*\pm g_t^*,\qquad
 \vartheta_{\rm j}^*=\mp\bigl(\phi^{*+}(1)-\phi^{*-}(1)\bigr)\Bigl(1-\frac{|\tau|}\pi\Bigr)
 \label{app:wedge-coefficient}
\end{equation}
on the lower and upper face, so that $\norm{\vartheta_{\rm j}^*}_\infty=\varDelta_{\rm te}$.

\emph{Reference angles of the target.} The angles of the target differ at the trailing edge, so that they do not satisfy~\eqref{app:reference-conditions}. Define
\begin{equation}
 \phi_\circ^\pm=\phi^{*\pm}\mp\tfrac12\bigl(\phi^{*+}(1)-\phi^{*-}(1)\bigr)\Bigl(1-\frac{\arccos\eta}{\pi}\Bigr),
 \label{app:reference-angles}
\end{equation}
that is, $\phi_\circ=\phi^*-\tfrac12\vartheta_{\rm j}^*$ on each face. These functions are H\"older continuous on $[-1,1]$, because $\phi^{*\pm}$ and $\arccos\eta$ are; $\phi_\circ^\pm(-1)=\phi^{*\pm}(-1)=\pm\pi/2$; and $\phi_\circ^+(1)=\phi_\circ^-(1)=\tfrac12(\phi^{*+}(1)+\phi^{*-}(1))$. Lemma~\ref{lem:rounded-bounds} therefore applies to them, and we put
\begin{equation}
 \mathcal N_{\rm c}^*=\mathcal N_\circ\quad\text{for the reference angles~\eqref{app:reference-angles}, with $\mathcal N_\circ$ from~\eqref{app:RH-bound}.}
 \label{app:target-constant}
\end{equation}
This constant depends on the target geometry and the flow data only.

\emph{Part~(i).} Here the difference is taken with the operator $\mathsf A$ of the second geometry, whose angles need not be continuous. Subtracting the two systems,
\[
 \mathsf A\begin{pmatrix}d-d^*\\ \ell-\ell^*\end{pmatrix}=(\mathsf A^*-\mathsf A)\begin{pmatrix}d^*\\ \ell^*\end{pmatrix}+\begin{pmatrix}F-F^*\\ G-G^*\end{pmatrix}
 =\begin{pmatrix}-\delta k\,(2U\cos\alpha+\HH\ell^*)-\delta g\,d^*\\ \delta g\,(2U\cos\alpha+\HH\ell^*)+\delta k\,d^*\end{pmatrix},
\]
which is the difference of part~(i) of Corollary~\ref{thm:geometry-stability} with the target solution on the right. The slope $g_t$ is unbounded at the nose, so that this right-hand side is measured on the faces. By~\eqref{wp:rows}, $\mathsf A=\mathsf S\mathsf N$; applying $\mathsf S^{-1}$, that is, taking the half-difference and the negative half-sum of the two entries, and using $u^{*\pm}=\Re\,\mathsf P_\pm f^*=\tfrac12(\HH\ell^*\pm d^*)$ with $f^*=d^*-i\ell^*$, gives the face rows
\[
 \Re\bigl[c_\pm\,\mathsf P_\pm(f-f^*)\bigr]=(\delta k\pm\delta g)\,\bigl(U\cos\alpha+u^{*\pm}\bigr),\qquad f=d-i\ell,
\]
with $c_\pm$ the coefficients~\eqref{wp:angles} of the second geometry. Dividing by $|c_\pm|=\sqrt{(a-k_c\mp g_t)^2+b^2}$ and using~\eqref{app:normalized-rows},
\begin{equation}
 \bigl|\Im\bigl[e^{i\phi^\pm}\mathsf P_\pm(f-f^*)\bigr]\bigr|=\frac{|\delta k\pm\delta g|}{\sqrt{(a-k_c\mp g_t)^2+b^2}}\,\bigl|U\cos\alpha+u^{*\pm}\bigr| ,
 \label{app:target-defect}
\end{equation}
and the sum over the two faces of the squares of the norms in $Y$ of these functions is $\norm{\delta}_*^2$. By~\eqref{app:reference-angles}, the angles $\phi^\pm$ of the second geometry satisfy
\[
 2\bigl(\phi^\pm-\phi_\circ^\pm\bigr)=2\bigl(\phi^\pm-\phi^{*\pm}\bigr)\pm\bigl(\phi^{*+}(1)-\phi^{*-}(1)\bigr)\Bigl(1-\frac{\arccos\eta}{\pi}\Bigr),
\]
that is, $2(\phi-\phi_\circ)=2(\phi-\phi^*)+\vartheta_{\rm j}^*$ on each face, so that the quantity~\eqref{app:reference-distance} is $\varDelta_\circ=\varDelta_{\rm ang}<\pi/2$. Lemma~\ref{lem:rounded-bounds}, applied to $f-f^*\in Y$, gives~\eqref{stab:integral-bound}.

\emph{Part~(ii): the approximants.} Put $\widetilde P^*(\tau)=P^*(\cos\tau)$. This is an even, $2\pi$-periodic function with $\widetilde P^*(0)=0$ and $\widetilde P^*(\pi)=P^*(-1)>0$. Its derivative $\widetilde P^{*\prime}=-wP^{*\prime}$ is odd and bounded, continuous on $(0,\pi)$, and has finite one-sided limits at $0$ and $\pi$; put $C_0=\norm{\widetilde P^{*\prime}}_\infty$. Let $\mathcal F_N\ge0$ be the Fej\'er kernel, normalized by $\int_{-\pi}^{\pi}\mathcal F_N\,\dd\tau=1$, and define the polynomials $P_N$ by
\[
 P_N(\cos\tau)=\widetilde P_N(\tau),\qquad \widetilde P_N=\mathcal F_N*\widetilde P^*-\bigl(\mathcal F_N*\widetilde P^*\bigr)(0).
\]
Since the Fej\'er mean $\mathcal F_N*\widetilde P^*$ is an even trigonometric polynomial, $P_N$ is a polynomial, and $P_N(1)=0$. The following properties hold~\cite[Chapter~III]{Zygmund2002}.
\begin{enumerate}
 \item[(a)] $\norm{\widetilde P_N-\widetilde P^*}_\infty\to0$, because $\widetilde P^*$ is continuous and $(\mathcal F_N*\widetilde P^*)(0)\to\widetilde P^*(0)=0$. In particular $P_N(-1)\to P^*(-1)>0$ and $\norm{T_N-T^*}_\infty\to0$.
 \item[(b)] $\widetilde P_N'=\mathcal F_N*\widetilde P^{*\prime}$, so that $\norm{\widetilde P_N'}_\infty\le C_0$ and $|\widetilde P_N(\tau)|\le C_0|\tau|$.
 \item[(c)] $\widetilde P_N'\to\widetilde P^{*\prime}$ uniformly on every closed subinterval of $(0,\pi)$, on which $\widetilde P^{*\prime}$ is continuous.
 \item[(d)] For every $\varepsilon>0$ there are $\varrho>0$ and $N_0$ such that, for $N\ge N_0$ and $0<\tau<\varrho$, $\widetilde P_N'(\tau)$ differs by at most $\varepsilon$ from a number between $-\varepsilon\,\widetilde P^{*\prime}(0^+)$ and $\widetilde P^{*\prime}(0^+)$. Indeed, write $\widetilde P^{*\prime}=\widetilde P^{*\prime}(0^+)\operatorname{sgn}+\widetilde P^{*\prime}_{\rm r}$, where $\operatorname{sgn}$ denotes the $2\pi$-periodic function equal to $\operatorname{sgn}\tau$ on $(-\pi,\pi)$ and $\widetilde P^{*\prime}_{\rm r}$ is bounded and continuous at $0$ with value $0$ there. Choose $\varrho<\pi/2$ with $|\widetilde P^{*\prime}_{\rm r}|\le\varepsilon/2$ on $(-2\varrho,2\varrho)$. Then $|\mathcal F_N*\widetilde P^{*\prime}_{\rm r}(\tau)|\le\varepsilon/2+\norm{\widetilde P^{*\prime}_{\rm r}}_\infty\int_{\varrho\le|\tau'|\le\pi}\mathcal F_N\,\dd\tau'\le\varepsilon$ for large $N$, and $\mathcal F_N*\operatorname{sgn}(\tau)=2\int_{\tau-\pi}^{\tau}\mathcal F_N\,\dd\tau'-1$ lies in $[-\varepsilon,1]$ for large $N$, since $\int_{\tau-\pi}^{\tau}\mathcal F_N\ge\tfrac12-\int_{\pi/2}^{\pi}\mathcal F_N$.
\end{enumerate}
For the mean slope, the camber vanishes at both endpoints, so that $\int_{-1}^1k_c^*\,\dd\eta=0$. By the Weierstrass approximation theorem there are polynomials $\widetilde k_N$ with $\norm{\widetilde k_N-k_c^*}_\infty\to0$; then $k_N=\widetilde k_N-\tfrac12\int_{-1}^1\widetilde k_N\,\dd\eta$ has zero mean, so that it is the slope of a polynomial camber of the form~\eqref{total:camber-family}, and $\norm{k_N-k_c^*}_\infty\le2\norm{\widetilde k_N-k_c^*}_\infty\to0$.

\emph{Part~(ii): the slopes.} For a thickness $wP$ and $\widetilde P(\tau)=P(\cos\tau)$, $0<\tau<\pi$, the relation $\dd\eta=-\sin\tau\,\dd\tau$ gives
\begin{equation}
 g_t=-\frac1{\sin\tau}\frac{\dd}{\dd\tau}\bigl[\sin\tau\,\widetilde P\bigr]=-\cot\tau\;\widetilde P-\widetilde P' .
 \label{app:slope-in-tau}
\end{equation}
Three regions are considered. \emph{Interior.} On $[\varrho,\pi-\varrho]$, properties~(a) and~(c) give $g_{t,N}\to g_t^*$ uniformly, and hence, with $k_N\to k_c^*$, $\phi_N^\pm\to\phi^{*\pm}$ uniformly. \emph{Nose.} For $\pi-\varrho<\tau<\pi$ and $\varrho$ small, $-\cot\tau\ge1/(2(\pi-\tau))$, and by~(a) and~(b), for large $N$, $g_{t,N}\ge P^*(-1)/(4(\pi-\tau))-C_0$, and the same holds for $g_t^*$. Hence $|\mathcal B_N|$ and $|\mathcal B^*|$ exceed $P^*(-1)/(8(\pi-\tau))$ there, $|\mathcal B^*|/|\mathcal B_N|$ is bounded uniformly in $N$, and the slope angles of both sections are within $O(\pi-\tau)$ of $\pm\pi/2$, uniformly in $N$. \emph{Trailing edge.} For $0<\tau<\varrho$, property~(d) and $\widetilde P_N(\tau)=\int_0^\tau\widetilde P_N'$ show that $\widetilde P_N/\tau$ has the same property as $\widetilde P_N'$ in~(d), and $\tau\cot\tau\in[1-\varrho^2,1]$. By~\eqref{app:slope-in-tau}, and since $g_t^*(1)=-2\widetilde P^{*\prime}(0^+)$, the slope $g_{t,N}(\tau)$ differs by at most $C\varepsilon$ from a number between $-\varepsilon\,g_t^*(1)$ and $g_t^*(1)$, with a constant $C$ that depends on $C_0$ only, when $\varrho$ is small enough. Thus, near the trailing edge, the thickness slope of the approximant lies between zero and that of the target, up to $C\varepsilon$. It follows that $k_N\pm g_{t,N}$ lies between $k_c^*(1)$ and $k_c^*(1)\pm g_t^*(1)$, up to an error that tends to zero with $\varepsilon$ and $\varrho$, and, since the arctangent is increasing and Lipschitz continuous, that $\phi_N^\pm$ lies between $\phi_{\rm c}=\arctan((k_c^*(1)-a)/b)$ and $\phi^{*\pm}(1)$, up to such an error.

\emph{Part~(ii): the two limits.} The integrand of~\eqref{stab:target-weighted-norm} is the square of~\eqref{app:target-defect}, that is, of $\mathsf e_N\,|U\cos\alpha+u^{*\pm}|$ with $\mathsf e_N=|\mathcal B_N-\mathcal B^*|/|\mathcal B_N+ib|$. By the three cases above, $\mathsf e_N$ is bounded uniformly in $N$ and $\tau$: in the interior and near the trailing edge the slopes are bounded and $|\mathcal B_N+ib|\ge b$, and near the nose $\mathsf e_N\le1+|\mathcal B^*+ib|/|\mathcal B_N|$. Moreover $\mathsf e_N\to0$ at every $\tau\ne0,\pm\pi$, by the interior case. Since $U\cos\alpha+u^{*\pm}$ belongs to $Y$ on each face, that is, to $L^2(-\pi,\pi)$ as a function of $\tau$, the dominated convergence theorem gives $\norm{\delta_N}_*\to0$.

For $\varDelta_{{\rm ang},N}=\norm{\vartheta_{\rm j}^*+2(\phi_N-\phi^*)}_\infty$: in the interior, $\phi_N-\phi^*\to0$ uniformly and $|\vartheta_{\rm j}^*|\le\varDelta_{\rm te}$; near the nose, $|\phi_N-\phi^*|$ and $|\vartheta_{\rm j}^*|$ are both $O(\varrho)$; near the trailing edge on the lower face, $\vartheta_{\rm j}^*$ is within $O(\varrho)$ of $\phi^{*-}(1)-\phi^{*+}(1)$, and $2(\phi_N^--\phi^{*-})$ lies, up to a small error, between $0$ and $2(\phi_{\rm c}-\phi^{*-}(1))$, so that $\vartheta_{\rm j}^*+2(\phi_N-\phi^*)$ lies, up to a small error, between $\phi^{*-}(1)-\phi^{*+}(1)$ and $2\phi_{\rm c}-\phi^{*+}(1)-\phi^{*-}(1)$. Both numbers are at most $\varDelta_{\rm te}$ in absolute value, the second because $\phi_{\rm c}$ lies between $\phi^{*+}(1)$ and $\phi^{*-}(1)$. The upper face is treated in the same way. Letting $\varepsilon$ and $\varrho$ tend to zero gives $\limsup_N\varDelta_{{\rm ang},N}\le\varDelta_{\rm te}$.

\emph{Part~(iii).} The sections of part~(ii) belong to the rounded-leading-edge family with $P_N(1)=0$ and $P_N(-1)>0$, so that Theorem~\ref{thm:rounded-family} gives their solutions $(d_N,\ell_N)\in Y\times Y$, and their slopes $k_N$ and $\sqrt{1+\eta}\,g_{t,N}$ are bounded. For large $N$, $\varDelta_{{\rm ang},N}\le(\varDelta_{\rm te}+\pi/2)/2<\pi/2$, and part~(i) gives $\norm{(d_N-d^*,\ell_N-\ell^*)}_Y\le\mathcal N_{\rm c}^*\norm{\delta_N}_*/\cos\varDelta_{{\rm ang},N}\to0$.
\end{proof}

\paragraph{Remarks on Theorem~\ref{thm:te-angle}.}
The closed-form solutions of Theorem~\ref{thm:rounded-family} are for sections whose faces have a common tangent at the trailing edge, whereas the sections of practical interest usually have a finite trailing-edge angle. Part~(iii) of Theorem~\ref{thm:existence} shows that the system is uniquely solvable for such a section. The supremum norm of Corollary~\ref{thm:geometry-stability} cannot measure the error of approximating it by the rounded-leading-edge family: every member of the family has zero thickness slope at the trailing edge, so that $\norm{\sqrt{1+\eta}\,\delta g}_\infty\ge\sqrt2\,|g_t^*(1)|$ for every approximant. The slope error is confined, however, to a neighbourhood of the trailing edge that shrinks as the degree increases. The following theorem shows that an integral norm of the slope error both tends to zero for suitable members of the family and bounds the distance between the two solutions.

The norm~\eqref{stab:target-weighted-norm} is an integral of the squared errors of the two face slopes. Its weight is fixed by the target: it is the square of the total chordwise velocity on the faces of the target section, which is integrable because the target solution belongs to $Y\times Y$. A slope error that remains bounded and is confined to a shrinking neighbourhood of the trailing edge therefore tends to zero in this norm, although it does not tend to zero in the supremum norm. The quantity~\eqref{stab:angle-distance} compares the slope angles of the two sections after the trailing-edge jump of the target has been removed; it only has to stay below $\pi/2$, not to be small, and it does so for the approximants of part~(ii) because near the trailing edge their slopes lie between the common tangent and the slopes of the target faces.

The hypothesis of part~(ii) holds for a finite trailing-edge angle, for which $P^*$ behaves like $\sqrt{1-\eta}$ at the trailing edge, and it permits the term proportional to $\sqrt{1+\eta}$ that the regular factor of the NACA thickness distributions contains at the nose. Part~(ii) is an existence statement for a particular sequence of approximants, constructed from Fej\'er means; it does not assert that the least-squares fits used in Section~\ref{sec:comparison-pressure-loads} are such a sequence, for which~\eqref{stab:target-weighted-norm} and~\eqref{stab:angle-distance} would have to be evaluated. No rate of convergence is given, and the thickness of the approximants is not asserted to be nonnegative, which Theorem~\ref{thm:rounded-family} does not require.

\subsection{Proofs of Theorems~\ref{thm:flat-approximation} and~\ref{thm:rounded-approximation}}\label{app:proof-approximation}
Both proofs use the Bernstein polynomials of a function $g$ on $[-1,1]$,
\[
 \mathbb B_ng(\eta)=\sum_{j=0}^n g\Bigl(-1+\frac{2j}n\Bigr)\binom nj\Bigl(\frac{1+\eta}2\Bigr)^j\Bigl(\frac{1-\eta}2\Bigr)^{n-j},
\]
and rest on two of their classical properties~\cite[Theorems~1.1.1 and~1.8.1]{Lorentz1986}: if $g\in C[-1,1]$, then $\norm{\mathbb B_ng-g}_\infty\to0$, and if $g\in C^1[-1,1]$, then also $\norm{(\mathbb B_ng)'-g'}_\infty\to0$. Moreover, $\mathbb B_ng\ge0$ whenever $g\ge0$, and $\mathbb B_ng(\pm1)=g(\pm1)$, both of which follow from the definition of $\mathbb B_ng$. For $0<\varrho<1$ let $\mathsf h_\varrho\in C^1[-1,1]$ be a cutoff with $0\le\mathsf h_\varrho\le1$, $\mathsf h_\varrho=1$ on $[-1+\varrho,1-\varrho]$, $\mathsf h_\varrho=0$ on $[-1,-1+\varrho/2]\cup[1-\varrho/2,1]$ and $\norm{\mathsf h_\varrho'}_\infty\le4/\varrho$.

\begin{recalled}[Approximation by flat-endpoint profiles]{Theorem~\ref{thm:flat-approximation}}
Let $w$ be the weight in~\eqref{total:coordinate-weights}, and let primes denote derivatives in $\eta$.

\emph{(i) Profiles.} For every real $T\in C[-1,1]$ with $T(-1)=T(1)=0$, there exist real polynomials $P_N$ such that
\begin{equation}
 T_N=w^3P_N,\qquad \|T_N-T\|_\infty\longrightarrow0.
 \tag{\ref{approx:flat-uniform}}
\end{equation}
If $T\ge0$, the polynomials can be chosen nonnegative on $[-1,1]$.

\emph{(ii) Slopes.} If, in addition, $T\in C^1[-1,1]$ and $T'(-1)=T'(1)=0$, the polynomials can be chosen so that
\begin{equation}
 \|T_N-T\|_{C^1}:=\|T_N-T\|_\infty+\|T_N'-T'\|_\infty
 \longrightarrow0,
 \tag{\ref{approx:flat-slopes}}
\end{equation}
with the same positivity property.
\end{recalled}

\begin{proof}[Proof of Theorem~\ref{thm:flat-approximation}]
\emph{Part (i).} Since $T$ is continuous and $T(\pm1)=0$, $\norm{T-\mathsf h_\varrho T}_\infty\le\sup_{1-|\eta|<\varrho}|T|\to0$ as $\varrho\to0$. For fixed $\varrho$ the function $\mathsf g_\varrho=\mathsf h_\varrho T/w^3$ is continuous on $[-1,1]$, because $\mathsf h_\varrho$ vanishes near the endpoints, and $0\le w^3\le1$ gives $\norm{w^3\mathbb B_n\mathsf g_\varrho-\mathsf h_\varrho T}_\infty\le\norm{\mathbb B_n\mathsf g_\varrho-\mathsf g_\varrho}_\infty\to0$ as $n\to\infty$. For each $N\ge1$ choose $\varrho_N$ such that $\norm{T-\mathsf h_{\varrho_N}T}_\infty\le1/N$, and then $n_N$ such that $\norm{\mathbb B_{n_N}\mathsf g_{\varrho_N}-\mathsf g_{\varrho_N}}_\infty\le1/N$. With $P_N=\mathbb B_{n_N}\mathsf g_{\varrho_N}$ the triangle inequality gives $\norm{w^3P_N-T}_\infty\le2/N$, which is~\eqref{approx:flat-uniform}. If $T\ge0$, then $\mathsf g_\varrho\ge0$ and $P_N\ge0$.

\emph{Part (ii).} Let $T\in C^1[-1,1]$ with $T(\pm1)=T'(\pm1)=0$, and put $\iota_\varrho=\sup_{1-|\eta|<\varrho}|T'|$, which tends to zero with $\varrho$. By the mean value theorem $|T|\le\varrho\,\iota_\varrho$ for $1-|\eta|<\varrho$. Hence
\[
 \norm{(T-\mathsf h_\varrho T)'}_\infty\le\norm{(1-\mathsf h_\varrho)T'}_\infty+\norm{\mathsf h_\varrho'T}_\infty\le\iota_\varrho+\frac4\varrho\,\varrho\,\iota_\varrho=5\iota_\varrho,
\]
so that $\mathsf h_\varrho T\to T$ in $C^1$. For fixed $\varrho$, $\mathsf g_\varrho=\mathsf h_\varrho T/w^3$ belongs to $C^1[-1,1]$, and
\[
 \bigl(w^3\mathbb B_n\mathsf g_\varrho-\mathsf h_\varrho T\bigr)'=w^3\bigl(\mathbb B_n\mathsf g_\varrho-\mathsf g_\varrho\bigr)'-3\eta w\bigl(\mathbb B_n\mathsf g_\varrho-\mathsf g_\varrho\bigr)
\]
tends to zero uniformly as $n\to\infty$, since $w^3\le1$ and $|3\eta w|\le3$, while $\norm{w^3\mathbb B_n\mathsf g_\varrho-\mathsf h_\varrho T}_\infty\to0$ as in part~(i). For each $N\ge1$ choose $\varrho_N$ such that $\norm{T-\mathsf h_{\varrho_N}T}_{C^1}\le1/N$, and then $n_N$ such that $\norm{w^3\mathbb B_{n_N}\mathsf g_{\varrho_N}-\mathsf h_{\varrho_N}T}_{C^1}\le1/N$. With $P_N=\mathbb B_{n_N}\mathsf g_{\varrho_N}$ this gives $\norm{w^3P_N-T}_{C^1}\le2/N$, which is~\eqref{approx:flat-slopes}, and positivity is preserved as in part~(i).
\end{proof}

\begin{recalled}[Approximation by rounded-leading-edge profiles]{Theorem~\ref{thm:rounded-approximation}}
Let $w$ be the weight in~\eqref{total:coordinate-weights}, and let primes denote derivatives in $\eta$.

\emph{(i) Profiles.} For every real $T\in C[-1,1]$ with $T(-1)=T(1)=0$, there exist real polynomials $P_N$ with $P_N(1)=0$ such that
\begin{equation}
 T_N=wP_N,\qquad \|T_N-T\|_\infty\longrightarrow0.
 \tag{\ref{approx:rounded-uniform}}
\end{equation}
If $T\ge0$, the polynomials can be chosen nonnegative on $[-1,1]$.

\emph{(ii) Slopes.} If, in addition, $P_T=T/w$ extends from $(-1,1)$ to a function in $C^1[-1,1]$ with $P_T(1)=0$, the polynomials can be chosen so that $P_N(1)=0$, $P_N(-1)=P_T(-1)$ and
\begin{equation}
 \|T_N-T\|_\infty+
 \|w(T_N'-T')\|_\infty\longrightarrow0,
 \tag{\ref{approx:rounded-slopes}}
\end{equation}
with the same positivity property. The weighted derivatives are understood by their continuous extensions to the endpoints.
\end{recalled}

\begin{proof}[Proof of Theorem~\ref{thm:rounded-approximation}]
\emph{Part (i).} Repeat the proof of part~(i) of Theorem~\ref{thm:flat-approximation} with $\mathsf g_\varrho=\mathsf h_\varrho T/[(1-\eta)w]$ and $P_N=(1-\eta)\mathbb B_{n_N}\mathsf g_{\varrho_N}$. Then $P_N(1)=0$, the function $\mathsf g_\varrho$ is continuous on $[-1,1]$ because $\mathsf h_\varrho$ vanishes near both endpoints, and $0\le(1-\eta)w\le2$ gives $\norm{w(1-\eta)\mathbb B_n\mathsf g_\varrho-\mathsf h_\varrho T}_\infty\le2\norm{\mathbb B_n\mathsf g_\varrho-\mathsf g_\varrho}_\infty\to0$, so that the choice of $\varrho_N$ and $n_N$ made there gives~\eqref{approx:rounded-uniform}. If $T\ge0$, then $P_N\ge0$.

\emph{Part (ii).} Take $P_N=\mathbb B_NP_T$. Then $P_N(1)=P_T(1)=0$ and $P_N(-1)=P_T(-1)$. If $T\ge0$, then $P_T=T/w\ge0$ and $P_N\ge0$. Since $P_T\in C^1[-1,1]$, $\delta P_N=P_N-P_T$ tends to zero in $C^1[-1,1]$. Since $ww'=-\eta$,
\[
 T_N-T=w\,\delta P_N,\qquad w(T_N'-T')=(1-\eta^2)\,\delta P_N'-\eta\,\delta P_N,
\]
so that $\norm{T_N-T}_\infty+\norm{w(T_N'-T')}_\infty\le2\norm{\delta P_N}_\infty+\norm{\delta P_N'}_\infty\to0$, which is~\eqref{approx:rounded-slopes}.
\end{proof}

\begin{remark}[Asymmetric profiles]\label{rem:asymmetric-density}\label{sec:asymmetric-approximation}
Theorems~\ref{thm:flat-approximation} and~\ref{thm:rounded-approximation} extend to cambered sections. Let $f^\pm\in C[-1,1]$ with $f^+\ge f^-$ and $f^\pm(\pm1)=0$, and split them into the camber $f_c$ and the thickness $f_t\ge0$ of~\eqref{geo:profile-slopes}. By the Weierstrass theorem, after a cutoff near the endpoints, $f_c$ can be approximated uniformly by polynomials of the form~\eqref{total:camber-family}, or~\eqref{total:flat-camber} for flat endpoints, and $f_t$ by either thickness family. Both faces are then approximated uniformly, since the error of each face is at most the sum of the two errors. Flat endpoints, or $P_N(-1)>0$ and $P_N(1)=0$ for the rounded family, can be imposed in the process. This concerns the profile heights only. Convergence of the slopes, which is what Corollary~\ref{thm:geometry-stability} requires, holds under additional regularity: if the normalized camber $f_c/(1-\eta^2)^j$, with $j=2$ for flat endpoints and $j=1$ otherwise, extends to a $C^1$ function, polynomial approximation of its derivative followed by integration gives camber approximants converging in $C^1$, and combined with the slope statements of Theorems~\ref{thm:flat-approximation} and~\ref{thm:rounded-approximation} this gives slope convergence for both families.
\end{remark}

\subsection{Weighted Cauchy integrals}
\label{app:weightedCauchyIntegrals}
The transform identities~\eqref{total:transform-identities} follow from the classical Cauchy integrals of the endpoint weights~\cite{Muskhelishvili1953,King2009},
\[
 \int_{-1}^1\frac{r(\eta)}{\zeta -\eta}\,\dd\eta=\pi[1-K(\zeta )],\qquad
 \int_{-1}^1\frac{w(\eta)}{\zeta -\eta}\,\dd\eta=\pi[\zeta -W(\zeta )],\qquad \zeta \notin[-1,1],
\]
together with the Plemelj--Sokhotski formulas, which give $\HH r=1$ and $\HH w=\eta_1$, and the partial-fraction identity $[(\zeta -\eta)(\eta_1-\eta)]^{-1}=(\zeta -\eta_1)^{-1}[(\eta_1-\eta)^{-1}-(\zeta -\eta)^{-1}]$. The branches are those of the main text, $W(\zeta )\sim \zeta $ at infinity and $K=(\zeta -1)/W$. Differentiation with respect to $\zeta $ commutes with these integrals for $\zeta $ off the chord, which gives~\eqref{total:confluent-transforms}; the constant $-1$ survives only when $\nu=1$.

\subsection{Proof of Theorem~\ref{thm:flat-family}}\label{app:proof-flat-family}
\begin{recalled}[Computable confluent rational solution for flat-endpoint profiles]{Theorem~\ref{thm:flat-family}}
Under the assumptions stated at the beginning of Section~\ref{sec:flat-solutions}, the following hold.

\emph{(i) Pole count.} The function $\Psi_{\rm f}$ in~\eqref{total:flat-poles} has exactly ${\mathsf q}+1$ zeros counted with multiplicity, all off $[-1,1]$. Hence $\sum_{j=1}^J\mathsf r_j={\mathsf q}+1$, and $\bm M_{\rm f}$ in~\eqref{total:flat-real-system} is square.

\emph{(ii) Amplitudes.} The matrix $\bm M_{\rm f}$ is invertible. Thus the amplitudes exist uniquely and satisfy
\begin{equation}
 2\Re\sum_{(j,\nu)\in\mathcal I_{\rm f}}A_{j\nu}E_{1,j\nu}
 =2(U\sin\alpha-aU\cos\alpha),\qquad
 2\Re\sum_{(j,\nu)\in\mathcal I_{\rm f}}A_{j\nu}E_{2,j\nu}
 =-bN_0.
 \tag{\ref{total:flat-amplitudes}}
\end{equation}

\emph{(iii) Solution.} The jumps~\eqref{total:flat-ansatz} solve~\eqref{total:coupled-system}. They are real and integrable, satisfy $d,n=O(r)$ at the leading edge, and vanish at the trailing edge; their transforms are bounded there. The amplitudes and jumps are unique within this fixed-pole confluent representation.

\emph{(iv) Existence and uniqueness in $X\times X$.} With $\ell$ as in~\eqref{stab:system}, the pair $(d,\ell)$ belongs to $X\times X$, defined in~\eqref{stab:space}, and solves~\eqref{stab:system}; it is the unique solution in $X\times X$ given by part~(ii) of Theorem~\ref{thm:existence}, so the jumps~\eqref{total:flat-ansatz} are unique in all of $X\times X$, not only within the fixed-pole representation.
\end{recalled}

\begin{proof}
Throughout, $k_c$ is an arbitrary real polynomial. By Remark~\ref{rem:camber-transfer}, the asymmetric problem is exactly this system; no step below requires zero undeformed camber.
For the jumps in~\eqref{total:flat-ansatz}, the transform identities~\eqref{total:confluent-transforms} give
\begin{equation}
 \begin{aligned}
 \HH d&=2\Re\sum_{(j,\nu)\in\mathcal I_{\rm f}}
 A_{j\nu}\Phi_\nu(\zeta _j,\eta),\\
 \HH n&=-2b\,U\cos\alpha+2\Re\sum_{(j,\nu)\in\mathcal I_{\rm f}}
 sA_{j\nu}\Phi_\nu(\zeta _j,\eta).
 \end{aligned}
 \label{app:flat-solution-transforms}
\end{equation}

\emph{Pole count.} The function $\Psi_{\rm f}=QW-b-i\lambda$ is analytic off the slit. On its banks,
\begin{equation}
 \Psi_{{\rm f},\pm}(\eta_1)=-b+i[\pm\sqrt{1-\eta_1^2}\,Q(\eta_1)-\lambda(\eta_1)],
 \qquad -1\le \eta_1\le1.
\end{equation}
Both boundary values have real part $-b<0$, including their endpoint limits. A sufficiently close contour around the slit therefore has zero image winding about the origin. At infinity,
\begin{equation}
 \Psi_{\rm f}(\zeta )=\Psi_{\rm f}^\infty\zeta ^{{\mathsf q}+1}(1+o(1)),\qquad
 \Psi_{\rm f}^\infty\ne0.
\end{equation}
If ${\mathsf p}={\mathsf m}+1$, the leading coefficient is the sum of a nonzero real contribution from $QW$ and a purely imaginary contribution from $ik_c$, so it cannot cancel. The argument principle between the slit contour and a large circle gives exactly ${\mathsf q}+1$ off-chord zeros, counted with multiplicity. The candidate polynomial has no roots on the chord either: its imaginary part there forces $\lambda=0$, after which $(\eta_1^2-1)Q(\eta_1)^2=b^2$ is impossible. Hence $\sum_{j=1}^J\mathsf r_j={\mathsf q}+1$, since ${\mathsf q}+1=\max\{{\mathsf m}+1,{\mathsf p}\}$.

Here the inner slit contour can be chosen so that $\Re\Psi_{\rm f}<-b/2$ on it, uniformly near the two endpoints as well as on the banks. Its winding number is therefore zero. On a sufficiently large outer circle, the quotient $\Psi_{\rm f}/(\Psi_{\rm f}^\infty\zeta ^{{\mathsf q}+1})$ is uniformly close to one, so that image has winding number ${\mathsf q}+1$. The argument principle on the intervening domain gives ${\mathsf q}+1$ zeros because $\Psi_{\rm f}$ has no poles there. This count concerns the unsquared analytic branch. Squaring only generates candidates and may add roots of the opposite branch of $W$; it cannot replace the branch test or determine its multiplicities.

\emph{Pole cancellation and confluent residuals.} For a simple-pole mode with independent coefficients $A,B$, the pole equations are
\begin{equation}
 \begin{pmatrix}
 a^2+b^2-ak_c(\zeta )-bQ(\zeta )W(\zeta )&k_c(\zeta )-a\\
 aQ(\zeta )W(\zeta )-bk_c(\zeta )&b-Q(\zeta )W(\zeta )
 \end{pmatrix}\binom AB=0.
 \label{total:flat-residue-matrix}
\end{equation}
Its determinant is $b[(b-QW)^2+\lambda^2]$, which vanishes at every zero of $\Psi_{\rm f}$, and there the direction $B=sA$ lies in its kernel.

For example, in the second row of~\eqref{total:flat-residue-matrix}, substitute $QW=b+i\lambda$, $k_c=a-\lambda$ and $B=(a-ib)A$. The coefficient of $A$ becomes
\[
 a(b+i\lambda)-b(a-\lambda)
 +(b-b-i\lambda)(a-ib)=0.
\]
The first row vanishes by the same substitution.

The contributions of a unit mode $d=r/(\zeta -\eta)$, $n=sr/(\zeta -\eta)$ to the first equation and to $b/r$ times the second equation are, respectively,
\begin{equation}
 \begin{aligned}
 \mathcal R_1(\zeta ,\eta)&=E_1(\zeta ,\eta)-\frac{K(\zeta )\Psi_{\rm f}(\zeta )}{\zeta -\eta},\\
 \mathcal R_2(\zeta ,\eta)&=E_2(\zeta ,\eta)-\frac{(s-a)\Psi_{\rm f}(\zeta )}{\zeta -\eta}.
 \end{aligned}
 \label{app:flat-confluent-residual}
\end{equation}
These identities follow by subtracting each numerator's value at $\eta=\zeta $ and using $(\zeta +1)K(\zeta )=W(\zeta )$ and $(\zeta -1)Q(\zeta )=K(\zeta )Q(\zeta )W(\zeta )$. If $\zeta _j$ has multiplicity $\mathsf r_j$ in $\Psi_{\rm f}$, then $\partial_\zeta ^l\Psi_{\rm f}(\zeta _j)=0$ for $0\le l<\mathsf r_j$. Apply $\mathcal L_{\zeta ,\nu}$ to~\eqref{app:flat-confluent-residual} for $1\le\nu\le \mathsf r_j$ and evaluate at $\zeta _j$. Every term containing $\Psi_{\rm f}$ vanishes by the product rule.

In fact, for any analytic factor $\mathcal X(\zeta ,\eta)$ and $k<\mathsf r_j$,
\[
 \left.\partial_\zeta ^k[\mathcal X(\zeta ,\eta)\Psi_{\rm f}(\zeta )]\right|_{\zeta _j}
 =\sum_{l=0}^k\binom{k}{l}
   (\partial_\zeta ^{k-l}\mathcal X)(\zeta _j,\eta)\Psi_{\rm f}^{(l)}(\zeta _j)=0.
\]
For real interior $\eta$, the factors involving $(\zeta -\eta)^{-1}$ are analytic near $\zeta _j$ because the pole is off the chord. Taking $k=\nu-1$ justifies every order through $\mathsf r_j$. The coefficient $s$ does not depend on $\zeta$. Thus the confluence construction is an exact derivative identity, not a limiting assumption about separated simple roots. The remaining residuals are exactly $E_{1,j\nu}$ and $E_{2,j\nu}$, and~\eqref{total:confluent-transforms} gives the higher-order transform formulas.

The background term $n=rN_0$ cancels the full thickness forcing in the second equation. Indeed $\HH(rN_0)=-2b\,U\cos\alpha$ by~\eqref{total:velocity-background}, so the background contributes $2\lambda\,U\cos\alpha$ to the left-hand side of the first equation and
\[
 rN_0-\frac{g_t}b\HH(rN_0)=rN_0+2g_t\,U\cos\alpha=rN_0+G
\]
to that of the second. In the first equation the remaining forcing is therefore
\begin{equation}
 F-2\lambda\,U\cos\alpha=2(U\sin\alpha-a\,U\cos\alpha),
\end{equation}
and in the second equation, after moving $rN_0$ to the right and multiplying by $b/r$, it is $-bN_0$. Thus both equations reduce to~\eqref{total:flat-amplitudes}. These forcings are the constants $2U(\sin\alpha-a\cos\alpha)$ and $-bN_0=2b^2U\cos\alpha$. Divided differences give degree at most ${\mathsf q}$, and derivatives in $\zeta $ do not increase the degree in $\eta$.

No camber term is discarded in this reduction: the product $k_c\,U\cos\alpha$ in $F$ is cancelled exactly by the background. Furthermore $\mathcal D_\zeta [(1\mp\eta)Q]$ has degree at most ${\mathsf m}$, while $\mathcal D_\zeta k_c$ has degree at most ${\mathsf p}-1$ (or is zero). Their maximum is exactly the degree bound ${\mathsf q}$, and the two forcings are constants. Separating the real and imaginary parts of the amplitudes and matching coefficients gives~\eqref{total:flat-real-system}.

\emph{Invertibility.} A real vector in $\ker\bm M_{\rm f}$ reconstructs homogeneous jumps with zero forcing and $N_0=0$. The preceding residual identities show that these jumps satisfy the homogeneous integral equations. They are $r$ times rational functions without poles on the closed chord, so that $d$ and $\ell$ belong to $X$. Since $k_c$ and $g_t$ are H\"older continuous on $[-1,1]$ and $g_t(\pm1)=0$, part~(ii) of Theorem~\ref{thm:existence} shows that the homogeneous system has only the zero solution in $X\times X$. Hence $d=n=0$. To pass from zero jumps to zero amplitudes, put
\begin{equation}
 \mathcal Z=\sum_{(j,\nu)\in\mathcal I_{\rm f}}\frac{A_{j\nu}}{(\zeta _j-\eta)^\nu},\qquad
 \frac dr=2\Re\mathcal Z,\qquad
 \frac{n-ad}{br}=2\Re\Bigl[\frac{s-a}b\,\mathcal Z\Bigr]=2\Im\mathcal Z.
\end{equation}
Hence $\mathcal Z$ vanishes on the open chord. It is a rational function, so it vanishes identically, and its principal parts at the distinct off-chord points $\zeta _j$ vanish; every $A_{j\nu}$ is zero. Thus $\bm M_{\rm f}$ has trivial kernel. It is square: by the pole count there are ${\mathsf q}+1$ complex amplitudes, hence $2{\mathsf q}+2$ real unknowns, and the two residual polynomials of degree at most ${\mathsf q}$ give $2{\mathsf q}+2$ real coefficient equations. Injectivity of this square real map implies surjectivity, so the forced amplitudes exist uniquely. Off-chord principal parts are bounded on the closed chord; multiplication by $r$ establishes the stated integrability and trailing-edge decay. The rational transforms are bounded at both endpoints. This proves parts (i)--(iii).

\emph{Part (iv).} For a flat-endpoint profile, $k_c$ and $g_t$ are H\"older continuous on $[-1,1]$ and $g_t(\pm1)=0$, so that the hypotheses of part~(ii) of Theorem~\ref{thm:existence} hold, with zero edge angles. The jumps~\eqref{total:flat-ansatz} are $r$ times functions bounded on the closed chord. Hence $d$ and $\ell$ lie in $X$ and solve~\eqref{stab:system}, which is equivalent to~\eqref{total:coupled-system}. The uniqueness in part~(ii) of Theorem~\ref{thm:existence} identifies them with that solution.
\end{proof}

\subsection{Proof of Theorem~\ref{thm:rounded-family}}\label{app:proof-rounded-family}
\begin{recalled}[Computable confluent solution for rounded-leading-edge profiles]{Theorem~\ref{thm:rounded-family}}
Under the assumptions stated at the beginning of Section~\ref{sec:rounded-solutions}, the following hold; the construction includes repeated off-chord roots.

\emph{(i) Pole count.} The function $\Psi_{\rm r}$ in~\eqref{total:rounded-poles} has exactly ${\mathsf q}_1+1$ off-chord zeros counted with multiplicity, so that $|\mathcal I_{\rm r}|=\sum_{j=1}^J\mathsf r_j={\mathsf q}_1+1$.

\emph{(ii) Amplitudes.} The system~\eqref{total:rounded-real-system} is consistent and has full column rank. More precisely, with ${\mathsf q}={\mathsf q}_1$ and ${\mathsf q}_2={\mathsf q}+1$, its $2{\mathsf q}+3$ coefficient equations have the single redundant relation $\mathcal T(1)=0$ in the second polynomial block. Representing that block in $(\eta-1)\mathbb R[\eta]_{\le {\mathsf q}}$ gives an invertible $(2{\mathsf q}+2)\times(2{\mathsf q}+2)$ system. The unique amplitudes satisfy
\begin{equation}
 2\Re\sum_{(j,\nu)\in\mathcal I_{\rm r}}A_{j\nu}E_{j\nu}=F_{\rm r},\qquad
 2\Re\sum_{(j,\nu)\in\mathcal I_{\rm r}}A_{j\nu}\mathcal T_{j\nu}=G_{\rm r}.
 \tag{\ref{total:rounded-amplitudes}}
\end{equation}

\emph{(iii) Solution.} The jumps~\eqref{total:rounded-ansatz} solve~\eqref{total:coupled-system}. The jumps are real and satisfy $d,n=O(w)$ at both endpoints. The amplitudes and jumps are unique within this complete confluent representation. No separate consistency, full-column-rank or simplicity assumption is needed under the stated endpoint conditions. No completeness claim in larger endpoint classes is made.

\emph{(iv) Existence and uniqueness in $Y\times Y$.} The profile satisfies~\eqref{stab:rounded-class}. With $\ell$ as in~\eqref{stab:system}, the pair $(d,\ell)$ belongs to $Y\times Y$, defined in~\eqref{stab:space-Y}, and is the unique solution in $Y\times Y$ given by part~(iii) of Theorem~\ref{thm:existence}; no smallness condition on the slopes is required.
\end{recalled}

\begin{proof}
Use the total polynomial mean slope $k_c$. Polynomial camber is bounded with bounded slope at the nose and does not change $R(-1)=P(-1)>0$ or $R(1)=0$. Thus it changes $\lambda$ but neither endpoint structure used below.
For the jumps in~\eqref{total:rounded-ansatz}, set
\begin{equation}
 \mathcal Y=\sum_{(j,\nu)\in\mathcal I_{\rm r}}
 A_{j\nu}\Omega_\nu(\zeta _j,\eta).
\end{equation}
Equation~\eqref{total:confluent-transforms} yields
\begin{equation}
 \HH d=2\Re\mathcal Y,\qquad \HH n=2\Re(s\mathcal Y).
 \label{app:rounded-solution-transforms}
\end{equation}

\emph{Exact pole count.} Set ${\mathsf q}={\mathsf q}_1=\max\{{\mathsf m}-1,{\mathsf p}\}$, so ${\mathsf q}_2={\mathsf q}+1$ and ${\mathsf q}\ge1$.
Use the disk coordinate $\omega=\zeta -W(\zeta )$, with
\begin{equation}
 \zeta =\frac{\omega+\omega^{-1}}2,\qquad
 W=\frac{1-\omega^2}{2\omega}.
\end{equation}
Define the meromorphic disk function
\begin{equation}
 \mathcal M(\omega)=\frac{R(\zeta (\omega))}{W(\zeta (\omega))}+i\lambda(\zeta (\omega)),\qquad
 \mathcal M_\upsilon(\omega)=b+\upsilon \mathcal M(\omega),\quad 0<\upsilon\le1.
 \label{app:rounded-count-homotopy}
\end{equation}
Since $R(1)=0$, the apparent singularity at $\omega=1$ is removable. At $\omega=0$, $\mathcal M$ has a pole of order exactly ${\mathsf q}$: the leading contributions from $R/W$ and $i\lambda$ are respectively real and purely imaginary, so they cannot cancel when their degrees agree. At $\omega=-1$,
\begin{equation}
 \mathcal M(\omega)=-\frac{P(-1)}{1+\omega}+O(1),
 \label{app:rounded-nose-pole}
\end{equation}
since $R(-1)=P(-1)>0$. There are no other poles in the closed disk.

To verify the two endpoint claims directly, write
\[
 \zeta -1=\frac{(1-\omega)^2}{2\omega},\qquad
 \zeta +1=\frac{(1+\omega)^2}{2\omega},\qquad
 W=\frac{(1-\omega)(1+\omega)}{2\omega}.
\]
At $\omega=1$, $R(1)=0$ makes $R(\zeta )=O((1-\omega)^2)$, so $R/W=O(1-\omega)$. At $\omega=-1$, $W=-(1+\omega)+O((1+\omega)^2)$ and $R(\zeta )=P(-1)+O((1+\omega)^2)$, giving the displayed nonzero residue. At $\omega=0$, $\zeta \sim W\sim(2\omega)^{-1}$; thus $R/W$ has pole order ${\mathsf m}-1$, and $i\lambda$ has order ${\mathsf p}$ when ${\mathsf p}>0$. Their leading coefficients are respectively real and purely imaginary, which proves order ${\mathsf q}$ even when the degrees coincide. For $\omega=e^{i\tau}\ne\pm1$, $\zeta =\cos \tau $ and $W=-i\sin \tau $, so
\begin{equation}
 \Re \mathcal M_\upsilon(e^{i\tau})=b>0.
\end{equation}
The same identity holds by continuity at $\omega=1$. Thus no zero crosses the unit circle for $\upsilon>0$; $\omega=-1$ remains a pole with nonzero residue, and $\omega=0$ remains a pole of order ${\mathsf q}$. The number of zeros in the disk, counted with multiplicity, is constant on $0<\upsilon\le1$.

For a precise continuation argument, restrict first to $\upsilon\in[\upsilon_0,1]$ with $\upsilon_0>0$. The leading coefficients of the poles at $0$ and $-1$ are then uniformly bounded away from zero. Multiplication by their fixed pole factors produces analytic local functions nonzero at those points, so zeros cannot disappear into either pole. They cannot cross the remaining unit circle because its boundary real part is $b$. The argument principle on a disk with small neighborhoods of the poles removed therefore keeps the zero count fixed. Since $\upsilon_0$ is arbitrary, this holds throughout the asserted interval.

For small $\upsilon$, exactly ${\mathsf q}$ zeros approach the pole at zero: multiplying by $\omega^{\mathsf q}$ gives the local equation $b\omega^{\mathsf q}+\upsilon\mathcal M_0+O(\upsilon\omega)=0$, with $\mathcal M_0\ne0$. One further zero approaches $-1$ and satisfies
\begin{equation}
 \omega=-1+\frac{\upsilon P(-1)}b+O(\upsilon^2),
\end{equation}
so it lies inside the disk. These counts follow from Rouch\'e's theorem on fixed small circles around the two poles; on the remaining compact set $\mathcal M_\upsilon\to b$ uniformly, so there are no further zeros. Consequently $\mathcal M_1$ has exactly ${\mathsf q}+1$ disk zeros. Since $\mathcal M_1=\Psi_{\rm r}/W$ and the slit-plane map is conformal away from the endpoints, these correspond exactly, with their multiplicities, to the ${\mathsf q}+1$ off-chord zeros of $\Psi_{\rm r}$. Neither endpoint supplies a mode. Since $\Re\mathcal M_1=b$ on the unit circle, $\Psi_{\rm r}=W\mathcal M_1$ has no zeros on the banks of the chord, $\Psi_{\rm r}(-1)=R(-1)=P(-1)>0$, and $\Psi_{\rm r}(1)=0$ only through the factor $W$. Correspondingly, with $R=(1-\eta)R_1$, the candidate polynomial in~\eqref{total:rounded-candidate-system} factors as $(\zeta-1)\{(\zeta-1)R_1(\zeta)^2-[b+i\lambda(\zeta)]^2(\zeta+1)\}$. The second factor equals $-2[b+i\lambda(1)]^2\ne0$ at $\zeta=1$ and $-2R_1(-1)^2\ne0$ at $\zeta=-1$, and it has no roots in $(-1,1)$: there, multiplication by $\zeta-1$ shows that a root would satisfy $(\zeta-1)^2R_1^2=[b+i\lambda]^2(\zeta^2-1)$, whose imaginary part forces $\lambda=0$ and then $(\zeta-1)^2R_1^2=b^2(\zeta^2-1)<0$, which is impossible.

The small-parameter counts can also be made explicit. Near zero, $\omega^q\mathcal M_\upsilon=b\omega^{\mathsf q}+\upsilon\mathcal M_0+O(\upsilon\omega)$; on a fixed sufficiently small circle the first term dominates for sufficiently small $\upsilon$, giving ${\mathsf q}$ zeros by Rouch\'e's theorem. Near $-1$, put $\delta\omega =1+\omega$ and multiply by $\delta\omega $. The equation is
\[
 \delta\omega \mathcal M_\upsilon=b\delta\omega-\upsilon P(-1)+O(\upsilon \delta\omega ).
\]
Its derivative in $\delta\omega $ at $(\delta\omega ,\upsilon)=(0,0)$ is $b\ne0$, so it has exactly one local root $\delta\omega =\upsilon P(-1)/b+O(\upsilon^2)$. Its positive real leading term puts $\omega=-1+\delta\omega $ inside the unit circle for small positive $\upsilon$, including when the second-order correction is complex. The remaining compact region contains no zeros because $\mathcal M_\upsilon\to b$ uniformly there. These local facts supply the count needed by the continuation argument.

\emph{Pole cancellation and confluent residuals.}
For independent simple-pole coefficients in $d$ and $n$, the first equation and the second equation multiplied by $w$ give the pole matrix
\begin{equation}
 \begin{pmatrix}
 (a^2+b^2-ak_c(\zeta ))W(\zeta )+bR(\zeta )&(k_c(\zeta )-a)W(\zeta )\\
 bk_c(\zeta )W(\zeta )+aR(\zeta )&-[bW(\zeta )+R(\zeta )]
 \end{pmatrix}.
 \label{total:rounded-residue-matrix}
\end{equation}
Its determinant is $-b[(R+bW)^2+\lambda^2W^2]$. It vanishes at every zero of $\Psi_{\rm r}=R+(b+i\lambda)W$, and there the direction $s$ lies in its kernel.

For a unit mode $d=w/(\zeta -\eta)$, $n=sw/(\zeta -\eta)$, put $Y_\zeta =-1+W(\zeta )/(\zeta -\eta)$. Using $s-a=-ib$, the first equation and $w$ times the second reduce, after real-part reconstruction, to the complex mode residuals
\begin{equation}
 \begin{aligned}
 \mathcal R_E(\zeta ,\eta)&=[b+i\lambda(\eta)]Y_\zeta +\frac{R(\eta)}{\zeta -\eta}
 =E(\zeta ,\eta)+\frac{\Psi_{\rm r}(\zeta )}{\zeta -\eta},\\
 \mathcal R_\mathcal T(\zeta ,\eta)&=\frac{(1-\eta^2)[\lambda(\eta)-ib]}{\zeta -\eta}
                  +iR(\eta)Y_\zeta 
 =\mathcal T(\zeta ,\eta)+\frac{iW(\zeta )\Psi_{\rm r}(\zeta )}{\zeta -\eta}.
 \end{aligned}
 \label{app:rounded-confluent-residual}
\end{equation}
For the first identity, subtract $R(\zeta )+[b+i\lambda(\zeta )]W(\zeta )$ from the numerator and use the polynomial divided differences. For the second, the numerator's value at $\eta=\zeta $ is
\begin{equation}
 (1-\zeta ^2)[\lambda(\zeta )-ib]+iR(\zeta )W(\zeta )=iW(\zeta )\Psi_{\rm r}(\zeta ).
\end{equation}
Subtracting this value and dividing by $\zeta -\eta$ gives the stated $\mathcal T(\zeta ,\eta)$. These are identities at every off-chord $\zeta $, before imposing the pole condition.

If $\zeta _j$ is a zero of order $\mathsf r_j$ of $\Psi_{\rm r}$, apply $\mathcal L_{\zeta ,\nu}$ to~\eqref{app:rounded-confluent-residual} and evaluate at $\zeta _j$, for $1\le\nu\le \mathsf r_j$. All differentiated remainder terms vanish because derivatives of $\Psi_{\rm r}$ of orders less than $\mathsf r_j$ vanish. The resulting polynomial residuals are $E_{j\nu}$ and $\mathcal T_{j\nu}$. Equation~\eqref{total:confluent-transforms} gives the corresponding higher-order Hilbert transforms, including the constant $-1$ only for order one. Here $F=F_{\rm r}$ and $wG=2U\cos\alpha\,R=G_{\rm r}$ in~\eqref{total:rounded-forcings}. Consequently the integral equations are equivalent within this ansatz to~\eqref{total:rounded-amplitudes}.

Polynomial division gives $\deg E_{j\nu}\le {\mathsf q}$ and $\deg \mathcal T_{j\nu}\le {\mathsf q}+1$, since pole-parameter derivatives do not increase these degrees. Matching coefficients gives~\eqref{total:rounded-real-system}.

Indeed, $\mathcal D_\zeta R$ has degree at most ${\mathsf m}-1$ and $\mathcal D_\zeta \lambda$ has degree at most ${\mathsf p}-1$. Hence $E$ has degree at most $\max({\mathsf m}-1,{\mathsf p})={\mathsf q}$. In $\mathcal T$, the only potentially higher degrees arise from $R$ and $(1-\eta^2)\mathcal D_\zeta \lambda$, bounded by ${\mathsf m}$ and ${\mathsf p}+1$; both are at most ${\mathsf q}+1$. The pole derivatives act on coefficients and the off-chord parameter only. As in the flat proof, the Leibniz expansion of each differentiated remainder contains a derivative of $\Psi_{\rm r}$ of order at most $\nu-1<\mathsf r_j$, and therefore vanishes. This proves the polynomial identities for every confluent order before any real coefficient matching is performed.

\emph{Full column rank and compatibility.} A vector in $\ker\bm M_{\rm r}$ reconstructs homogeneous jumps by the residual identities above. These jumps are $w$ times rational functions without poles on the closed chord, so that $d$ and $\ell$ belong to $Y$. The profile satisfies~\eqref{stab:rounded-class}, as verified under part~(iv) below, and part~(iii) of Theorem~\ref{thm:existence} shows that the homogeneous system has only the zero solution in $Y\times Y$. Hence $d=n=0$. Since $\Im s=-b\ne0$, $\Re\mathcal Z=\Re(s\mathcal Z)=0$ implies $\mathcal Z=0$. Uniqueness of rational principal parts then forces every amplitude to vanish. Thus $\bm M_{\rm r}$ has full column rank $2({\mathsf q}+1)$, including confluent modes.

To see the amplitude step in real variables, note that the zero jumps imply $\Re\mathcal Z=0$ and $\Re[(a-ib)\mathcal Z]=a\Re\mathcal Z+b\Im\mathcal Z=0$. Since $b>0$, also $\Im\mathcal Z=0$. The resulting rational identity has unique principal parts within the selected unsquared branch. Thus every real and imaginary amplitude is zero, including amplitudes of higher pole orders.

For the missing compatibility relation, evaluate~\eqref{app:rounded-confluent-residual} at $\eta=1$. Its left-hand side $\mathcal R_\mathcal T(\zeta ,1)$ is zero because $1-\eta^2=R(1)=0$. Hence
\begin{equation}
 \mathcal T(\zeta ,1)=-\frac{iW(\zeta )\Psi_{\rm r}(\zeta )}{\zeta -1}.
\end{equation}
Applying $\mathcal L_{\zeta ,\nu}$ and using the unsquared multiplicity at $\zeta _j$ gives $\mathcal T_{j\nu}(1)=0$. Consequently the image of the coefficient matrix lies in
\begin{equation}
 \mathcal S=\mathbb R[\eta]_{\le {\mathsf q}}\times
 \{\Pi\in\mathbb R[\eta]_{\le {\mathsf q}+1}:\Pi(1)=0\},\qquad
 \dim\mathcal S=2{\mathsf q}+2.
\end{equation}
The domain has the same dimension by the exact pole count. Full column rank therefore makes the coefficient map an isomorphism onto $\mathcal S$. The prescribed right-hand side belongs to $\mathcal S$: $\deg F_{\rm r}=\deg k_c\le {\mathsf q}$, $\deg G_{\rm r}=\deg R\le {\mathsf q}+1$, and $G_{\rm r}(1)=0$ because $R(1)=0$. Thus the system is consistent and has a unique amplitude vector. Dividing the second polynomial block by $\eta-1$ identifies this map with an invertible square system of order $2{\mathsf q}+2$.

There are ${\mathsf q}+1$ coefficients in the first block and ${\mathsf q}+2$ in the second, hence $2{\mathsf q}+3$ rows before the endpoint relation. Evaluation at $1$ is a nonzero linear functional on the second polynomial space, so imposing its zero value removes exactly one dimension. The full-rank image already has dimension $2{\mathsf q}+2$, the same as $\mathcal S$; it must equal that entire subspace. There can be no additional independent compatibility condition. Since $R(1)=0$, the second forcing polynomial is divisible by $\eta-1$, and division maps it, and each second-block column, bijectively into $\mathbb R[\eta]_{\le {\mathsf q}}$. This gives an explicit square implementation of the existence proof, with no least-squares consistency assumption. Off-chord poles make the principal parts bounded on the chord, giving $d,n=O(w)$ at both endpoints. This proves parts (i)--(iii), including repeated roots.

\emph{Part (iv).} Polynomial $k_c$ is Lipschitz continuous. By~\eqref{total:rounded-profile-slope}, $R(1)=-P(1)=0$, so $R=(1-\eta)R_1$ with a polynomial $R_1$, and $R_1(-1)=R(-1)/2=P(-1)/2>0$. Thus~\eqref{stab:rounded-class} holds. The jumps~\eqref{total:rounded-ansatz} are $w$ times functions bounded on the closed chord and therefore lie in $Y$, and $\ell$ does as well; by the uniqueness in part~(iii) of Theorem~\ref{thm:existence} they are that solution.
\end{proof}

\end{document}